%% file: paper.tex
\documentclass[conference]{IEEEtran}
\usepackage{graphicx}
\usepackage{amssymb}
\usepackage{pifont}
\usepackage{comment}
\usepackage{subfig}
\usepackage{algorithm}
\usepackage[noend]{algpseudocode}
\usepackage[dvipsnames]{xcolor}
\usepackage{comment}
\usepackage{framed}
\usepackage{balance}
\input{includes}
\input{macros}
\usepackage{float}
\usepackage{caption}

\input{pygments}
\usepackage{amsthm}
\usepackage[numbers]{natbib}
\newtheorem{definition}{Definition}[section]

\usepackage{amssymb}

\let\emptyset\varnothing

\begin{document}
\title{Effective Seed Scheduling for Fuzzing with \\ Graph Centrality Analysis} 
\author{\IEEEauthorblockN{Dongdong She, Abhishek Shah and 
Suman Jana}
\IEEEauthorblockA{Columbia University\\
\
}
}

\maketitle
\thispagestyle{plain}
\vspace{-5cm}

\pagestyle{plain}
\vspace{-10in}

\input{abstract.tex}
\input{intro.tex}
\input{background.tex}
\input{overview.tex}

\input{methodology.tex}

\input{implementation}
\input{evaluation.tex}
\input{related.tex}
\input{conclusion.tex}

\section*{Acknowledgements} 

We thank the anonymous reviewers for their constructive and valuable feedback. Abhishek Shah is supported by an NSF Graduate Fellowship. This work is sponsored in part by NSF grants CNS-18-42456, CNS-18-01426; a NSF CAREER award; a
Google Faculty Fellowship; a JP Morgan Faculty Fellowship; and a Capital One Research
Grant.

\bibliographystyle{plainnat}
\newpage
\bibliography{paper}
\input{appendix.tex}
\normalsize
\end{document}

%% file: includes.tex
\input{math_commands.tex}

\usepackage{hyperref}
\usepackage{cleveref}
\usepackage{url}
\usepackage{multirow}
\usepackage{booktabs}
\usepackage{amsmath}
\usepackage{listings}
\usepackage{graphicx}
\usepackage{amsmath}
\usepackage{xspace}
\usepackage{longfbox}

\newcommand{\rom}[1]{\uppercase\expandafter{\romannumeral #1\relax}}

\definecolor{javared}{rgb}{0.6,0,0} 
\definecolor{javagreen}{rgb}{0.25,0.5,0.35} 
\definecolor{javapurple}{rgb}{0.5,0,0.35} 
\definecolor{javadocblue}{rgb}{0.25,0.35,0.75} 

\lstdefinestyle{customc}{
  belowcaptionskip=\baselineskip,
  breaklines=true,
  xleftmargin=\parindent,
  language=java,
  showstringspaces=false,
  basicstyle=\footnotesize\ttfamily,
  keywordstyle=\bfseries\color{javapurple},
  commentstyle=\itshape\blue,
 numbers=left,
 numbersep=.1cm, 
 linewidth=\columnwidth,
 xleftmargin=2.2em,
 framexleftmargin=1.1em,
 frame=single,
 float=H, 
 aboveskip=\baselineskip
}

\lstdefinestyle{CStyle}{
    backgroundcolor=\color{backgroundColour},   
    commentstyle=\color{mGreen},
    keywordstyle=\color{magenta},
    numberstyle=\tiny\color{mGray},
    stringstyle=\color{mPurple},
    basicstyle=\footnotesize,
    breakatwhitespace=false,         
    breaklines=true,                 
    captionpos=b,                    
    keepspaces=true,                 
    numbers=left,                    
    numbersep=5pt,                  
    showspaces=false,                
    showstringspaces=false,
    showtabs=false,                  
    tabsize=2,
    language=C
}

\definecolor{codegreen}{rgb}{0,0.6,0}
\definecolor{codegray}{rgb}{0.5,0.5,0.5}
\definecolor{codepurple}{rgb}{0.58,0,0.82}
\definecolor{backcolour}{rgb}{0.95,0.95,0.92}

\lstdefinestyle{braystyle}{
  commentstyle=\Red,
  keywordstyle=\blue,
  numberstyle=\tiny\color{codegray},
  stringstyle=\color{codepurple},
  basicstyle=\scriptsize,
  breakatwhitespace=false,         
  breaklines=true,                 
  captionpos=b,                    
  keepspaces=true,                 
  numbers=left,                    
  numbersep=4pt,                  
  showspaces=false,                
  showstringspaces=false,
  showtabs=false,                  
  tabsize=2,
  belowskip=-3
  \baselineskip,
  frame=None,
  language=C,
  float=H
}

\lstdefinestyle{braystyle1}{
  commentstyle=\Red,
  keywordstyle=\blue,
  numberstyle=\tiny\color{codegray},
  stringstyle=\color{codepurple},
  basicstyle=\scriptsize,
  breakatwhitespace=false,         
  breaklines=true,                 
  captionpos=b,                    
  keepspaces=true,                 
  numbers=left,                    
  numbersep=4pt,                  
  showspaces=false,                
  showstringspaces=false,
  showtabs=false,                  
  tabsize=2,
  belowskip=-3
  \baselineskip,
  frame=single,
  language=C,
  float=H
}

%% file: math_commands.tex
\usepackage{amsmath,amsfonts,bm}

\def\eqref#1{equation~\ref{#1}}

\def\1{\bm{1}}

\DeclareMathAlphabet{\mathsfit}{\encodingdefault}{\sfdefault}{m}{sl}
\SetMathAlphabet{\mathsfit}{bold}{\encodingdefault}{\sfdefault}{bx}{n}



%% file: macros.tex
\makeatletter
\newcommand{\LINEFORALL}[3][default]{%
  \ALC@it\algorithmicforall\ #2\ %
  \ALC@com{#1}\ #3\ %
}
\makeatother

\newcommand*{\blue}[1]{\textcolor{blue}{#1}}

\usepackage{amsmath}

\renewcommand{\baselinestretch}{0.99}
\newcommand{\ToolNameShort}[0]{\texttt{K-Sched}}
\newcommand{\ToolName}[0]{\texttt{K-Scheduler}}

\newcommand{\CamReady}[1] {\textcolor{black}{#1}}

%% file: pygments.tex
\newcommand\Red[1]{\textcolor[rgb]{1.00,0.00,0.00}{\textbf{#1}}}

%% file: abstract.tex
\begin{abstract}
Seed scheduling, the order in which seeds are selected, can greatly affect the performance of a fuzzer. Existing approaches schedule seeds based on their historical mutation data, but ignore the structure of the underlying Control Flow Graph (CFG). Examining the CFG can help seed scheduling by revealing the potential edge coverage gain from mutating a seed. 

An ideal strategy will schedule seeds based on a count of all reachable and feasible edges from a seed through mutations, but computing feasibility along all edges is prohibitively expensive. Therefore, a seed scheduling strategy must approximate this count. We observe that an approximate count should have 3 properties \textemdash (i) it should increase if there are more edges reachable from a seed; (ii) it should decrease if mutation history information suggests an edge is hard to reach or is located far away from currently visited edges; and (iii) it should be efficient to compute over large CFGs. 

We observe that centrality measures from graph analysis naturally provide these three properties and therefore can efficiently approximate the likelihood of reaching unvisited edges by mutating a seed. We therefore build a graph called the edge horizon graph that connects seeds to their closest unvisited nodes and compute the seed node's centrality to measure the potential edge coverage gain from mutating a seed.

We implement our approach in \ToolName{} and compare with many popular seed scheduling strategies. We find that \ToolName{} increases feature coverage by 25.89\% compared to Entropic and edge coverage by 4.21\% compared to the next-best AFL-based seed scheduler, in arithmetic mean on 12 Google FuzzBench programs. It also finds 3 more previously-unknown bugs than the next-best AFL-based seed scheduler.

\end{abstract}

%% file: intro.tex
\section{Introduction}

Fuzzing is a popular security testing technique that has found numerous vulnerabilities in real-world programs~\cite{schumilo2017kafl, redqueen, angora, AFLplusplus-Woot20, mutationalFuzz, mopt, ankou, Wang2020NotAC, xufuzz, sensitive,zheng2019firm, Wang2020NotAC}. Fuzzers automatically search through the input space of a program for specific inputs that result in potentially exploitable buggy behaviors. However, the input spaces of most real-world programs are too large to explore exhaustively. Therefore, most existing fuzzers follow an edge-coverage-guided evolutionary approach for guiding the input generation process to ensure that the generated inputs explore different control flow edges of the target program~\cite{afl, libfuzzer, honggafuzz}. Starting from a seed input corpus, a coverage-guided fuzzer repeatedly selects a seed from the corpus, mutates it, and adds only those mutated inputs back to the corpus that generate new edge coverage. The performance of such fuzzers have been shown to heavily depend on seed scheduling, the order in which the seeds are selected for mutation~\cite{seed_select}. 

The main challenge in seed scheduling is to identify which seeds in a corpus, when mutated, are more likely to explore many new edges. Performing more mutations on such promising seeds can achieve higher edge coverage. Most prior work on seed scheduling identifies and prioritizes the promising seeds based on the historical distribution of edge/path coverage across prior mutations of the seeds. For example, a fuzzer can prioritize the seeds whose mutations, in the past, resulted in a higher path coverage~\cite{ecofuzz} or triggered rarer edges~\cite{lemieux2017fairfuzz}. However, these existing approaches ignore the structure of the underlying Control Flow Graph (CFG). For example, consider a seed \texttt{s1} whose execution path is close to many unvisited edges and a seed \texttt{s2} whose execution path is close to only one unvisited edge. Existing coverage-guided fuzzers might schedule seed \texttt{S2} before \texttt{S1} based on historical patterns. However, examining the structure of the CFG will reveal that \texttt{S1} is indeed more promising than \texttt{S2} as mutating it can potentially result in exploration of many unvisited edges that are close to the \texttt{S1}'s execution path.

The naive strategy of scheduling seeds simply based on the counts of all potentially reachable edges in the CFG for each seed is unlikely to be effective. Such a naive approach assumes that all CFG edges are equally likely to be reachable through mutations which does not hold true for most real-world programs. In fact, some shallow edges tend to be reachable by a large number of mutated inputs while other deep edges are only reached by a few, if any at all (as many branches might be infeasible)~\cite{tfuzz}. An ideal strategy would schedule seeds based on the count of all reachable and feasible edges from a seed by mutations. The seeds with higher edge counts will be mutated more. However, computing the feasibility along all edges is impractical as it will incur prohibitive computational cost. 

Therefore, a seed scheduling strategy must approximate the feasible edge count. We observe that such an approximation should have 3 properties. First, the approximate count should increase if there are many edges reachable from a seed. Second, the count should decrease if mutation history information suggests that an edge is hard to reach or is located far away from currently visited edges. Empirical evidence from prior work has shown that reaching child nodes through input mutations is typically harder than reaching parent nodes \cite{tfuzz} because the number of inputs that can reach a child, for a given path, is strictly less than or equal to the number of inputs that can reach the parent. Third, the approximate count must be efficient to compute for large CFGs as real-world CFGs can be quite large (e.g., inter-procedural CFGs might contain thousands of nodes).  

Our key observation is that {\it centrality} measures from {\it graph influence analysis} naturally provide the aforementioned properties while measuring a node's influence on the graph. Influence analysis is often used to identify a graph's (e.g., a social network's) most influential nodes and graph centrality measures each node's influence on other nodes with three properties as described below. 
First, centrality measures additively scale up a node's influence proportional to the number of edges that are reachable from the node. Each sequence of edges of the same length is treated equally independent of its order.
Second, centrality measures can easily incorporate external contribution (e.g., based on past mutation history) to a node's influence and can decay contributions from farther away nodes to the node's influence. Contributions decay multiplicatively with the increase in distance (i.e., more intermediate nodes) to reduce contributions from longer paths. 
Finally, centrality can be efficiently approximated on large graphs using iterative methods~\cite{Jacob2005}. 

In this paper, we introduce a new approach for seed scheduling based on centrality analysis of the seeds on the CFG. We prioritize scheduling seeds with the largest centrality, i.e., approximate counts of unvisited but potentially reachable CFG edges from a seed through mutations. To measure a seed's influence with centrality, we modify the CFG to construct an {\it edge horizon graph} containing the eponymous {\it horizon} nodes. The horizon nodes form the boundary between the visited and unvisited regions of the CFG for a given fuzzing corpus. 

Since horizon nodes delineate between the visited and unvisited regions of the CFG, we first classify CFG nodes as visited or unvisited based on the coverage of a fuzzer's current corpus. We then define horizon nodes as unvisited nodes with a visited parent node.
These nodes are crucial to fuzzing because a fuzzer must first visit a horizon node before going further into the unvisited region of the CFG. The centrality of horizon nodes reachable by mutations on a seed therefore measures the seed's ability to discover new edge coverage. Hence, we introduce one node corresponding to each seed and connect the nodes to their corresponding horizon nodes. 
We do not keep any visited node in the edge horizon graph to avoid inflating a seed's centrality score with contributions from already visited nodes. 

To compute centrality over the edge horizon graph, we use Katz centrality because it provides all the three desired approximation properties described earlier in this section and can operate on directed graphs like CFGs. 
We also use historical mutation data to bias the influence of horizon nodes to a value between 0 and 1 where values closer to 0 mean the node is harder to reach by mutations. The bias value 
estimates the hardness to reach a node by counting how many mutations reach a node's parents but fail to reach the node itself. 
Using the centrality scores for all seeds, a fuzzer can prioritize the seed with the highest centrality. We also periodically re-compute the edge horizon graph and centrality scores during a fuzzing campaign.
%

We implement our centrality-analysis-based seed scheduling technique as part of \ToolName{} (K stands for Katz centrality).  Our evaluation shows that \ToolName{} increases feature coverage by 25.89\% compared to Entropic and edge coverage by 4.21\% compared to the next-best AFL-based seed scheduler, in arithmetic mean on 12 Google FuzzBench programs. It also finds 3 more previously-unknown bugs than the next-best AFL-based seed scheduler.
We also conduct preliminary experiments to show the utility of \ToolName{} in non-fuzzing seed scheduling settings such as concolic execution and measure the impact of \ToolName{}'s design choices. 
Our main contributions are described below:

\begin{itemize}

\item We model seed scheduling in fuzzing as a graph centrality analysis problem.
\item We construct an edge horizon graph and use Katz centrality to compute centrality scores that approximate the number of reachable and feasible unvisited CFG edges from a seed.
\item We implement our approach in \ToolName{} and integrate it into Libfuzzer and AFL to show the generic utility of our approach. We release our implementation on \textcolor{blue}{\url{https://github.com/Dongdongshe/K-Scheduler}}. 
\item We demonstrate that using \ToolName{} increases feature coverage by 25.89\% compared to Entropic and edge coverage by 4.21\% compared to the next-best AFL-based seed scheduler, in arithmetic mean on 12 Google FuzzBench programs. It also finds 3 more previously-unknown bugs than the next-best AFL-based seed scheduler.

\end{itemize}







%% file: background.tex
\section{Graph Influence Analysis Background}
\label{background}


\subsection{Centrality Measures for Influence Analysis}
Identifying a graph's most influential nodes is a common and important task in graph analysis. Many different centrality  measures exist in the literature to estimate a node's influence~\cite{Newman2016}. For example, degree centrality measures a node's influence by counting its direct neighbors. This technique can identify a node with local influence over its neighbors. Eigenvector centrality, in contrast, can identify nodes with global influence over the entire graph. However, eigenvector centrality can fail to produce useful scores on directed graphs~\cite{LU20161, dynKatz}. Because program CFGs are directed graphs and we want to measure the global influence of a node to reach other nodes in a graph, we use Katz centrality, a variant of eigenvector centrality for directed graphs. We believe that Pagerank centrality, another eigenvector centrality variant, is not suitable for our setting because it dilutes node influence by the number of its direct neighbors. Such artificial dilutions will undesirably decrease a node's influence in a program's CFG. We conduct experiments to experimentally support this claim in Section \ref{evaluation}. 

For directed graphs like a program CFG, a node's neighbors can be defined by incoming or outgoing edges. Therefore, centrality measures are classified as out-degree if they use outgoing edges or in-degree if they use incoming edges during the computation. Their actual usage depend on the target domain. For example, academic citation graphs use in-degree centrality measures because influential papers are highly cited. In our setting, we use out-degree Katz centrality because we want to measure a node's ability to reach as many unvisited CFG edges (with respect to the current fuzzing corpus) as possible. We describe the details of the out-degree Katz centrality measure below.

\begin{figure}[!t]
\centering
\includegraphics[scale=0.5]{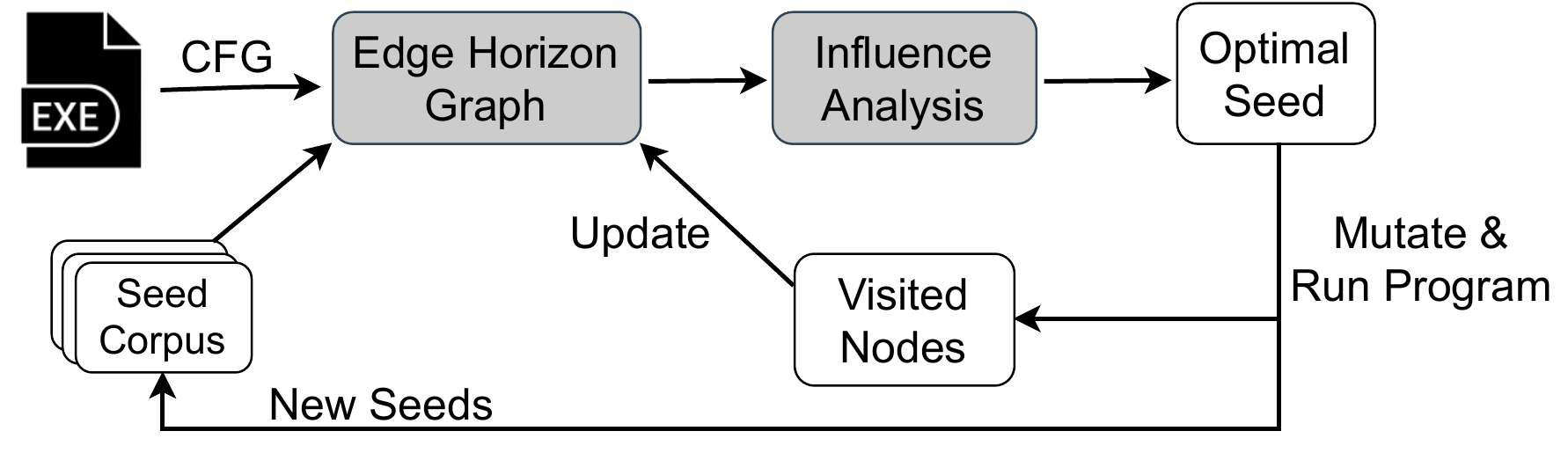}
\caption{\textbf{\small Fuzzer workflow with \ToolName{}.}}
\label{fig:overview}
\end{figure}


\subsection{Katz Centrality}

\input{overvie_fig}

Let $A$ denote an $n$ by $n$ adjacency matrix of a graph with $n$ nodes. If there is an edge connecting node i to node j, element $A_{ij} = 1$. Otherwise, $A_{ij} = 0$. Let $\mathbf{c}$ denote the Katz centrality vector of size $n$. The element corresponding to node $i$, $c_i$, is defined as follows,
\begin{equation}
    \label{katz2}
    c_i = \alpha\sum_{j=1}^{n}A_{ij}c_{j} + \beta_i
\end{equation}
where $\alpha \in [0, 1]$ and $\beta_i$ is the i-th element of $\boldsymbol{\beta}$, a vector of size $n$ consisting of non-negative elements. Conceptually, the left equation term captures that node centrality additively depends on its neighbors centrality and assigns each neighbor equal weight. 
Because the sum operator is commutative, the centrality score is independent of the order in which nodes are reached. 
The right term $\boldsymbol{\beta}$ captures the minimum centrality of a node, which we will later use in Section~\ref{methodology} to bias the centrality of horizon nodes based on historical mutation data. The $\alpha$ term represents the decay factor, so that long paths are weighted less than short paths as we show in Section~\ref{methodology}. 

In matrix form, equation~\ref{katz2} can be written as
\begin{equation}
    \label{katz3}
    \mathbf{c} = \alpha A \mathbf{c} + \boldsymbol{\beta}
\end{equation} 
To compute $\mathbf{c}$, the Katz centrality vector, one can solve the linear system so that 
\begin{equation}
    \label{katz4}
    \mathbf{c} = (I - \alpha A)^{-1} \boldsymbol{\beta}
\end{equation}
However, computing the matrix inverse in Equation~\ref{katz4} is prohibitively expensive with $O(n^3)$ complexity for large graphs. In practice, an iterative approach called the power method is used to approximate $\mathbf{c}$ based on Equation~\ref{katz3}. After initially setting $\mathbf{c}(0) = \mathbf{\beta}$, the power method computes the t-th iteration with the following formula,
\begin{equation}
    \label{power}
    \mathbf{c}(t) = \alpha A \mathbf{c}(t-1) + \boldsymbol{\beta}
\end{equation}
 where $\mathbf{c}(t)$ denotes the t-th iteration. 
Each iteration increases the power of matrix $A$ which corresponds to considering neighbors farther away. Hence, Katz centrality measures global node influence over the entire graph. Each iteration also reduces the contribution of farther away nodes to a node's influence as we describe in Section~\ref{methodology}.
The power method converges to the centrality vector in Equation \ref{katz4} with $O(n)$ complexity under some reasonable assumptions about the graph topology~\cite{dynKatz} such as $\alpha$ having to be less than the multiplicative inverse of the largest eigenvalue. We refer the reader to \cite{LU20161, dynKatz} for more technical details. 

%% file: overvie_fig.tex
\begin{figure*}[!t]
\centering
\begin{tabular}{cc}
\begin{minipage}{.23\linewidth}
\vspace{-1cm}
\lstset{basicstyle=\small\ttfamily,breaklines=true}
\begin{lstlisting} %[xleftmargin=1cm,frame=single]

a, b=read_input();
if(a > 20){
  return 1;
} 
else if(a > 10){
  if (b > 20)
    return 2;
  else if (b > 10)
    return 3;
  else
    return 4;
}  
else 
  return 5;
\end{lstlisting} 
\end{minipage}
&
\begin{minipage}{0.77\linewidth}
\centering
\subfloat[Program CFG.]{\includegraphics[width=0.3\linewidth]{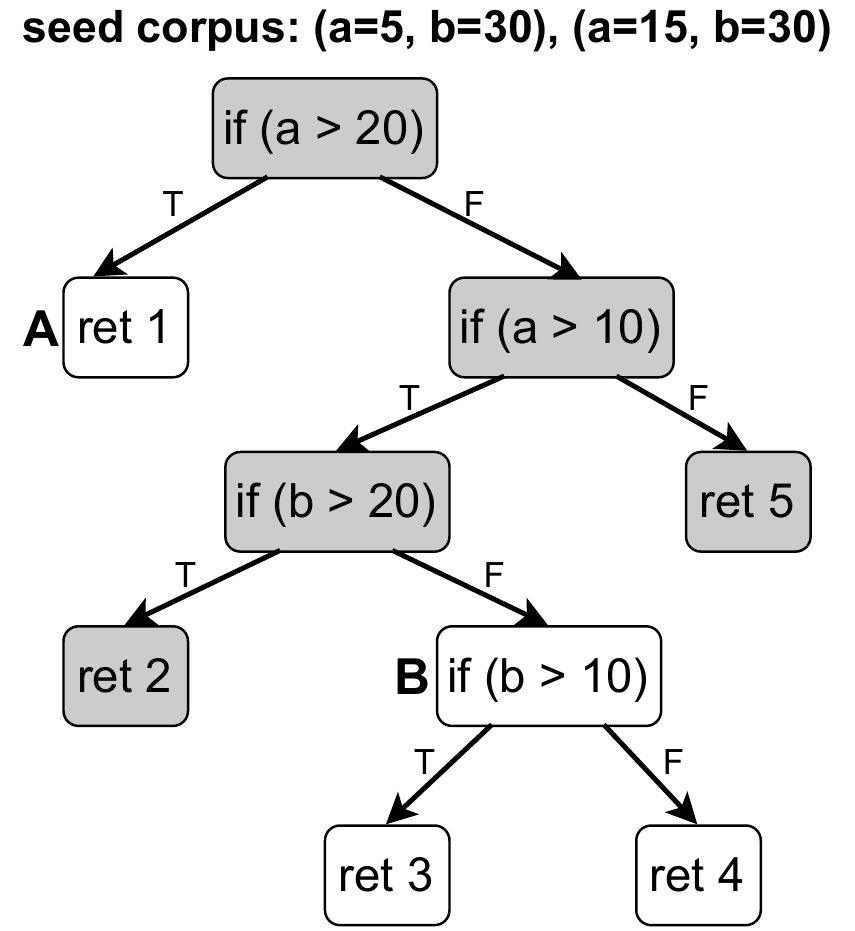}\label{subfig:code_cov}}
 \hspace{0.2cm}
\subfloat[Edge horizon graph]{\includegraphics[width=0.28\linewidth]{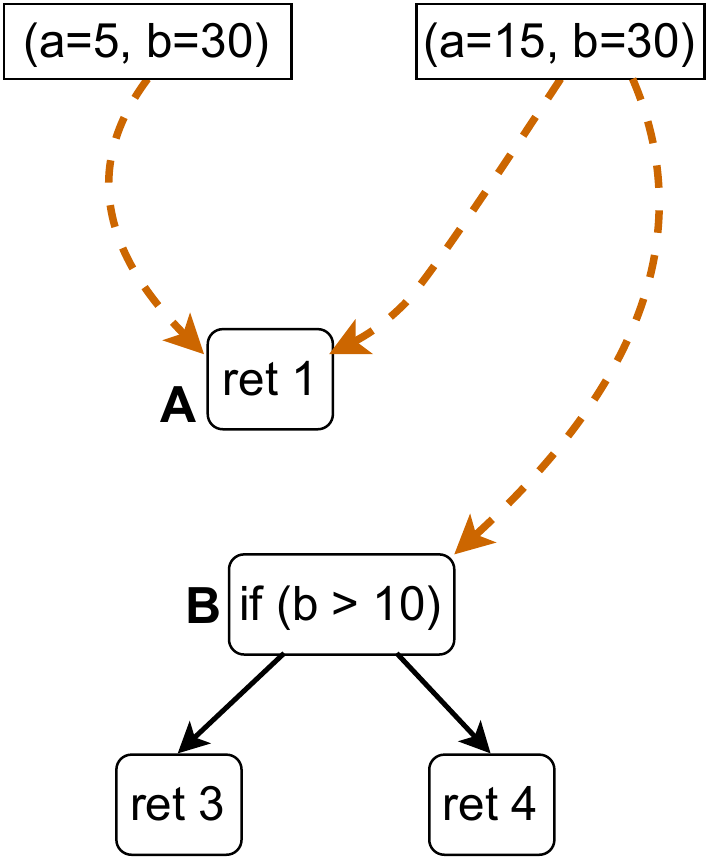}\label{subfig:seed1}}
 \hspace{0.2cm}
\subfloat[Computing Katz Centrality]{\includegraphics[width=0.32\linewidth]{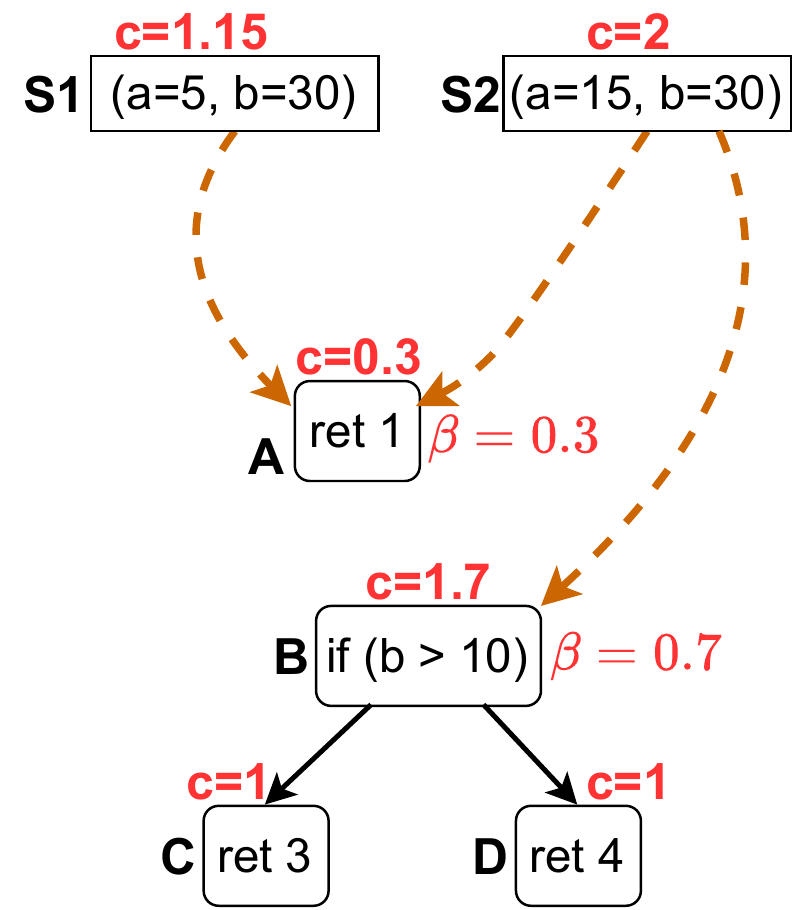}\label{subfig:seed2}}
\end{minipage}
\\
\end{tabular}
\caption{\textbf{\small  This figure shows how \ToolName{} is used for seed scheduling on a small program. Given the code example on the left, Figure \ref{subfig:code_cov} shows the corresponding CFG, colored as \textit{\textcolor{black}{gray}} if a node is visited and \textrm{white} if unvisited based on the fuzzer corpus. Figure \ref{subfig:seed1} shows the edge horizon graph. Figure~\ref{subfig:seed2} displays node Katz centrality scores computed by iterative power method illustrated in Table \ref{tab:movit}. A fuzzer will prioritize seed $(\texttt{a}=15, \texttt{b}=30)$ because it has the highest centrality score.
 }}
\label{fig:workflow}
\end{figure*}

%% file: overview.tex
\section{Overview of Our Approach}
\label{overview}

\medskip
\noindent
\textbf{Workflow.} Figure~\ref{fig:overview} depicts the workflow of \ToolName{}.
Given a program, seed corpus, and a target program's inter-procedural CFG, we modify the CFG to produce an edge horizon graph composed of only seed, horizon, and non-horizon unvisited nodes. We then use Katz centrality to perform centrality analysis on the edge horizon graph. A fuzzer prioritizes the seed with the highest centrality score. As a fuzzer's mutations reach previously unvisited nodes, we delete these newly visited nodes and re-compute Katz centrality on the updated edge horizon graph. 

\medskip
\noindent
\textbf{Motivating Example.} Figure~\ref{fig:workflow} shows a motivating example to explain our approach. The sample program (shown on the left) returns different values based on user input stored in variables \texttt{a} and \texttt{b}. Intuitively, we want to pick the seed node that can reach as many unvisited CFG edges as possible. In this case, this corresponds to seed node $(\texttt{a}=15, \texttt{b}=30)$. To do this, our approach \ToolName{} takes two steps. 

\medskip
\noindent
\textbf{Edge Horizon Graph.} First, we modify the CFG to build the edge horizon graph. We classify nodes in the program's CFG as visited or unvisited based on the coverage of a fuzzer's current corpus. 
Figure~\ref{subfig:code_cov} shows a classification of program's CFG nodes, where nodes in \textit{\textcolor{darkgray}{gray}} are visited and nodes in white are unvisited.
We next identify horizon nodes, which border the visited and unvisited CFG. In Figure~\ref{subfig:code_cov}, the horizon nodes are nodes \texttt{A} and \texttt{B} since they are unvisited nodes with a visited parent node.
We then insert seed nodes into the CFG and connect them to any horizon node whose parent is visited along the seed's execution path.
For example, seed $(\texttt{a}=5, \texttt{b}=30)$ takes both $False$ sides of the branch and hence its horizon node is node \texttt{A}. We connect this seed node to horizon node \texttt{A}. Finally, we delete all visited nodes in the CFG. Figure~\ref{subfig:seed1} shows the resulting edge horizon graph. 

\medskip
\noindent
\textbf{Katz centrality.} Second, we compute Katz centrality over the edge horizon graph.  We use the $\beta$ parameter in the centrality computation to estimate the hardness to reach a node by mutations. For this example, we assume that out of 100 mutations, $70$ reached the parent of horizon node \texttt{A}, so its $\beta = 1 - \frac{70}{100} = 0.3$ and 30 reached the parent of horizon node $B$, so its $\beta = 1 - \frac{30}{100} = 0.7$. This shows that horizon node \texttt{A} is harder to reach by mutations because a fuzzer failed to reach it with $70\%$ of its mutations. The remaining nodes default to $\beta=1$ as described in Section \ref{methodology}. Katz centrality also decays the contribution from further away nodes when computing a node's centrality with an $\alpha$ parameter. For this example, we assume $\alpha=0.5$.
\input{katz_iter_table}

\medskip
\noindent
\textbf{Detailed Katz centrality computation.} To see how Katz centrality is computed by the power method from Section \ref{background}, we show $\mathbf{c}(t=0), \mathbf{c}(t=1),...$ until it converges when $t=3$ in Table \ref{tab:movit}, where the rows indicate the centrality score for a node and the columns indicate time. To explain the intuition behind Katz centrality, we walk through the iteration for a single seed node \texttt{s2} to explain the computation. Initially, $c_{s2}(0) = 1$. Using Equation \ref{power} from Section \ref{background}, $c_{s2}(1) = \alpha(c_a(0)+c_b(0)) + \beta_{s2} = 0.5*(0.3+0.7) + 1 = 1.5$. Then, the next iteration is $c_{s2}(2) = \alpha(c_a(1)+c_b(1)) + \beta_{s2} = 0.5*(0.3+1.7) + 1 = 2$ and $c_{s2}(3) = c_{s2}(2)$ due to convergence. 
This computation illustrates how Katz centrality decays contributions from further away nodes. The number of edges reachable from \texttt{s2} is 4 but its Katz centrality score is 2 due to this decay. Moreover, the computation reflects that Katz centrality increases if there are more edges reachable from a node. Compared to \texttt{s2}, \texttt{s1} can only reach 1 edge and hence its centrality of $1.15$ is lower. 
Based on the results of Katz centrality, a fuzzer will prioritize seed $(\texttt{a}=15, \texttt{b}=30)$ because it has the highest centrality score among seed nodes.





%% file: katz_iter_table.tex
\begin{table}[!t]
    \small
    \caption{\small\textbf{Katz centrality computation by the iterative power method for the edge horizon graph in Figure \ref{subfig:seed2}. Each row corresponds to a node's centrality value and each column indicates the current iteration. The power method converges in 3 steps on this simple graph. Assume $\alpha=0.5$ and $\mathbf{\beta}=\mathbf{c}(0)$.}}
    \centering
    \renewcommand{\arraystretch}{1.1}
    \label{tab:movit}
\begin{tabular}{l|l|l|l|l}
\toprule
  & t=0 & t=1 & t=2 & t=3 \\
  \midrule
$c_a$ &  0.3 & 0.3 & 0.3  &  0.3 \\
$c_b$ &  0.7 & 1.7 & 1.7  &  1.7 \\
$c_c$&  1 & 1 & 1  & 1 \\
$c_d$ &  1 & 1 & 1  & 1 \\
$c_{s1}$ & 1 & 1.15  & 1.15  &1.15  \\
$c_{s2}$& 1 & 1.5 &  2 &  2\\
\bottomrule
\end{tabular}
\end{table}

%% file: methodology.tex
\section{Methodology}
\label{methodology}

In this section, we detail our approach to seed selection with influence analysis. We first describe how we build an edge horizon graph from a program's CFG and then how we compute Katz centrality  on the edge horizon graph. Lastly, we describe how our approach can be integrated into a coverage-guided fuzzer. 

\subsection{Edge Horizon Graph Construction}

\begin{figure}[!]
\centering
\includegraphics[scale=0.4]{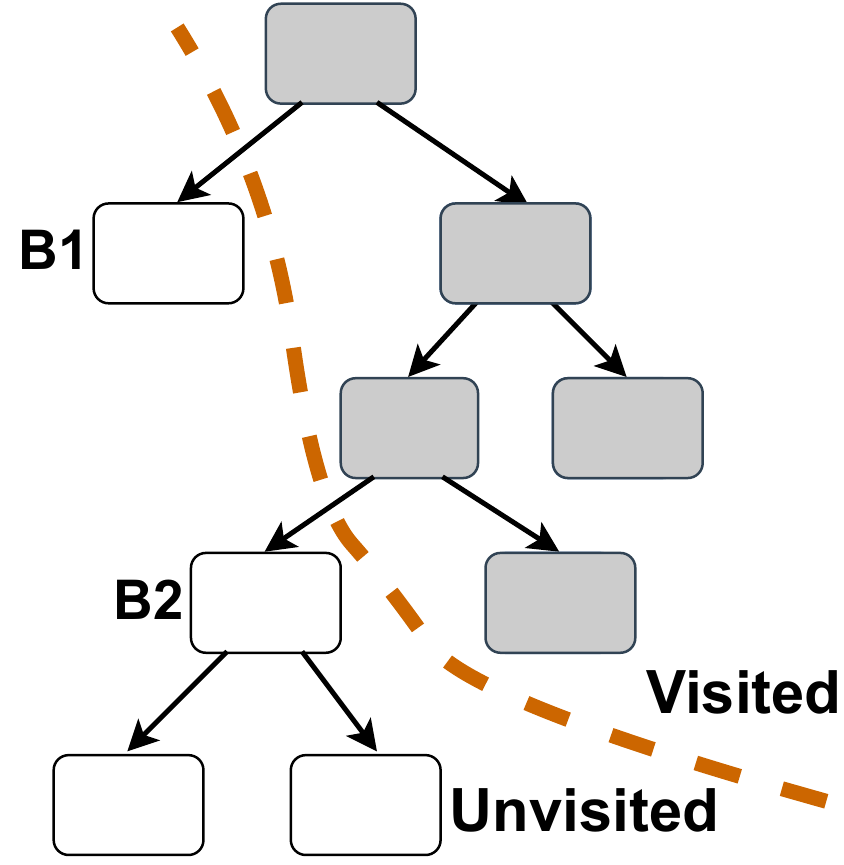}
\caption{\textbf{\small A target program's CFG with visited nodes colored in \textit{\textcolor{black}{gray}} and unvisited nodes colored \textrm{white}. The \textit{\textcolor{black}{dashed-brown}} line shows the boundary between the visited and unvisited regions of the CFG. Horizon nodes \texttt{B1} and \texttt{B2} sit at the border and are defined as unvisited nodes with a visited parent node.}}
\label{fig:cut}
\end{figure}


We construct the target program's directed inter-procedural control-flow graph $CFG=(N, E)$, where $N$ is the set of nodes representing the basic blocks and $E$ is the set of edges capturing control-flow transitions through branches, jumps, etc. In the rest of the paper, for clarity, we use CFG to refer to the inter-procedural CFG unless otherwise noted. 
\begin{algorithm}[!]
\footnotesize
\caption{Edge Horizon Graph Construction.} 
\label{alg:frontiergraph} 
\lstset{basicstyle=\ttfamily\footnotesize, breaklines=true}
\begin{tabular}{|lp{2.3in}|}\hline
\textbf{Input}:
    & \textit{$G$} $\leftarrow$ Inter-procedural CFG \\
    & \textit{$S$} $\leftarrow$ Seed corpus \\
    & \textit{$P$} $\leftarrow$ Program \\
\hline
\end{tabular}
\begin{algorithmic}[1] 
\State \textcolor{purple}{/* Classify Nodes as Visited/Unvisited */}
\State $V, U = \{\}, \{\}$
\For{$s \in S$}
    \State $visited\_nodes =$ \textsf{GetCoverage}$(P, s)$
    \State $V = V \cup visited\_nodes$ \Comment{\textcolor{purple}{Union $visited\_nodes$ with $V$}}
\EndFor
\State $U = G.nodes \setminus V $ \Comment{\textcolor{purple}{Compute the complement set of $V$}}

\State 
\State \textcolor{purple}{/* Identify Horizon Nodes */}
\State $H = \{\}$ 
\For{$u \in U$}
    \For{$p \in u.parents$}  
        \If{ $p \in V$}
            \State $H = H \cup u$ \Comment{\textcolor{purple}{Union $u$ with $H$}}
        \EndIf
    \EndFor
\EndFor
\State
\State  \textcolor{purple}{/* Insert Seed Nodes */}
\For{$s \in S$}
    \State $seed\_node = G.$\textsf{AddNode}$(s)$
    \State $visited\_nodes = $ \textsf{GetCoverage}$(P, s)$
    \For{$v \in visited\_nodes$} 
        \For{$n \in v.children$} 
            \If{ $n \in H$}
                \State $G.$\textsf{AddEdge}$(seed\_node, n)$
            \EndIf
        \EndFor
    \EndFor
\EndFor
\State
\For{$v \in V$}
    \State $G.$\textsf{RemoveNode}$(v)$ \Comment{\textcolor{purple}{Remove visited nodes}}
\EndFor
\State $G.$\textsf{RemoveLoops}$()$ \Comment{\textcolor{purple}{Convert $G$ to directed acyclic graph}}
\end{algorithmic}
\end{algorithm}
Directly computing centrality over the original CFG is not useful for seed selection because the graph lacks any reference to seed nodes. Hence, we modify the CFG to construct an edge horizon graph that contains seed nodes. We can then compute a seed's centrality for seed selection. 
At a high level, we classify original CFG nodes as visited or unvisited and connect newly-inserted seed nodes to their corresponding horizon nodes, which are unvisited nodes with a visited parent node. Such connections ensure that a seed's centrality measures its ability to discover new edge coverage. We also delete visited nodes from the CFG to avoid their contributions increasing a seed's centrality score. Lastly, we convert the CFG to a directed acyclic graph to mitigate the undesirable effects of loops on centrality. 
We present the algorithm for constructing the edge horizon graph in Algorithm \ref{alg:frontiergraph} and discuss each step in more detail below. 

\medskip
\noindent
\textbf{Classifying Nodes as Visited or Unvisited.}
We first classify all CFG nodes as visited or unvisited based on the coverage of a fuzzer's current corpus. A CFG node is visited if it is reached by the execution path of any seed in the corpus, or elsewise unvisited. 
We denote the set of visited nodes as $V$ and the set of unvisited nodes as $U$.
More formally, 
\begin{equation}
V=\{n | n \in N, visited(n) = 1\}
\end{equation}
\begin{equation}
U=\{n | n \in N, visited(n) = 0\}
\end{equation}
Lines 1 to 6 in Algorithm \ref{alg:frontiergraph} detail the classification process. Figure~\ref{fig:cut} colors visited nodes in gray and unvisited nodes in white based on the fuzzer's current corpus. 

\medskip
\noindent
\textbf{Identifying Horizon Nodes.}
We define a horizon node in terms of the prior graph partition of $V$ and $U$, the visited and unvisited nodes as shown below.  
\begin{equation}
H=\{u | (v, u) \in E, v \in V, u \in U\}
\end{equation}
In other words, a horizon node is an unvisited node with a visited parent node.
Conceptually, horizon nodes border the unvisited and visited region between $V$ and $U$. Figure~\ref{fig:cut} shows how horizon nodes \texttt{B1} and \texttt{B2} border the unvisited and visited regions of the CFG. Algorithm~\ref{alg:frontiergraph} computes this set of horizon nodes in lines 8-13. 
Horizon nodes are \emph{crucial} for fuzzing because a fuzzer must first reach a horizon node to increase edge coverage. 
This property can be seen in Figure~\ref{fig:cut} where a fuzzer must first reach horizon node \texttt{B1} or \texttt{B2} to discover new edge coverage. Therefore, a horizon node's centrality measures the number of edges that can potentially be reached by mutations after visiting a horizon node. 


Not all horizon nodes, however, have equal centrality. Some horizon nodes can increase edge coverage more than others. As shown in Figure~\ref{fig:cut}, horizon node \texttt{B2} reaches more edges in $U$ than horizon node \texttt{B1}. Hence, a fuzzer should prioritize seeds close to horizon node \texttt{B2} because \texttt{B2} can reach more edges in the unvisited CFG. 




\medskip
\noindent
\textbf{Inserting Seed Nodes.}
For each seed, we insert one node into the edge horizon graph and connect this seed node to a horizon node if the horizon node's parent is visited along the seed's execution path.
Lines 15 to 22 from Algorithm~\ref{alg:frontiergraph} specify how seed nodes are connected to horizon nodes and Figure~\ref{subfig:seed1} visualizes the connection between seed nodes and horizon nodes. 
Connecting seed nodes to their corresponding horizon nodes ensures that a seed node's centrality is the sum of its horizon nodes centrality (i.e., Equation \ref{katz2}). 
Therefore, a seed's centrality measures its ability to discover new edge coverage through mutations. 



\medskip
\noindent
\textbf{Deleting Visited Nodes.}
We delete visited nodes from the edge horizon graph because we do not want a seed's centrality score to include contributions from already visited nodes.
Note that we preserve the connectivity of the CFG when deleting visited nodes. For example, given a graph $\texttt{A} \rightarrow \texttt{B} \rightarrow \texttt{C}$, if $\texttt{$B$}$ was visited, we preserve the connectivity by adding an edge producing $\texttt{A} \rightarrow \texttt{C}$ . Although this deletion changes the distance between nodes (i.e., $A \rightarrow C$ now has distance 1), it does preserve the connectivity, which is the most critical when measuring centrality. 


\begin{figure}[!]
\centering
\subfloat[Original CFG]{%
  \includegraphics[width=0.35\columnwidth]{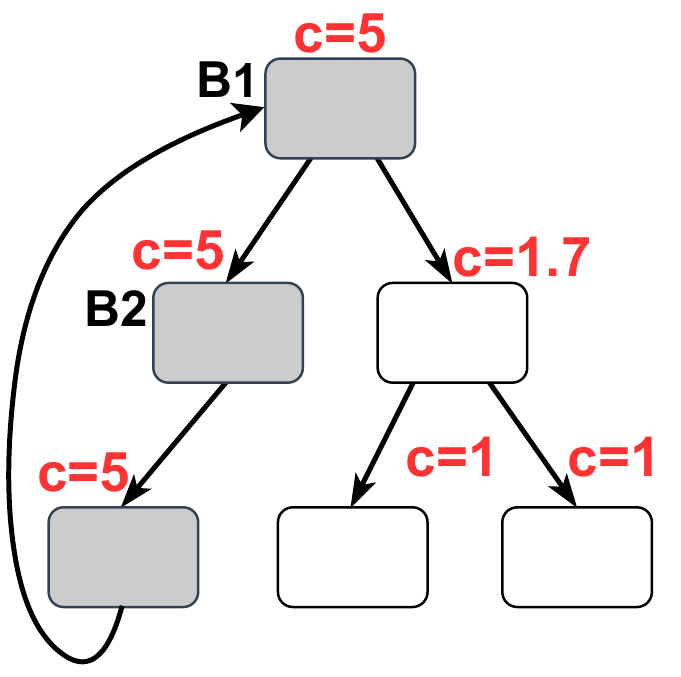}%
  \label{fig:loop_a}%
}\hspace{1em}
\subfloat[Transformed CFG]{%
  \includegraphics[width=0.35\columnwidth]{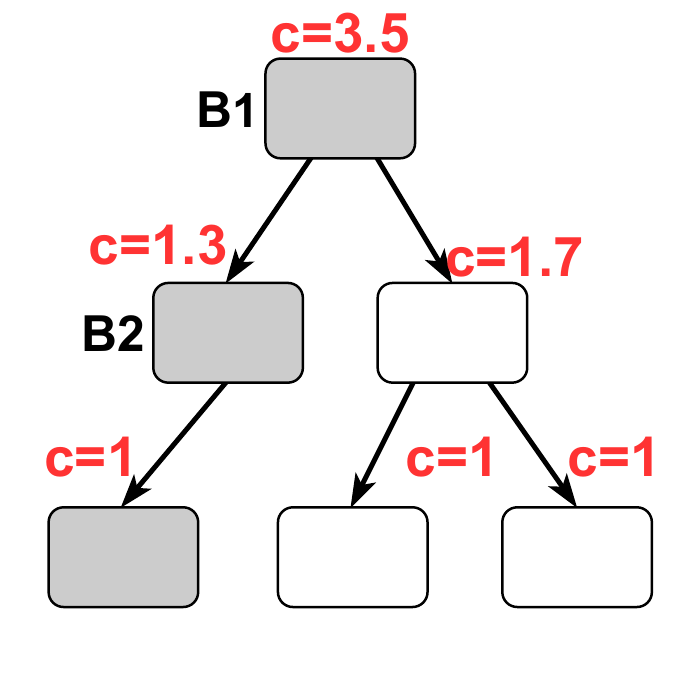}%
  \label{fig:loop_b}%
}
\caption{\textbf{\small{Figure \ref{fig:loop_a} shows that node \texttt{B1} has the same centrality as node \texttt{B2} as an artifact of the loop. However, \texttt{B1} should have higher centrality than \texttt{B2} because it can reach more edges. To resolve this, we remove loops from the CFG and Figure \ref{fig:loop_b} shows the graph after this transformation. }}}
\end{figure}

\medskip
\noindent
\textbf{Mitigating the effect of loops on centrality.} 
Loops in a CFG can hurt the utility of a seed's centrality score for seed selection. Figure~\ref{fig:loop_a} shows a level loop where node \texttt{B1} and its child node \texttt{B2} are assigned equal scores by a centrality analysis. However, nodes that initiate a loop should have more centrality than nodes in the loop body. In this case, the node that initiates the loop \texttt{B1} should have higher centrality because it can reach more edges. To solve this problem, we convert the CFG to a directed acyclic graph by removing loops in the CFG. Such loops originate in program constructs such as \texttt{while} or \texttt{for} statements as well as connections between caller and callee nodes (i.e., caller to callee edge and callee to caller backedge can form a loop). 

\subsection{Influence Analysis}
To compute a seed's centrality, we could count the number of potentially reachable edges from a seed node in the edge horizon graph. However, this count assumes that all edges independent of distance are equally reachable and feasible which does not hold true for most real-world programs \cite{qsym, tfuzz}. Ideally, we want to count all feasible and reachable edges from a seed through mutations, but this is impractical to compute as it requires computing feasibility along all edges. Instead we use Katz centrality to approximate this count. Katz centrality provides three properties that make it a natural fit to approximate this count. First, it increases its approximation additively if more edges can be reached from a seed node independent of the order as described in Section \ref{background}. Second, Katz centrality decreases its approximation if mutation frequency information suggests an edge is hard to reach or if edges are far away. Third, Katz centrality is efficient to compute with the power method as discussed in Section \ref{background}. 
Below, we explain how we set the mutation frequency information  mechanism in Katz centrality and why Katz centrality multiplicatively decays contributions from further-away edges.



\medskip
\noindent
\textbf{Using historical mutation data as a bias.} 
We observe that $\mathbf{\beta}$ is a generic way of biasing a node's centrality based on external information. We therefore use $\mathbf{\beta}$ to lower a node's centrality if a node appears harder to reach by mutations. 
We set each element of $\mathbf{\beta}$ to range from 0 to 1, where values closer to 0 mean the node is harder to reach through mutations. 
To measure this hardness, we use historical mutation data. We initialize $\mathbf{\beta}=1$ if there are no mutations and iteratively refine it as a fuzzer generates mutations. 
We use the following equation for node i,
\begin{equation}
    \label{beta}
    \beta_{i} = 1 - \frac{R_i}{T}
\end{equation}
where $R_i$ measures the number of mutations that reach node i's parents
and $T$ measures the total number of mutations for all seeds. 

Lastly, to set $\alpha$, which ranges from 0 to 1, from Equation \ref{powerunroll}, we observe that setting $\alpha=0$ means all nodes in the edge horizon graph will have the same centrality. This would not be useful for seed selection because we could not distinguish which seed node was more likely to discover new edge coverage with its centrality score. In contrast, setting $\alpha=1$ treats closer and further-away edges with equal contribution, which fails to reflect program behavior. In practice, we set $\alpha=0.5$ based on our experiments as described in Section~\ref{evaluation}.




\medskip
\noindent
\textbf{Decaying contributions from longer paths.}
Katz centrality multiplicatively decays the contribution from further away edges when computing a node's centrality . This decay corresponds to a well-known program behavior where further away edges are harder to reach by mutations~\cite{tfuzz}. We also verify this behavior with our own experiments in Appendix \ref{appd:far}. 
To see how Katz centrality reduces the contribution from further-away edges toward a node's centrality, consider Equation \ref{powerunroll} which shows the 2nd iteration of the power method from Section \ref{background}. 
\begin{equation}
    \label{powerunroll}
    \mathbf{c}(2) = ((\alpha)^0 I + (\alpha)^1 A + (\alpha)^2 A^2)\mathbf{\beta}
\end{equation}
Notice how the parameter $\alpha$, which ranges between 0 and 1, multiplicatively decays the contribution from higher matrix powers. As discussed in Section \ref{background}, higher matrix powers consider edges farther away. Thus, this equation shows Katz centrality reduces the contribution from further away edges with multiplicative decay.


\subsection{Seed Scheduling}
Algorithm \ref{alg:fuzzer} shows how to integrate \ToolName{} into a coverage-guided fuzzer. \ToolName{} first builds the edge horizon graph as shown in Algorithm~\ref{alg:frontiergraph} and computes the Katz centrality over it to measure each seed's centrality. A fuzzer then uses these scores for seed scheduling which consists of selecting a seed and allocating a corresponding mutation budget. Because popular fuzzers such as AFL and LibFuzzer differ greatly in these two components, we abstract them out in lines 10 and 11 and specify how to integrate our generic technique into them in Section~\ref{implementation}. 
Finally, \ToolName{} re-computes the edge horizon graph and its Katz centrality when the fuzzer discovers new edge coverage or a fixed time has elapsed. Periodically updating centrality (i.e. via $\beta$) ensures that \ToolName{} provides useful guidance even when a fuzzer fails to find new edge coverage.
\setlength{\textfloatsep}{0.5pt}
\begin{algorithm}
\footnotesize
\caption{Fuzzer integration with \ToolName{}.} 
\label{alg:fuzzer} 
\lstset{basicstyle=\ttfamily\footnotesize, breaklines=true}
\begin{tabular}{|lp{2.6in}|}\hline
\textbf{Input}:
    & \textit{$G$} $\leftarrow$ Inter-procedural CFG \\
    & \textit{$S$} $\leftarrow$ Seed corpus \\
    & \textit{$P$} $\leftarrow$ Program \\
\hline
\end{tabular}

\begin{algorithmic}[1]
\State $stats = \{\}$ \Comment{\textcolor{purple}{Store mutation statistics}}
\State $has\_new = False$ \Comment{\textcolor{purple}{Indicate new edge coverage}} 
\State $t = $ \textsf{CreateTimer}$(k)$ \Comment{\textcolor{purple}{Build horizon graph every $k$ seconds}}
\While{fuzzer is running}
    \If{ $has\_new = True$ \textbf{or} $stats = \emptyset$ \textbf{or} $t.$\textsf{timeout}$()$}
        \State $H = $ \textsf{GetHorizonNodes}$(G, S, P)$
        \State $Beta = $ \textsf{ComputeBeta}$(H,  stats)$ 
        \State $G_{horizon} =$ \textsf{GetHorizonGraph}$(G, S, P)$
        \State $C_{katz} = $ \textsf{KatzCentrality}$(G_{horizon}, Beta)$ 
        \State $t.$\textsf{reset}$()$ \Comment{\textcolor{purple}{Reset timer $t$}}
    \EndIf
    \State $seed = $ \textsf{ChooseSeed}$(S, C_{katz})$ 
    \State $energy = $ \textsf{ComputeEnergy}$(seed, C_{katz})$ 
    \State $has\_new = $ \textsf{Mutate}$(seed, energy)$ \Comment{\textcolor{purple}{Fuzz $seed$ with $energy$}}
    \State $stats.$\textsf{update}$()$
\EndWhile
\end{algorithmic}
\end{algorithm}

%% file: implementation.tex
\section{Implementation}
\label{implementation}
\ToolName{} consists of two components. First, to build the edge horizon graph, we construct the target program's inter-procedural CFG. We initially compile the program with \texttt{wllvm} ~\cite{wllvm} and use the LLVM's (version 11.0.1) \texttt{opt} tool to extract each function's intra-procedural CFG. In Python 3.7, we then merge each intra-procedural CFG together based on caller-callee relations to produce the inter-procedural CFG. We also implement all pieces from Algorithm~\ref{alg:frontiergraph} such as loop removal in Python. To classify CFG nodes as visited, we re-use a fuzzer's edge coverage information to identify visited basic blocks. Second, to compute Katz centrality, we use the power method provided by networkit~\cite{networkit}, a large-scale graph computing library.

We now describe how we integrate \ToolName{} into LibFuzzer~\cite{libfuzzer} and AFL~\cite{afl} to show our technique is generic and widely applicable. We run \ToolName{} as a standalone process that communicates with a fuzzer to set the fuzzer's seed ranking based on centrality and identify the mapping between a seed node and its corresponding horizon nodes. We measure how much overhead \ToolName{} adds to the fuzzing process in Section \ref{evaluation}.

\medskip
\noindent
\textbf{Libfuzzer Integration.} Libfuzzer~\cite{libfuzzer} computes an energy for each seed in the form of a probability and flips a coin with bias corresponding to the seed's energy to determine whether a seed should be selected for mutation. Higher energy probabilities indicate a seed will be chosen more frequently. To integrate into Libfuzzer, we follow the same integration as Entropic, a state-of-the-art seed scheduler for Libfuzzer, and set each seed's energy to its Katz centrality score normalized by the total centrality scores for all seeds.

\medskip
\noindent
\textbf{AFL Integration.}
Unlike Libfuzzer's probabilistic seed selection, AFL generally selects every seed for mutation. A seed's energy also determines its corresponding mutation budget. To integrate into AFL, we set each seed's energy directly to its Katz centrality score. 

%% file: evaluation.tex
\section{Evaluation}
\label{evaluation}
Our evaluation aims to answer the following questions. 
\begin{enumerate}
    \item \textbf{Comparison against seed schedulers:} How does \ToolName{} compare against other seed scheduling strategies?
    \item \textbf{Bug Finding:} Does \ToolName{} improve a fuzzer's ability to find bugs?
    
    \item \textbf{Runtime Overhead:} What is the performance overhead of \ToolName{}?
    \item \textbf{Impact of Design Choices:} How do \ToolName{}'s various design choices contribute to its performance?
    
    \item \textbf{Non-evolutionary fuzzing settings:} Does \ToolName{} show promise for seed scheduling in non-evolutionary fuzzing settings? 
\end{enumerate}

\subsection{Experimental Setup}

\subsubsection{Baseline Seed Scheduling Strategies}

We compare against popular seed scheduling strategies from industry and the academic community. These strategies are generally integrated into AFL or Libfuzzer. Directly comparing a seed scheduling strategy that uses AFL with another seed scheduling strategy that uses Libfuzzer can be misleading since the underlying fuzzers may cause the performance difference instead of the underlying seed scheduling strategy. Therefore, to be fair, we integrate \ToolName{} into both Libfuzzer and AFL separately and make comparisons about seed scheduling strategies when the underlying fuzzer is the same. Note this integration also demonstrates that \ToolName{} is generic and widely applicable.

For \ToolName{}’s comparison against Libfuzzer-based seed schedulers, we compare \ToolName{} against Entropic, a state-of-the-art seed scheduler in Libfuzzer~\cite{entropic}. 
To ensure a fair comparsion, we follow the same integration with Libfuzzer as Entropic. We also compare against Libfuzzer's default seed scheduler as a baseline and refer to it as Default. We use Libfuzzer and Entropic from LLVM 11.0.1 in our comparison.
For \ToolName{}’s comparison against AFL-based seed schedulers, we compare against strategies that prioritize seeds if they take paths rarely observed (RarePath), reach rarely observed edges (RareEdge) or discover new paths (NewPath). We also compare against a strategy that prioritizes seeds based on security-sensitive coverage (SecCov). To compare against RarePath, RareEdge, NewPath, and SecCov we use AFLFast~\cite{aflfast}, FairFuzz~\cite{lemieux2017fairfuzz}, EcoFuzz~\cite{ecofuzz}, and TortoiseFuzz~\cite{Wang2020NotAC} respectively. Since these fuzzers all modify AFL, we integrate \ToolName{} into AFL using their same modifications for a fair comparison. Moreover, we set each fuzzer to use the same mutation strategy to a enable a fair comparison. Hence, we disabled FairFuzz's custom mutation strategy. We also compare against AFL's default seed scheduling strategy as a baseline and refer to it as Default.

\subsubsection{Benchmark Programs}
In our seed scheduler comparison, we use the Google FuzzBench benchmark, a commonly used dataset to evaluate fuzzing performance on real-world programs. At the time of this writing, the benchmark consists of 40+ programs, so we decide to evaluate over a subset of them. We pick 12 diverse real-world programs from the benchmark that includes cryptographic and database programs as well as parsers as shown in Table~\ref{tab:libfuzzer_cov}. We plan to evaluate against the entire benchmark in the future. We also use the default seed corpus and configuration provided by the benchmark to enable a fair comparison. Note that Google FuzzBench configures all AFL-based fuzzers to use havoc mode by default~\cite{FuzzBench}, since AFL havoc mode has been shown to significantly outperform AFL deterministic mode~\cite{wu2022one}. 


For our bug-finding experiments, we select 12 real-world parsing programs commonly used to evaluate fuzzer's bug finding performance ~\cite{aflfast, lemieux2017fairfuzz, ecofuzz}. The $12$ programs cover $8$ file formats: \texttt{ELF, ZIP, PNG, JPEG, TIFF, TAR, TEXT and XML}. The list of programs and their details can be found in Table~\ref{tab:studied_programs}. Since these programs do not come with a default seed corpus, we make a corpus with small valid files.


\subsubsection{Environmental Setup}
We run all our evaluations on $4$ 64-bit machines running Ubuntu 20.04 with Intel Xeon E5-2623 CPUs (96 cores in total). We follow standard operating procedure in fuzzing evaluations~\cite{aflfast, entropic, lemieux2017fairfuzz} and bound each fuzzer to 1 CPU core. Because our current implementation runs \ToolName{} in a separate process, we assign fuzzers using \ToolName{} 2 cores, one for the fuzzer and one for the \ToolName{}.

\subsection{RQ1: Seed scheduling comparison}

For \ToolName{}’s comparison against Libfuzzer-based seed schedulers, we follow the original evaluation of Entropic~\cite{entropic} and use the same two metrics for comparison: edge coverage and feature coverage. Edge coverage measures how many branches were reached along an input's execution path, whereas feature coverage includes this information as well as branch hit count. For example, edge coverage would not distinguish coverage between two inputs that visit the same branch a different number of times, but feature coverage would distinguish them. 


We run \ToolName{}, Default (i.e., Libfuzzer's default seed scheduler), and Entropic on the 12 Google FuzzBench programs for 24 hours. We repeat each 24 hour run ten times for statistical power. In arithmetic mean over these 10 runs, Table ~\ref{tab:libfuzzer_cov_1h} and Table~\ref{tab:libfuzzer_cov} summarize the edge and feature coverage results for 1 hour and 24 hours, respectively. Appendix Table~\ref{tab:libfuzzer_covp_1h} and Table~\ref{tab:libfuzzer_covp} show the corresponding result from applying the Mann Whitney U test between \ToolName{} and the tested seed schedulers in terms of edge and feature coverage.
Within 1 hour, \ToolName{} improves upon next-best seed scheduling strategy Entropic by $20.11\%$ in median and $31.75\%$ in arithmetic mean over the 12 FuzzBench programs in feature coverage. For the 24 hour runs, \ToolName{} achieves $20.66\%$ in median and $25.89\%$ in arithmetic mean more feature coverage than Entropic. We attribute the increased improvement of \ToolName{} over Entropic within the first hour to \ToolName{}'s scheduling of promising seeds more frequently given a limited fuzzing budget (i.e., fuzzer only schedules a limited number of seeds). However, as the fuzzing budget increases to 24 hours, Entropic will eventually also schedule those promising seeds more frequently, which narrows the performance difference between \ToolName{}. Moreover, with a significance level of 0.05, our feature coverage over Entropic results are statistically significant for all programs for 24 hour runs and all programs except \texttt{zlib} for the 1 hour runs. Our results show that using the CFG structure for seed scheduling can improve fuzzing performance.
\begin{table}[!t]
    \caption{\small\textbf{\CamReady{Arithmetic mean} feature and edge coverage of Libfuzzer-based seed schedulers on 12 FuzzBench programs for 1 hour over 10 runs. We mark the highest number in bold. }}
    \centering
    \setlength{\tabcolsep}{2pt}
    \renewcommand{\arraystretch}{1.1}
   \begin{tabular}{lrr|rr|rr}
   \toprule
    \multicolumn{1}{c}{\multirow{2}{*}{\textbf{Programs}}} & \multicolumn{2}{c}{\textbf{\ToolName{}}} & \multicolumn{2}{c}{\textbf{Entropic}} & \multicolumn{2}{c}{\textbf{Default}} \\ \cline{2-7} 
\multicolumn{1}{c}{}& \multicolumn{1}{c}{\textbf {feature}} & \textbf{edge}     & \multicolumn{1}{c}{\textbf {feature}} & \textbf{edge}  & \multicolumn{1}{c}{\textbf {feature}} & \textbf{edge}    \\
        \midrule
        freetype & \textbf{51,184} & \textbf{10,886} &  46,698 & 10,691 & 40,040 & 9,446 \\
        libxml2 &  \textbf{39,240}  & \textbf{7,661} & 24,167 & 6,128  & 25,914 & 6,296 \\
        lcms & \textbf{2,886} & \textbf{1,497} &  1,707  & 1,004  & 1,392 & 874  \\
        harfbuzz & \textbf{35,017} & \textbf{9,112} & 23,349 & 7,551 & 23,455 & 7,588  \\
        libjpeg & \textbf{10,974} & \textbf{2,553} &  7,424 & 2,193 & 7,510 & 2,208 \\
        libpng & \textbf{5,001} & \textbf{1,501} & 4,604 & 1,469 & 4,525 & 1,476 \\
        openssl & \textbf{14,520} & \textbf{4,622} & 12,830 & 4,294 & 13,029 & 4,327 \\
        openthread & \textbf{6,525} & \textbf{3,318} & 5,397 & 3,044 & 5,150 & 2,947 \\
        re2 &\textbf{31,292} & \textbf{6,275} & 28,877 & 6,147 & 29,941 & 6,207 \\
        sqlite & \textbf{73,532} & \textbf{13,299} & 44,198 & 12,189 & 52,060 & 12,735 \\
        vorbis & \textbf{9,106} & \textbf{2,136} & 7,632 & 2,010 & 5,710 & 1,823 \\
        zlib & \textbf{2,711} & \textbf{790} & 2,572 & 784 & 2,408 & 782 \\     
        \midrule
        \multicolumn{3}{c|}{Arithmetic mean coverage gain} &  31.75\% & 12.51\% & 37.37\%  & 15.72\% \\
        \multicolumn{3}{c|}{Median coverage gain} & 20.11\%  & 8.32\% & 34.54\%&  13.91\%\\
        \bottomrule
    \end{tabular}
    \label{tab:libfuzzer_cov_1h}
\end{table}

\begin{table}[!t]
    \caption{\small\textbf{\CamReady{Arithmetic mean} feature and edge coverage of Libfuzzer-based seed schedulers on 12 FuzzBench programs for 24 hours over 10 runs. We mark the highest number in bold. }}
    \centering
    \setlength{\tabcolsep}{2pt}
    \renewcommand{\arraystretch}{1.1}
   \begin{tabular}{lrr|rr|rr}
   \toprule
    \multicolumn{1}{c}{\multirow{2}{*}{\textbf{Programs}}} & \multicolumn{2}{c}{\textbf{\ToolName{}}} & \multicolumn{2}{c}{\textbf{Entropic}} & \multicolumn{2}{c}{\textbf{Default}} \\ \cline{2-7} 
\multicolumn{1}{c}{}& \multicolumn{1}{c}{\textbf {feature}} & \textbf{edge}     & \multicolumn{1}{c}{\textbf {feature}} & \textbf{edge}  & \multicolumn{1}{c}{\textbf {feature}} & \textbf{edge}    \\
        \midrule
        freetype & 71,717 & 13,754 &  \textbf{75,370} & \textbf{14,120} & 67,510 & 12,870\\
        libxml2 &  \textbf{54,081}  & \textbf{9,869} & 36,958 & 7,038  & 39,247 & 7,310 \\
        lcms & \textbf{6,345} & \textbf{2,541} &  4,425  &2,082  & 3,413 & 1,784  \\
        harfbuzz & \textbf{48,105} & \textbf{10,358} & 32,799 & 8,808 & 33,499 & 8,912  \\
        libjpeg & \textbf{15,861} & \textbf{3,033} &  11,755 & 2,646 & 11,220 & 2,574 \\
        libpng & \textbf{5,312} & \textbf{1,535} & 5,002 & 1,501 & 4,992 &1,501 \\
        openssl & \textbf{16,644} & \textbf{4,971} & 15,137 & 4,731 & 15,173 & 4,738 \\
        openthread & \textbf{11,405} & \textbf{4,965} & 6,435 & 3,276 & 6,123 & 3,196 \\
        re2 &\textbf{ 33,797} & \textbf{6,482} & 32,401 & 6,347 & 32,725 & 6,367 \\
        sqlite & \textbf{92,493} & \textbf{15,540} & 75,723 & 14,351 & 83,228 & 14,710 \\
        vorbis & \textbf{10,417} & \textbf{2,247} & 9,906 & 2,208 & 8,873 & 2,115 \\
        zlib & \textbf{3,215} & \textbf{801} & 2,698 & 790 & 2,510 & 787 \\     
        \midrule
        \multicolumn{3}{c|}{Arithmetic mean coverage gain} &  25.89\% & 13.69\% & 31.43\%  & 16.34\% \\
        \multicolumn{3}{c|}{Median coverage gain} & 20.66\%  & 6.68\% & 22.75\%&  6.54\%\\
        \bottomrule
    \end{tabular}
    \label{tab:libfuzzer_cov}
    \vspace{0.3cm}
\end{table}

For \ToolName{}’s comparison against AFL-based seed schedulers, we only use edge coverage as a metric for comparison because AFL does not report feature coverage. We run \ToolName{}, Default (i.e., AFL's default seed scheduler), RarePath, RareEdge, NewPath, and SecurityCov on the same 12 Google FuzzBench programs for 24 hours, repeated ten times. 
In arithmetic mean over these 10 runs, Table ~\ref{tab:afl_cov_1h} and Table ~\ref{tab:afl_cov} summarize the edge coverage results for 1 hour and 24 hours respectively. Appendix Table~\ref{tab:afl_p_1h} and Table~\ref{tab:afl_p_24h} show the Mann-Whitney U test results.
Similar to the comparison against Libfuzzer-based seed schedulers, we observe a higher improvement of \ToolName{} over the other seed scheduling strategies within the first hour. \ToolName{} outperforms the next best seed scheduling strategy (RarePath) by 7.95\% in arithmetic mean and 3.62\% in median over the 12 FuzzBench programs. For the 24 hour runs, \ToolName{} achieves 4.21\% in arithmetic mean and 1.91\% in median more coverage than RarePath. We note that the improvement of \ToolName{} against AFL-based seed schedulers is not as significant as \ToolName{}'s comparison against Libfuzzer-based seed schedulers. We believe \ToolName{}'s diminished performance difference occurs because the underlying fuzzer, AFL, iterates over the seed queue multiple times during the 24 hours fuzzing campaign and therefore will schedule nearly all seeds frequently, reducing the effect of seed selection. 

\begin{table}[!t]
    \caption{\small\textbf{\CamReady{Arithmetic mean} edge coverage of AFL-based seed schedulers on 12 FuzzBench programs for 1 hour over 10 runs.}}
    \centering
    \setlength{\tabcolsep}{1pt}
    \renewcommand{\arraystretch}{1.1}
    \begin{tabular}{lrrrrrr}
        \toprule
                \textbf{} & \textbf{\ToolNameShort{}}  & \textbf{Default} & \textbf{RarePath} & \textbf{RareEdge} & \textbf{NewPath} &\textbf{SecCov}  \\ 
        \midrule
        Fuzzer & AFL  & AFL &  AflFast & FairFuzz & EcoFuzz & TortoiseFuzz \\ 
        \midrule
        freetype & \textbf{12,077} &  11,001 &10,707 & 11,319 & 8,925 & 10,532\\
        libxml2 &  \textbf{8,120} &  5,793  & 5,836 & 7,247 & 5,841 & 5,476 \\
        lcms & 1,882 &  \textbf{1,989}  &  1,540  & 1,343  & 1,117 & 1,327 \\
        harfbuzz & \textbf{9,169} &  8,864  & 9,022 & 8,767 & 7,629 & 8,773 \\
        libjpeg & \textbf{2,391} & 2,354 & 2,374 & 2,140  & 1,739 &  2,073 \\
        libpng & 1,470 & \textbf{1,488} &  1,460  & 1,430  & 1,428 &  1,456  \\
        openssl & \textbf{4,560} & 4,485 & 4,399 & 4,381  & 4,252 & 4,336  \\
        openthread &\textbf{5,245} & 5,063 & 5,064 & 5,047  & 5,047 & 5,012  \\
        re2 & \textbf{5,792} &  5,612  & 5,533 &  5,335  & 5,484 & 5,252  \\
        sqlite & 9,865 & 10,038 &  9,890 & \textbf{10,065} & 9,722 & 9,627 \\
        vorbis & \textbf{2,048} & 2,006 & 1,946 & 1,933  & 1,761 & 1,914 \\
        zlib & \textbf{761} & 758 & 752  & 746 & 745 & 752 \\
        \midrule
        \multicolumn{2}{c|}{Arithmetic mean gain} & 4.80\% & 7.95\%  & 8.39\% & 20.01\% & 13.03\%  \\
        \multicolumn{2}{c|}{Median gain} & 1.87\% & 3.62\% & 5.27\% & 11.77\% & 6.07\%   \\
        \bottomrule
    \end{tabular}
    \label{tab:afl_cov_1h}
\end{table}

\begin{table}[!h]
    \caption{\small\textbf{\CamReady{Arithmetic mean} edge coverage of AFL-based seed schedulers on 12 FuzzBench programs for 24 hours over 10 runs.}}
    \centering
    \setlength{\tabcolsep}{1pt}
    \renewcommand{\arraystretch}{1.1}
    \begin{tabular}{lrrrrrr}
        \toprule
        \textbf{} & \textbf{\ToolNameShort{}}  & \textbf{Default} & \textbf{RarePath} & \textbf{RareEdge} & \textbf{NewPath} &\textbf{SecCov}  \\ 
        \midrule
        Fuzzer & AFL  & AFL &  AflFast & FairFuzz & EcoFuzz & TortoiseFuzz \\ 
        \midrule
        freetype & \textbf{14,188} &  13,508 &13,646 & 13,486 & 11,965 & 13,206\\
        libxml2 &  \textbf{10,936} &  9,295  & 8,546 & 10,241 & 8,964 & 9,147 \\
        lcms & \textbf{2,325} &  2,247  &  2,160  & 2,190  & 1,892 & 2,162 \\
        harfbuzz & \textbf{10,061} &  9,980  & 10,019 & 9,804  & 9,946 & 9,882 \\
        libjpeg & \textbf{2,678} & 2,513 & 2,601 & 2,497  & 2,309 &  2,413 \\
        libpng & \textbf{1,536} & \textbf{1,536} &  1,535  & 1,524  & 1,528 &  1,528  \\
        openssl & \textbf{4,863} & 4,805 & 4,761 & 4,788  & 4,732 & 4,685  \\
        openthread &\textbf{5,766} & 5,704 & 5,646 & 5,666  & 5,527 & 5,636  \\
        re2 & \textbf{5,887} &  5,875  & 5,790 &  5,536  & 5,774 & 5,758  \\
        sqlite & 12,081 & \textbf{12,360} &  12,019 & 10,648 & 12,199& 11,810 \\
        vorbis & \textbf{2,215} & 2,195 & 2,202 & 2,100  & 2,171 & 2,184 \\
        zlib & \textbf{780} & \textbf{780} & 775  & 778 & 777   & 769   \\
        \midrule
        \multicolumn{2}{c|}{Arithmetic mean gain} & 2.89\% & 4.21\% & 4.81\% & 7.63\% & 5.11\%   \\
        \multicolumn{2}{c|}{Median gain} & 1.00\% &  1.91\% &  5.34\% & 2.38\% & 2.30\%  \\
        \bottomrule
    \end{tabular}
    \label{tab:afl_cov}
        \vspace{0.3cm}
\end{table}

\begin{figure*}[!]
\centering
\captionsetup[subfloat]{captionskip=-.01cm, labelformat=empty}

\includegraphics[scale=0.6]{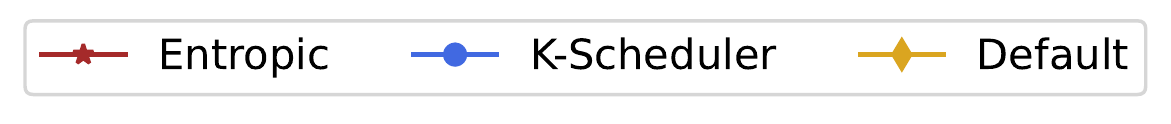}

\subfloat[freetype]{
\includegraphics[width=0.23\textwidth]{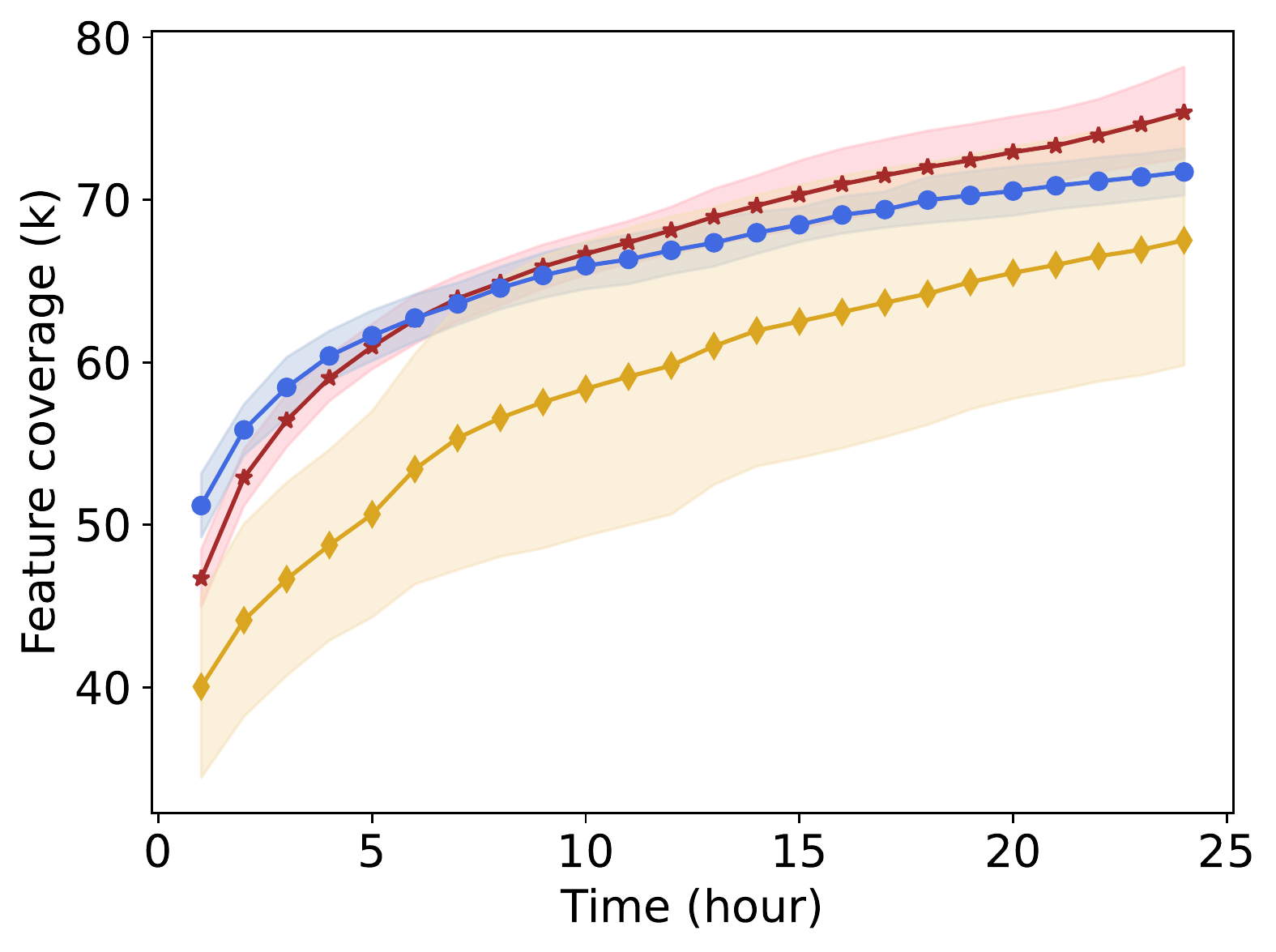}
\label{subfig:readelf_1h}}
\hspace{-.1cm}
\subfloat[harfbuzz]{
\includegraphics[width=0.23\textwidth]{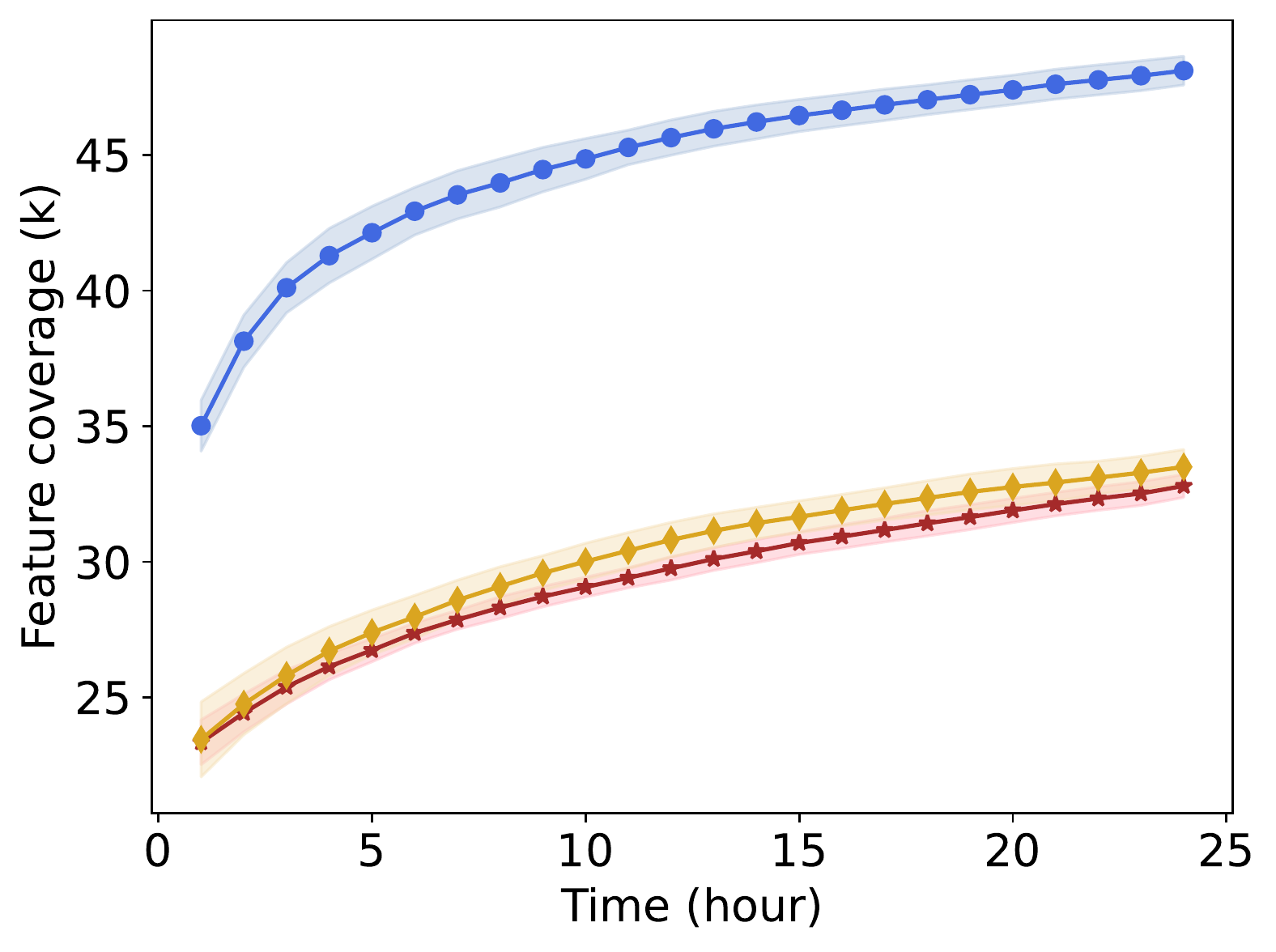}
\label{subfig:harfbuzz_1h}}
\hspace{-.1cm}
\subfloat[openthread]{
\includegraphics[width=0.23\textwidth]{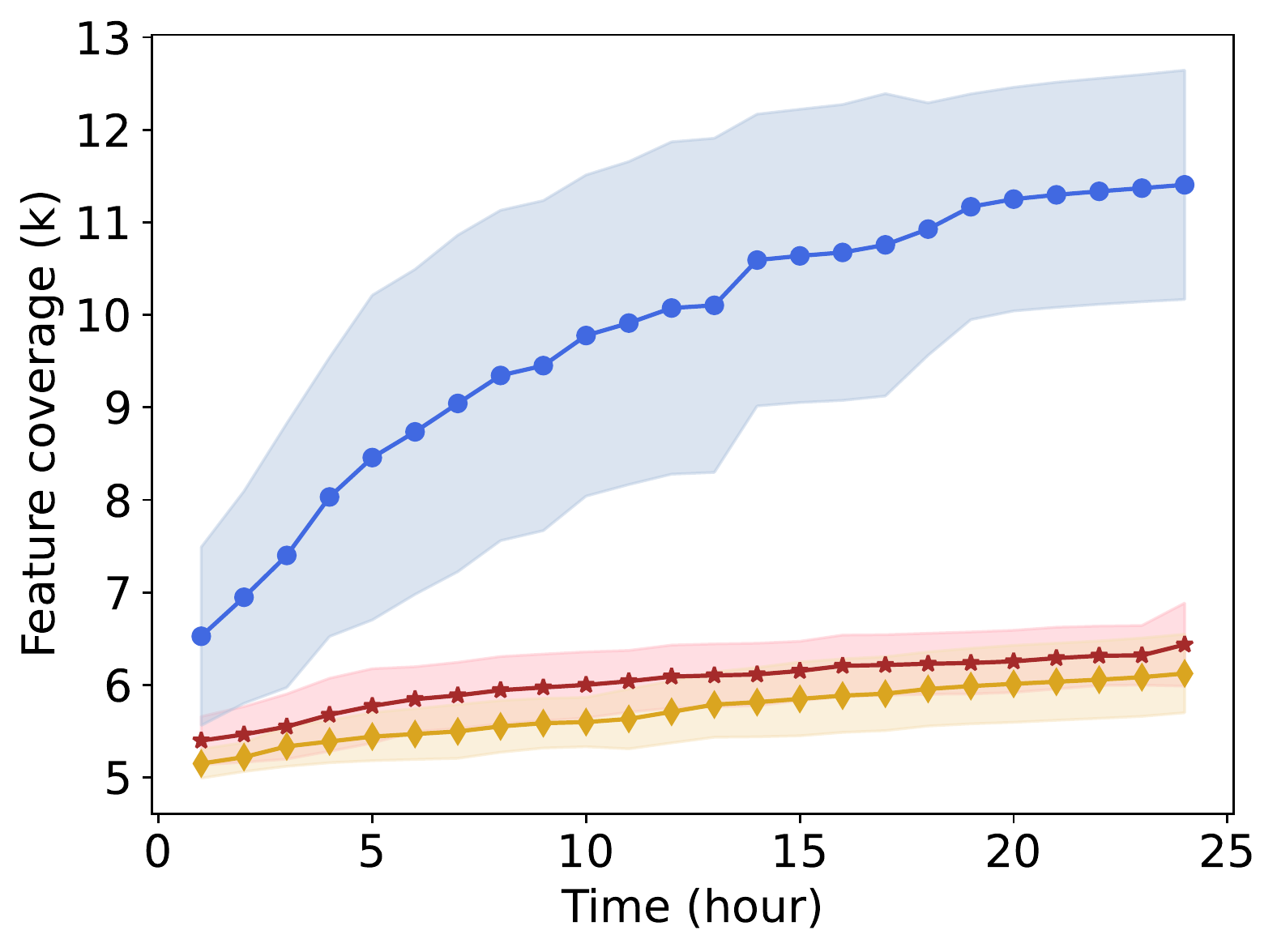}
\label{subfig:libjpeg_1h}}
\hspace{-.1cm}
\subfloat[libjpeg]{
\includegraphics[width=0.23\textwidth]{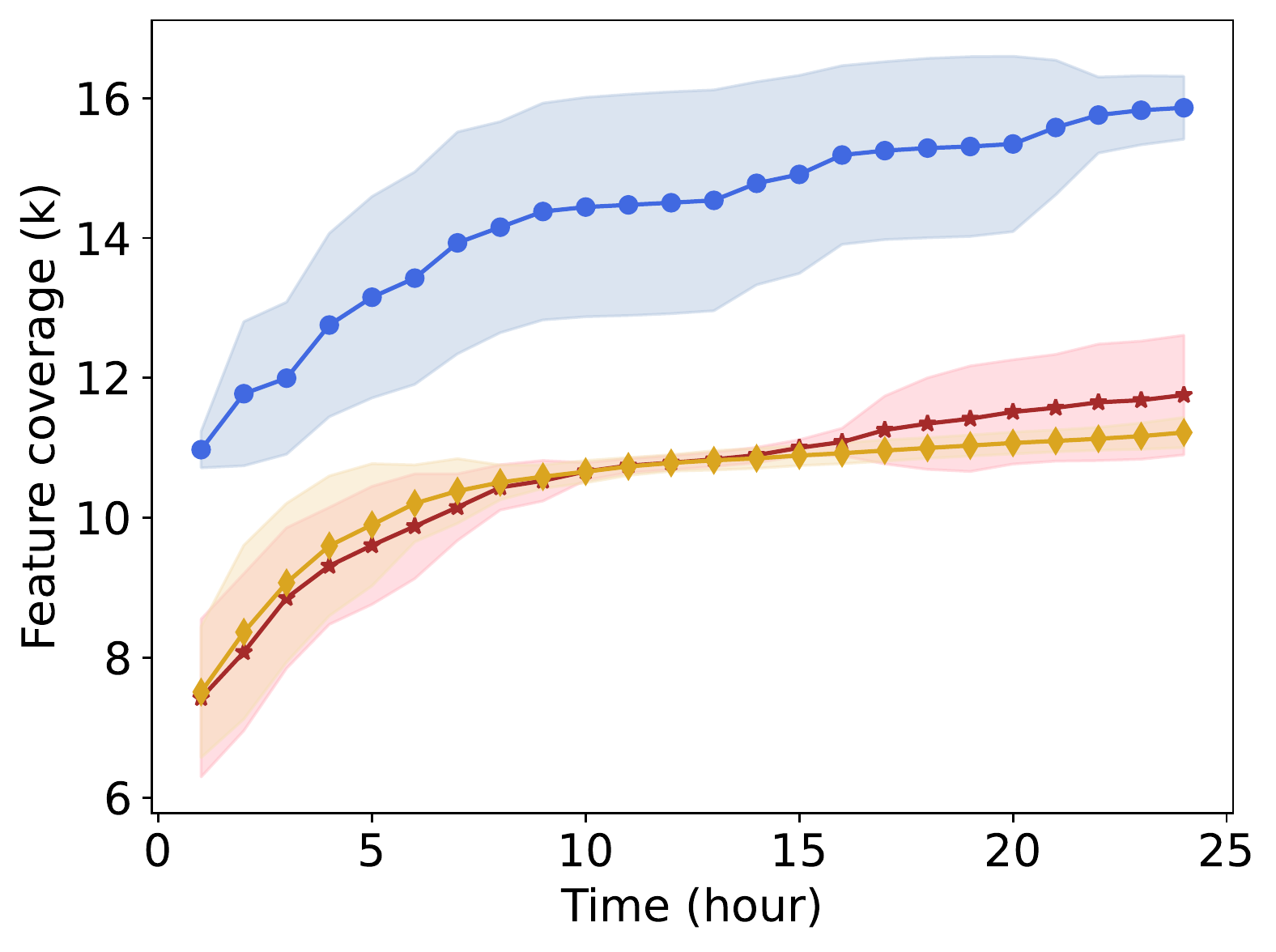}
\label{subfig:libjpeg_1h}}

\subfloat[lcms]{
\includegraphics[width=0.23\textwidth]{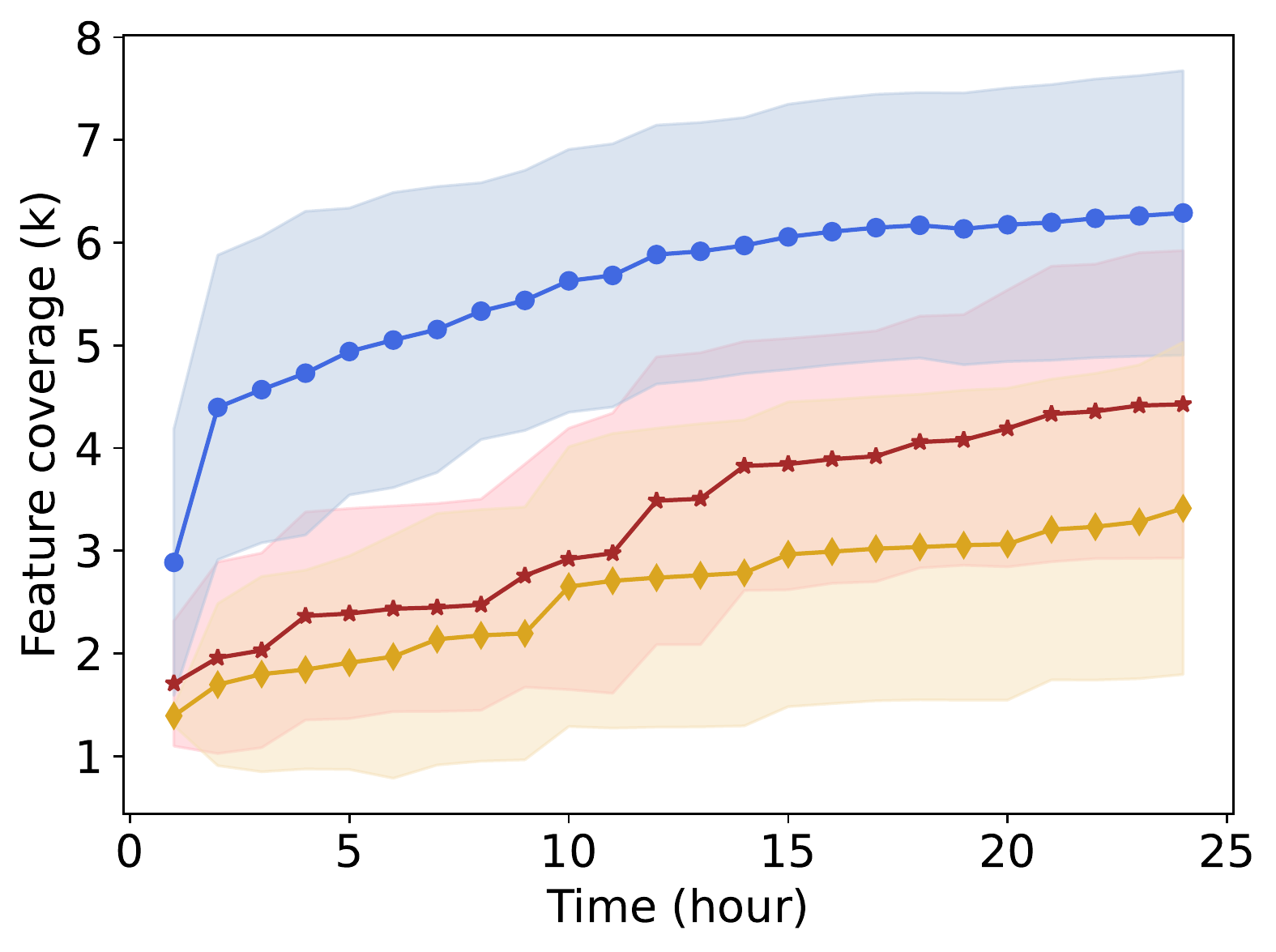}
\label{subfig:readelf_1h}}
\hspace{-.1cm}
\subfloat[libpng]{
\includegraphics[width=0.23\textwidth]{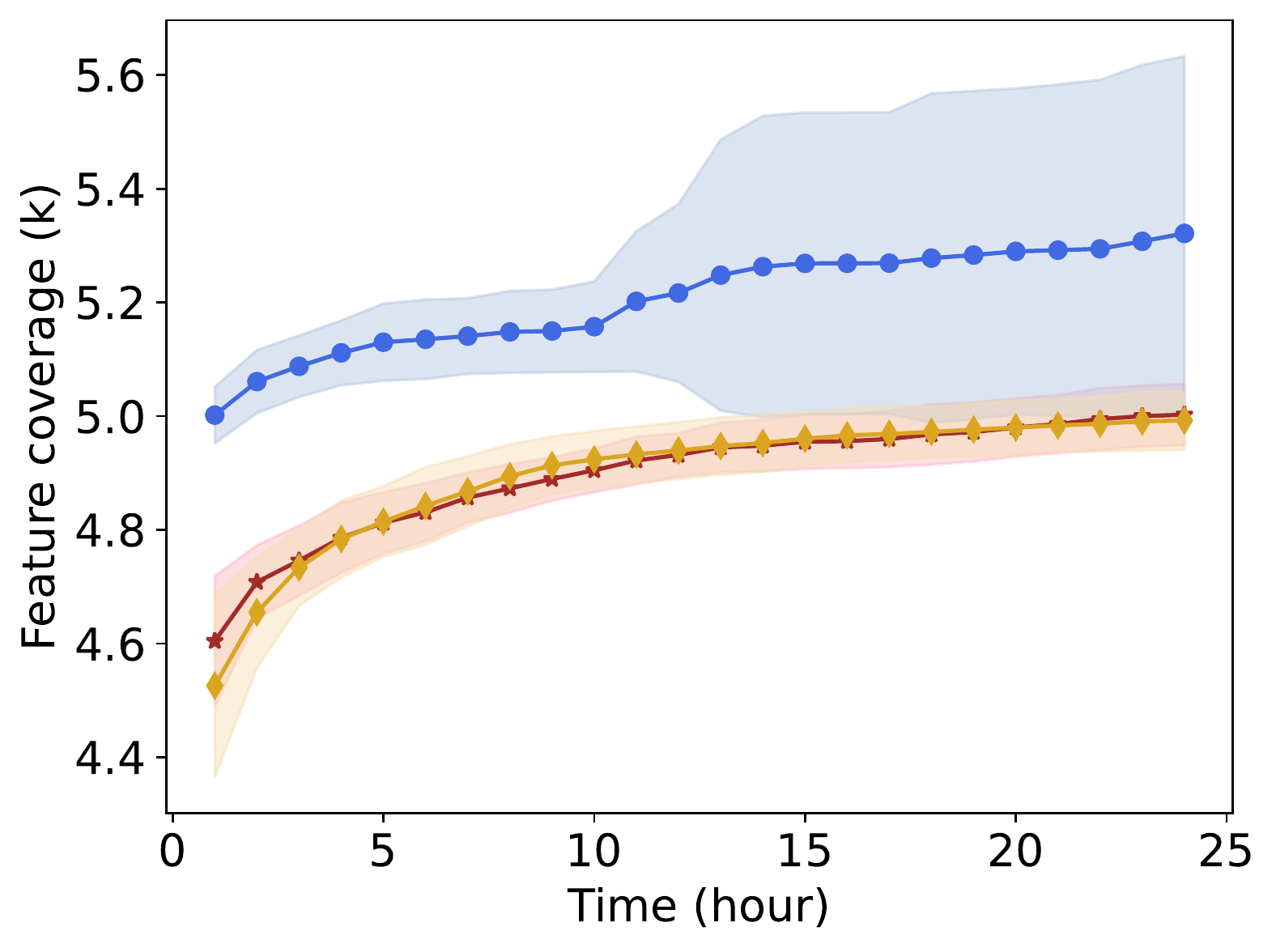}
\label{subfig:harfbuzz_1h}}
\hspace{-.1cm}
\subfloat[re2]{
\includegraphics[width=0.23\textwidth]{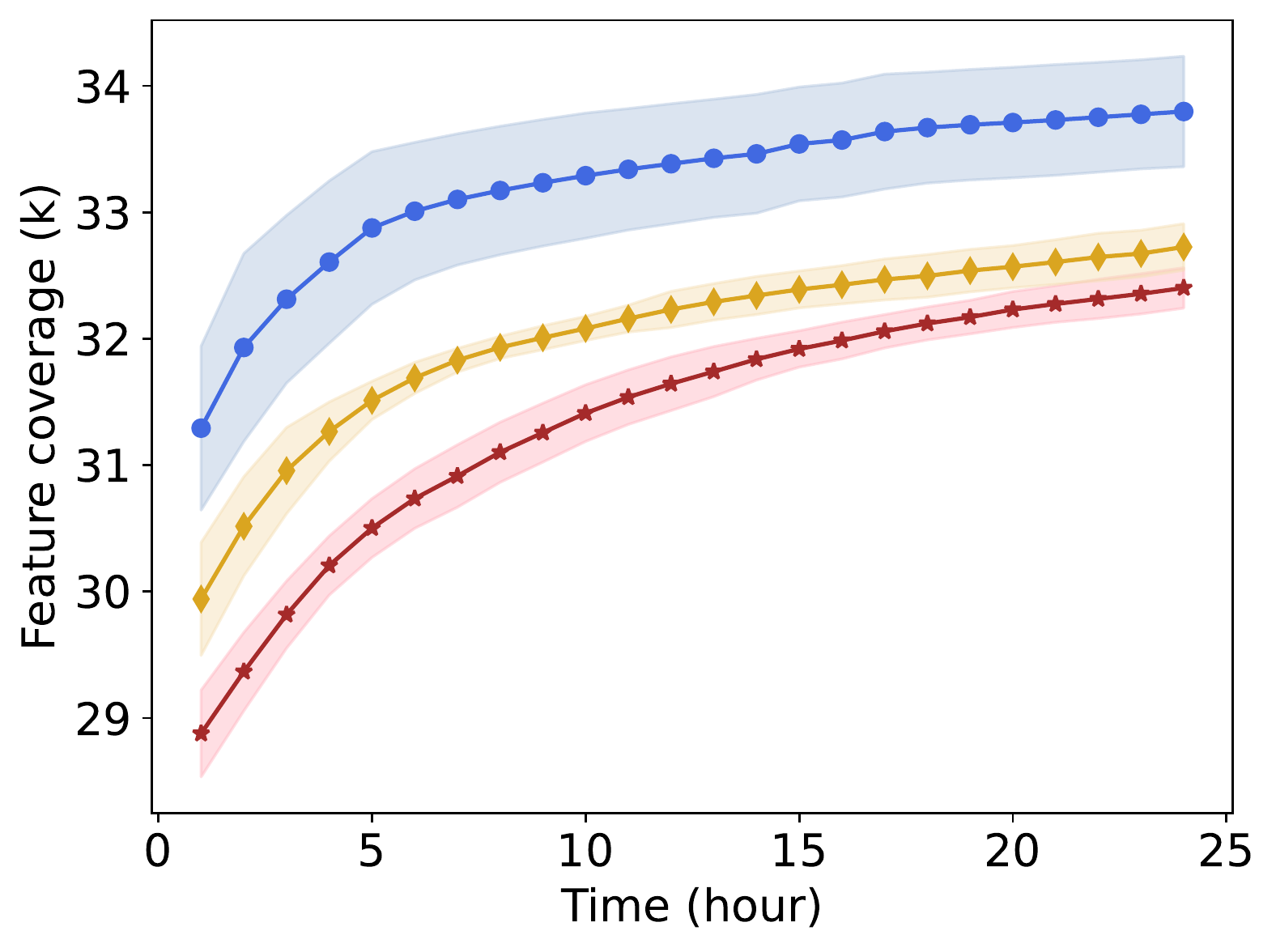}
\label{subfig:libjpeg_1h}}
\hspace{-.1cm}
\subfloat[sqlite]{
\includegraphics[width=0.23\textwidth]{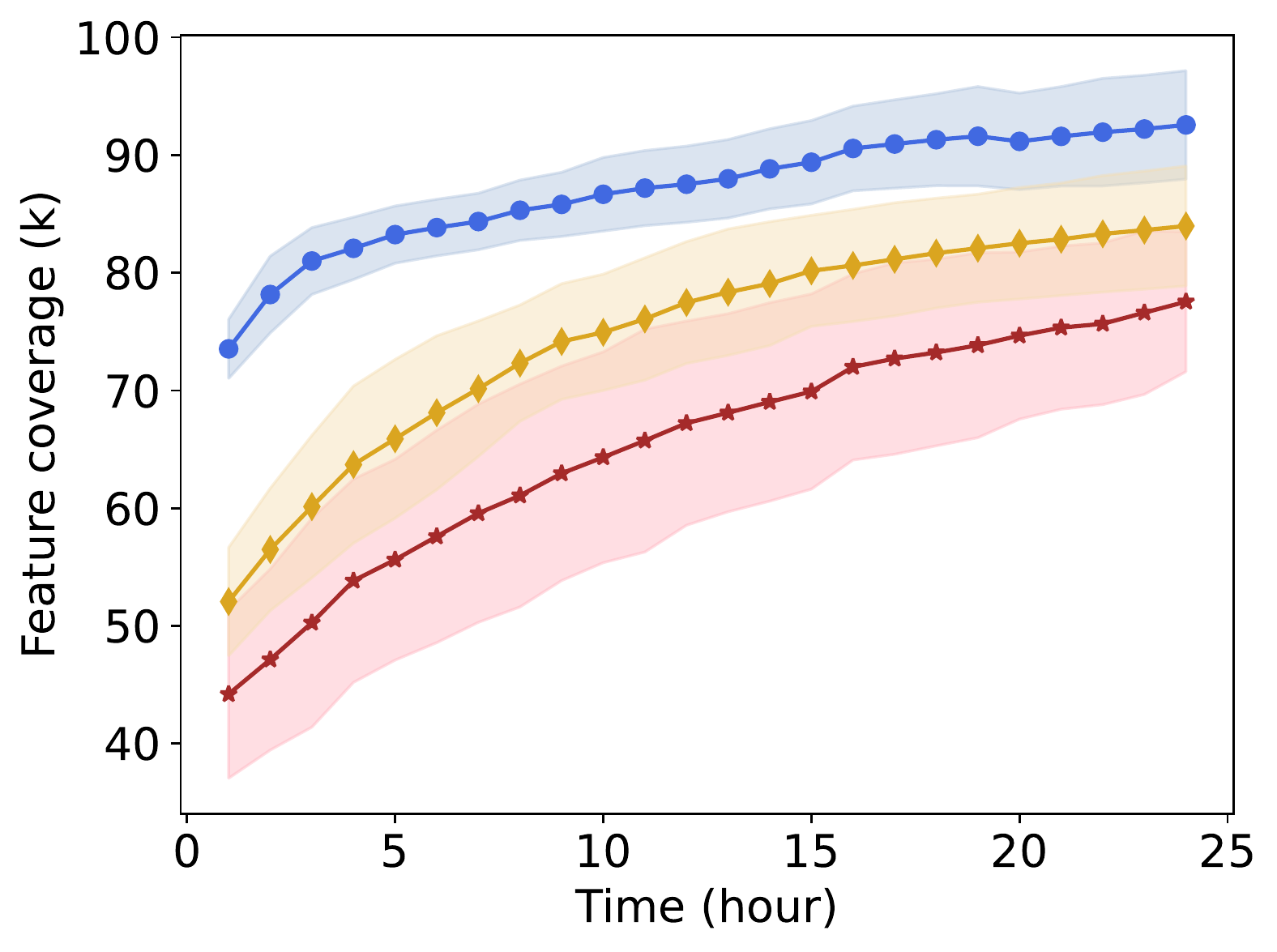}
\label{subfig:libjpeg_1h}}

\subfloat[openssl]{
\includegraphics[width=0.23\textwidth]{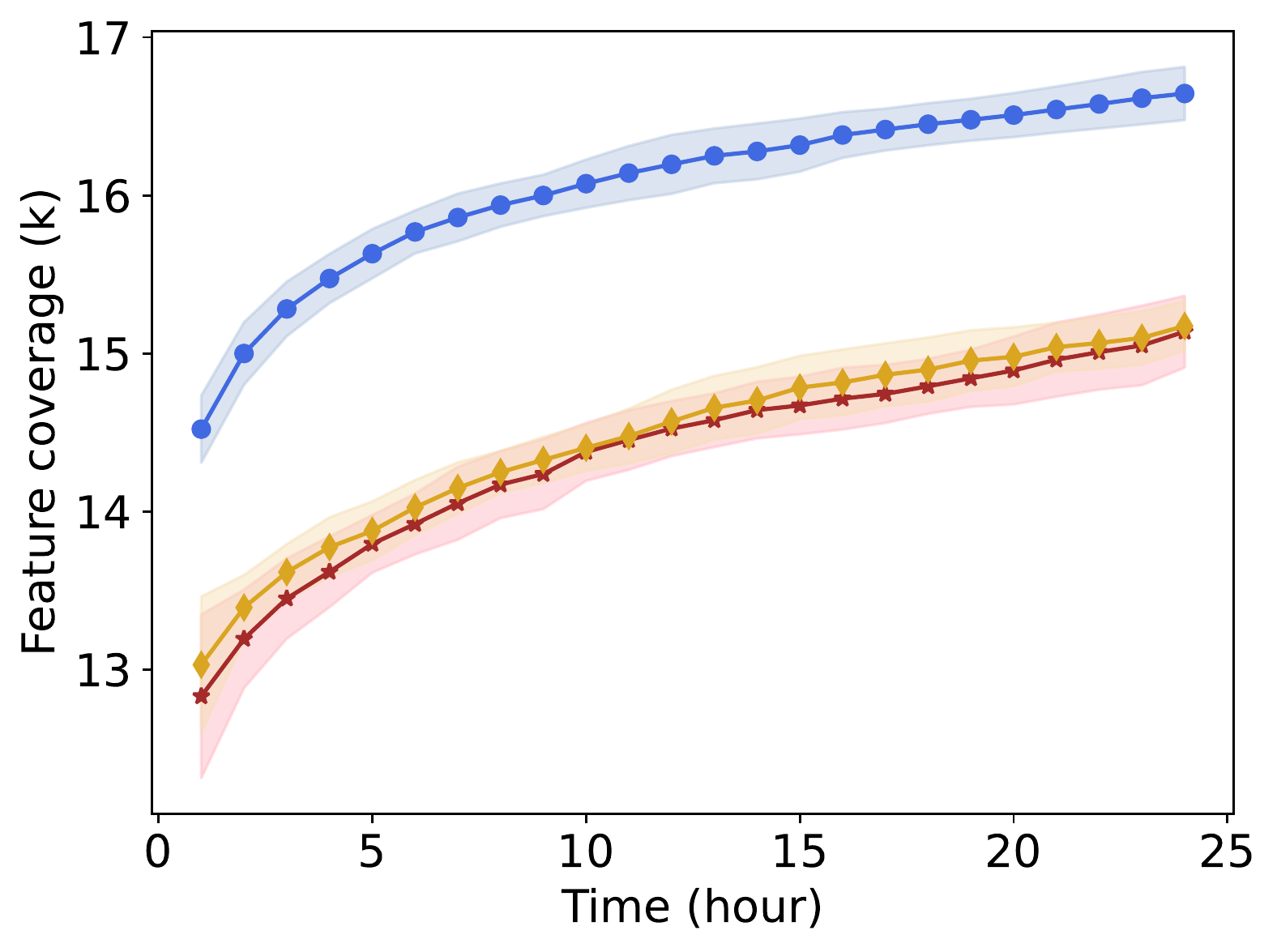}
\label{subfig:readelf_1h}}
\hspace{-.1cm}
\subfloat[vorbis]{
\includegraphics[width=0.23\textwidth]{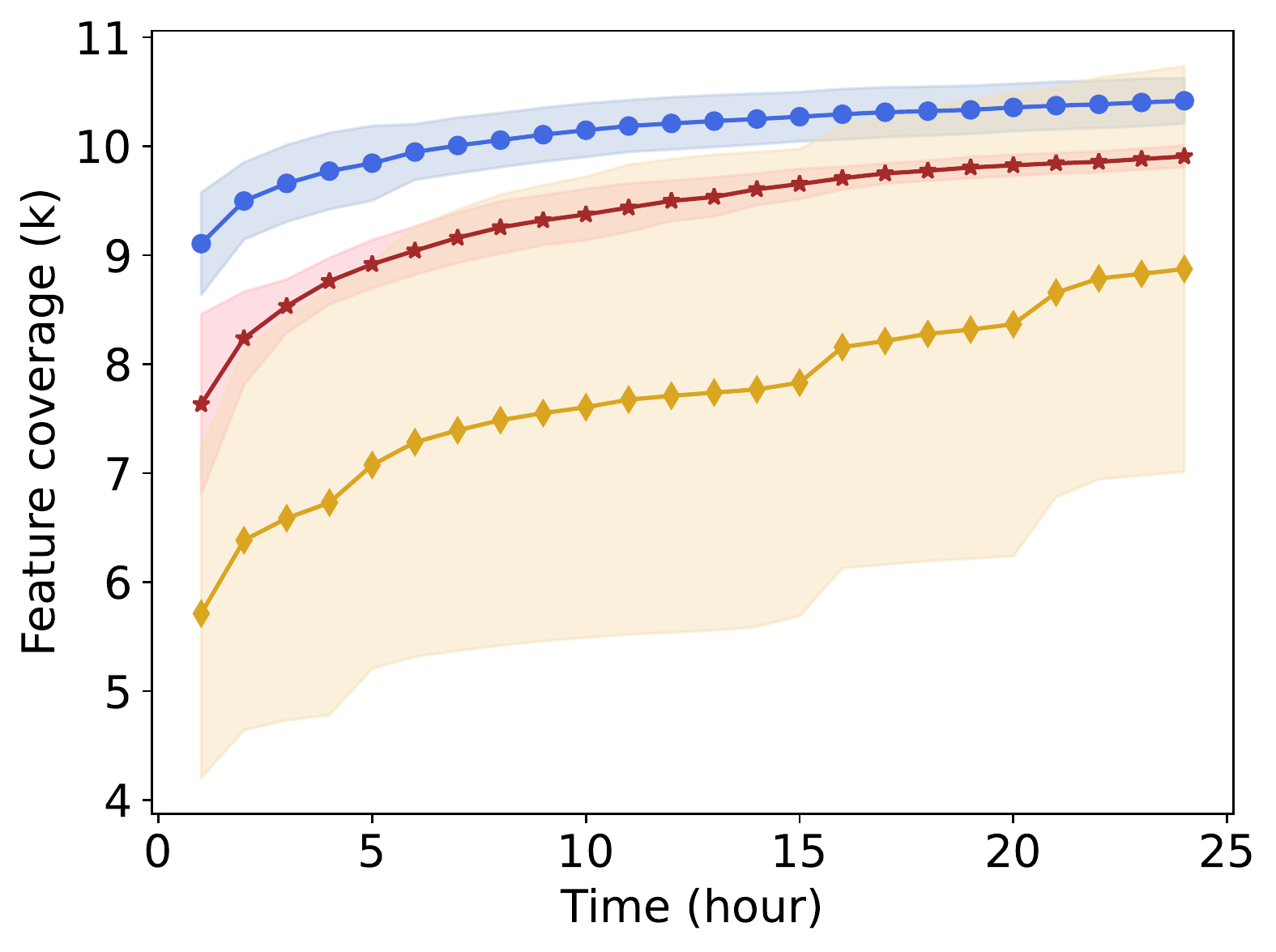}
\label{subfig:harfbuzz_1h}}
\hspace{-.1cm}
\subfloat[zlib]{
\includegraphics[width=0.23\textwidth]{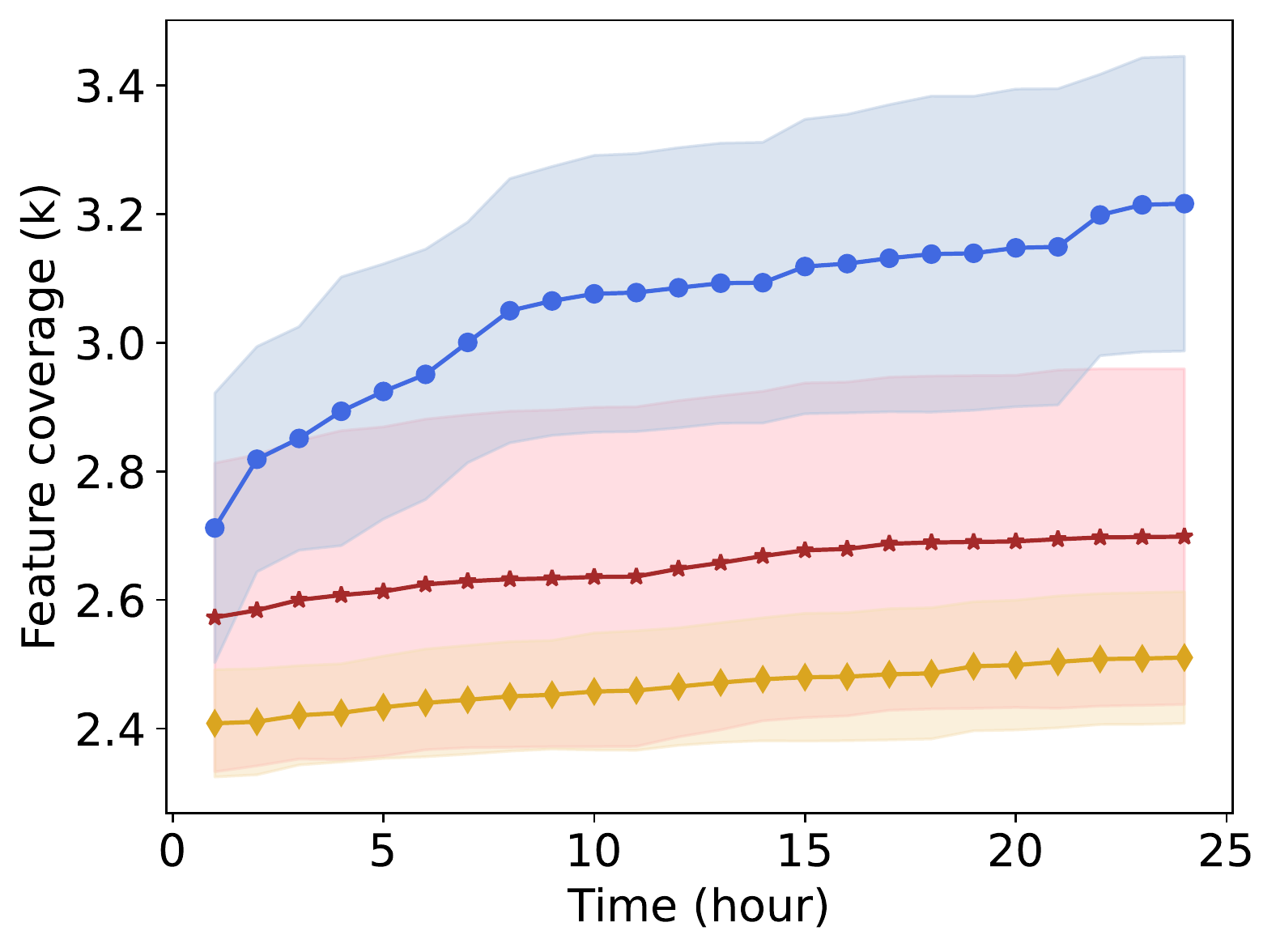}
\label{subfig:zlib}}
\hspace{-.1cm}
\subfloat[libxml2]{
\includegraphics[width=0.23\textwidth]{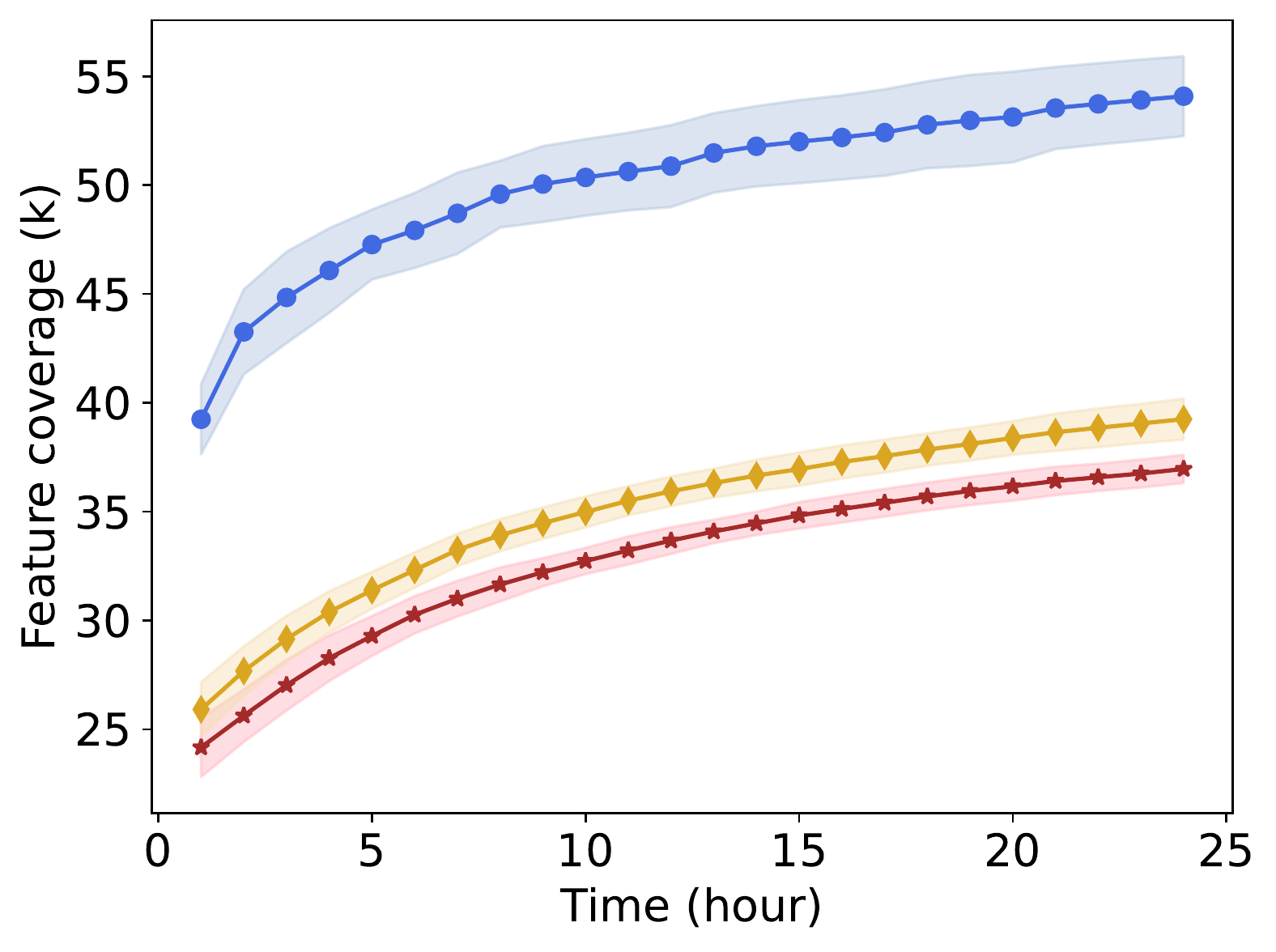}
\label{subfig:libxml}}

\caption{\textbf{\small The \CamReady{arithemic mean} feature coverage of Libfuzzer-based seed schedulers running for 24 hours and one standard deviation error bars over 10 runs. Default refers to the default seed scheduler in Libfuzzer.}}
\label{fig:libfuzzertime}
\end{figure*}

\begin{figure*}[!]
\centering
\captionsetup[subfloat]{captionskip=-.01cm, labelformat=empty}

\includegraphics[scale=0.6]{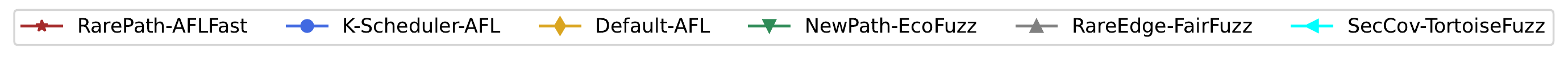}

\subfloat[freetype]{
\includegraphics[width=0.23\textwidth]{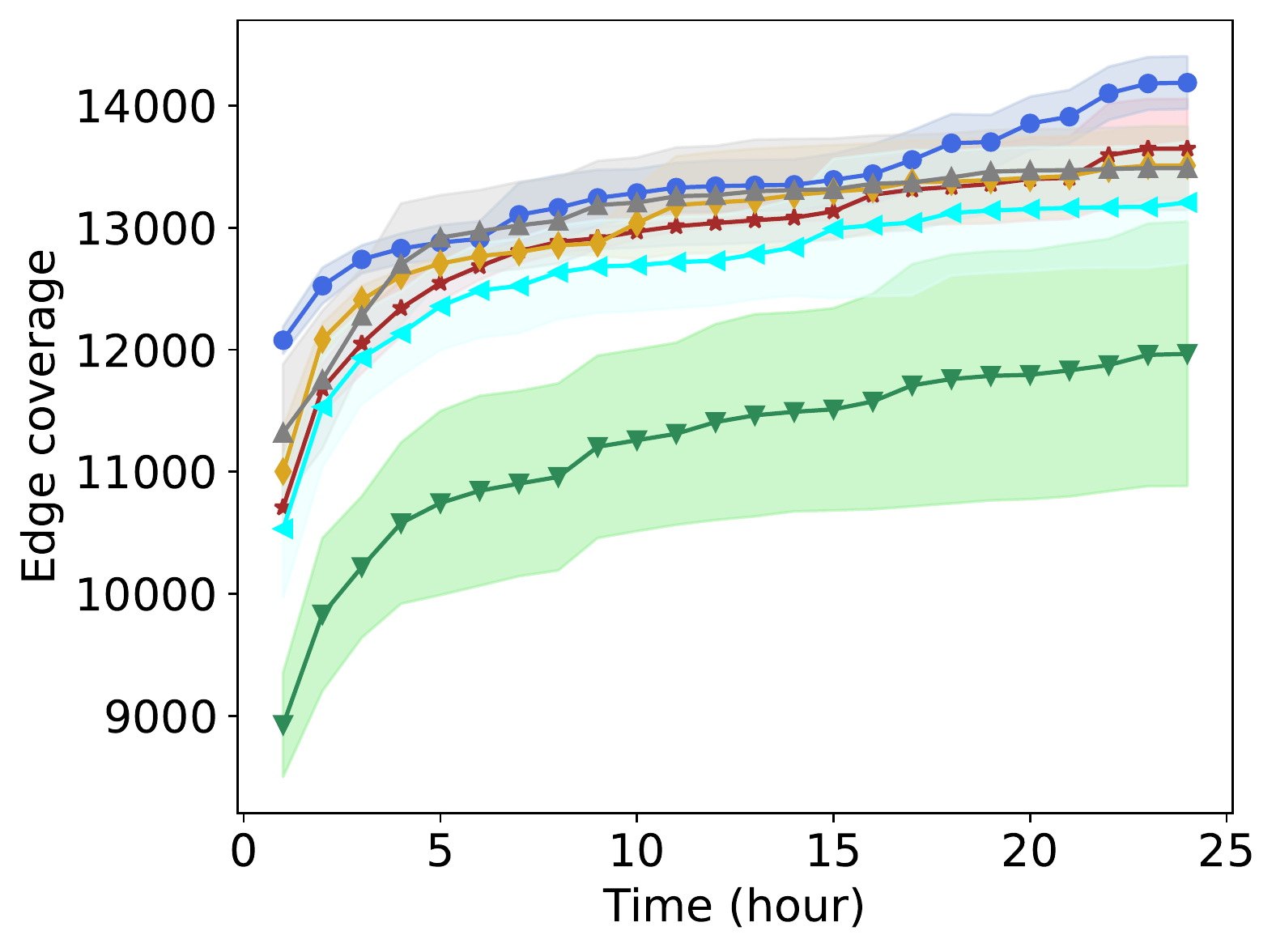}
\label{subfig:afl_freetype}}
\hspace{-.1cm}
\subfloat[harfbuzz]{
\includegraphics[width=0.23\textwidth]{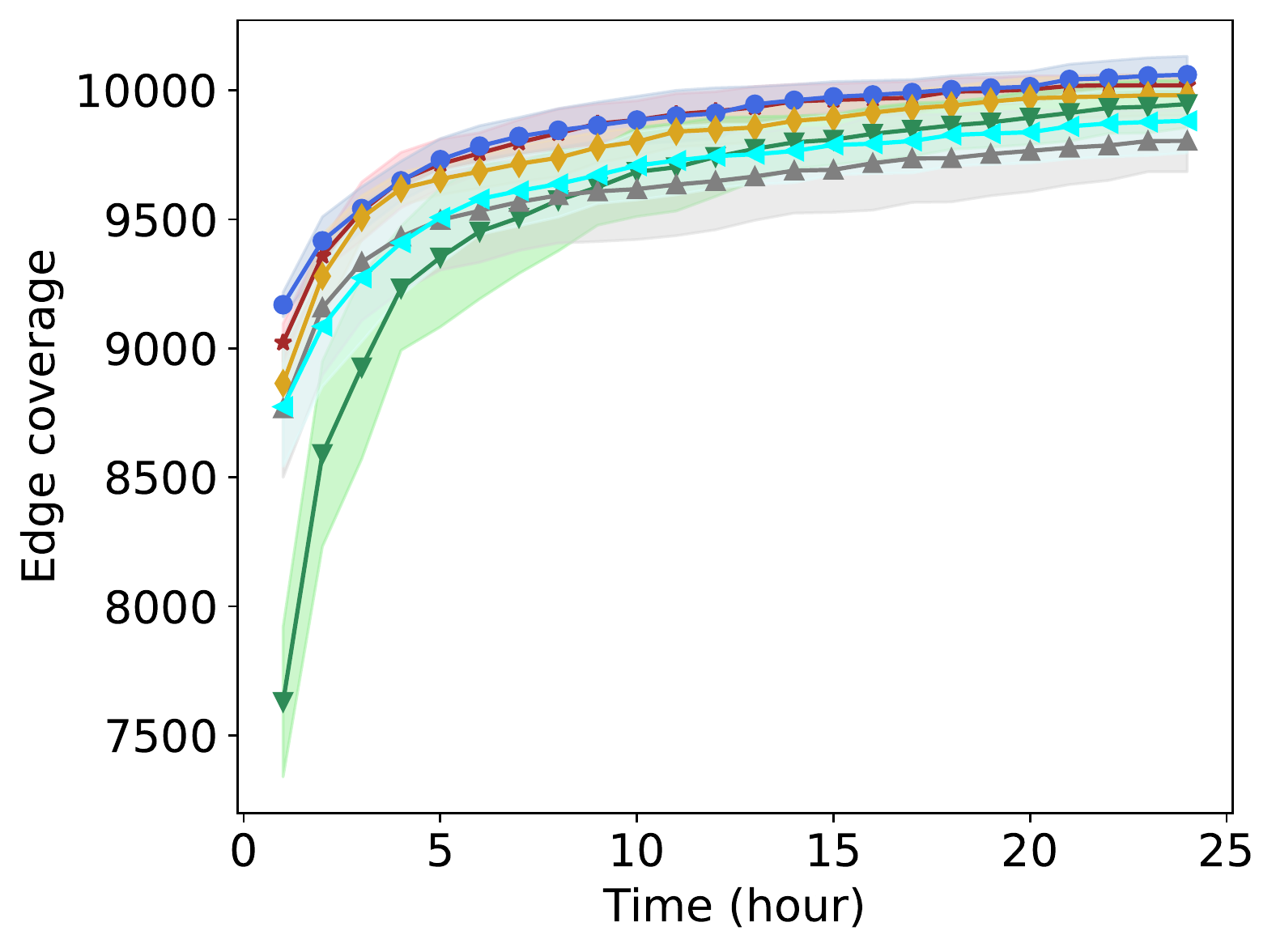}
\label{subfig:harfbuzz_afl}}
\hspace{-.1cm}
\subfloat[openthread]{
\includegraphics[width=0.23\textwidth]{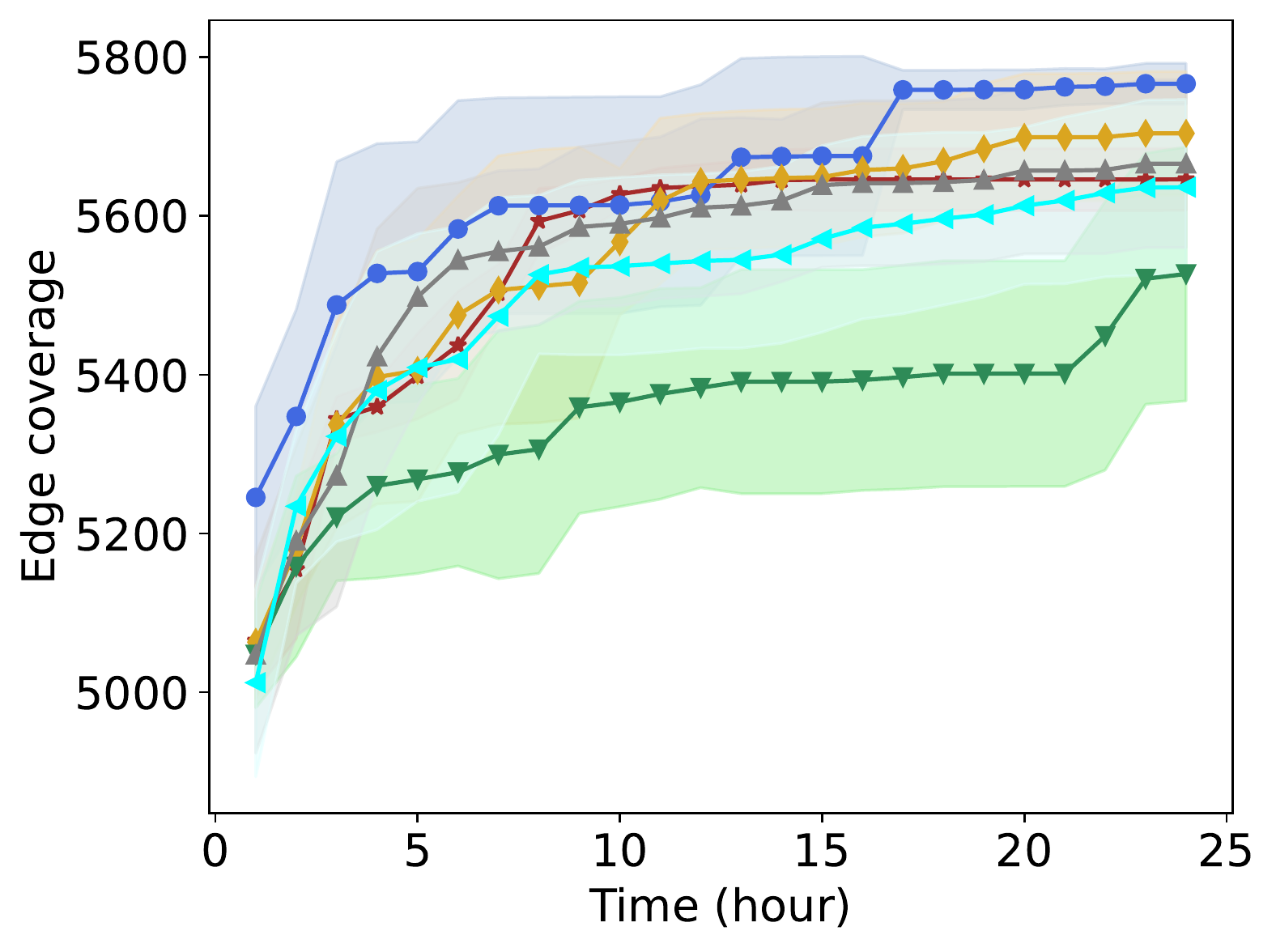}
\label{subfig:afl_radio}}
\hspace{-.1cm}
\subfloat[libjpeg]{
\includegraphics[width=0.23\textwidth]{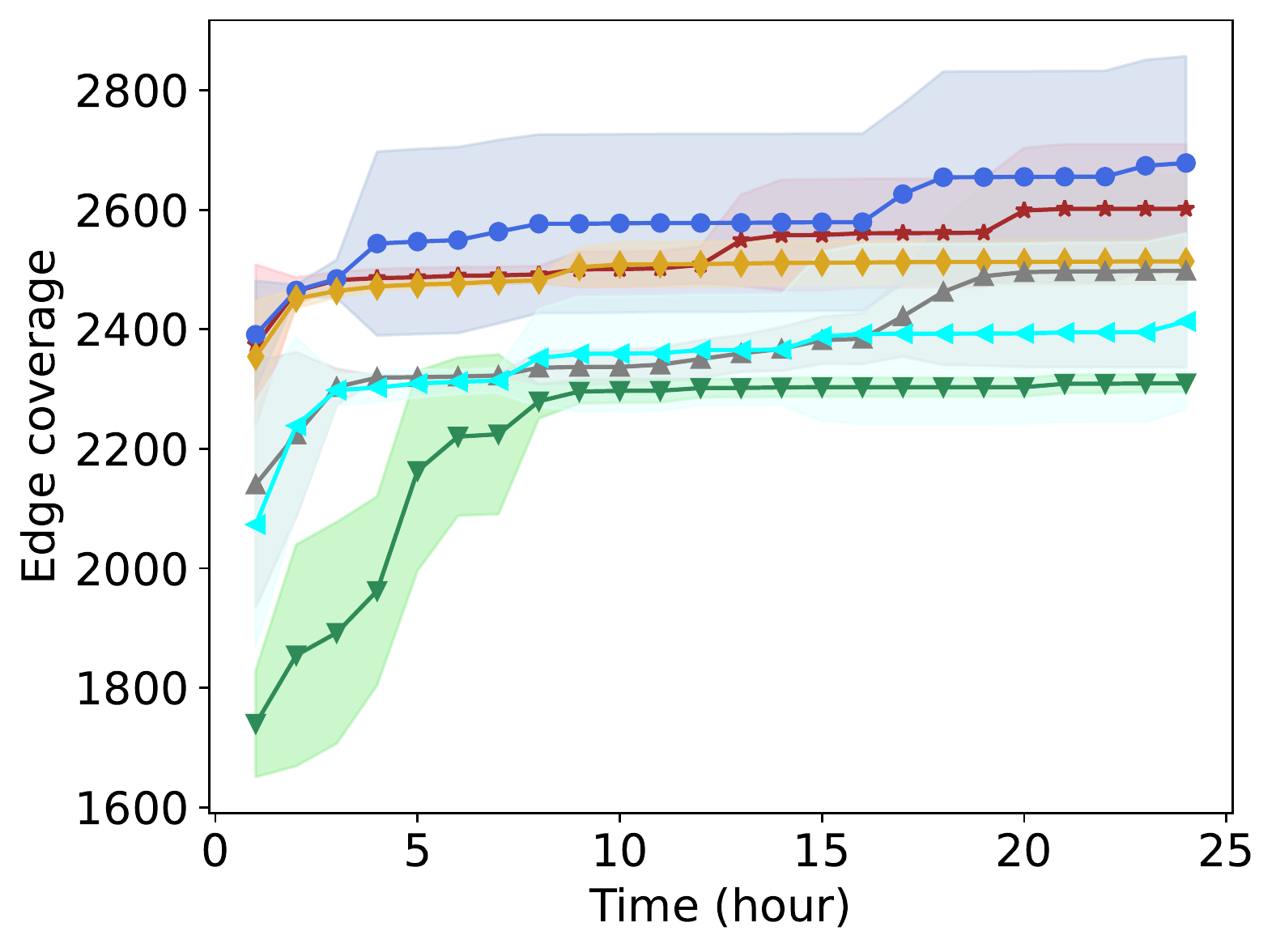}
\label{subfig:libjpeg_afl}}

\subfloat[lcms]{
\includegraphics[width=0.23\textwidth]{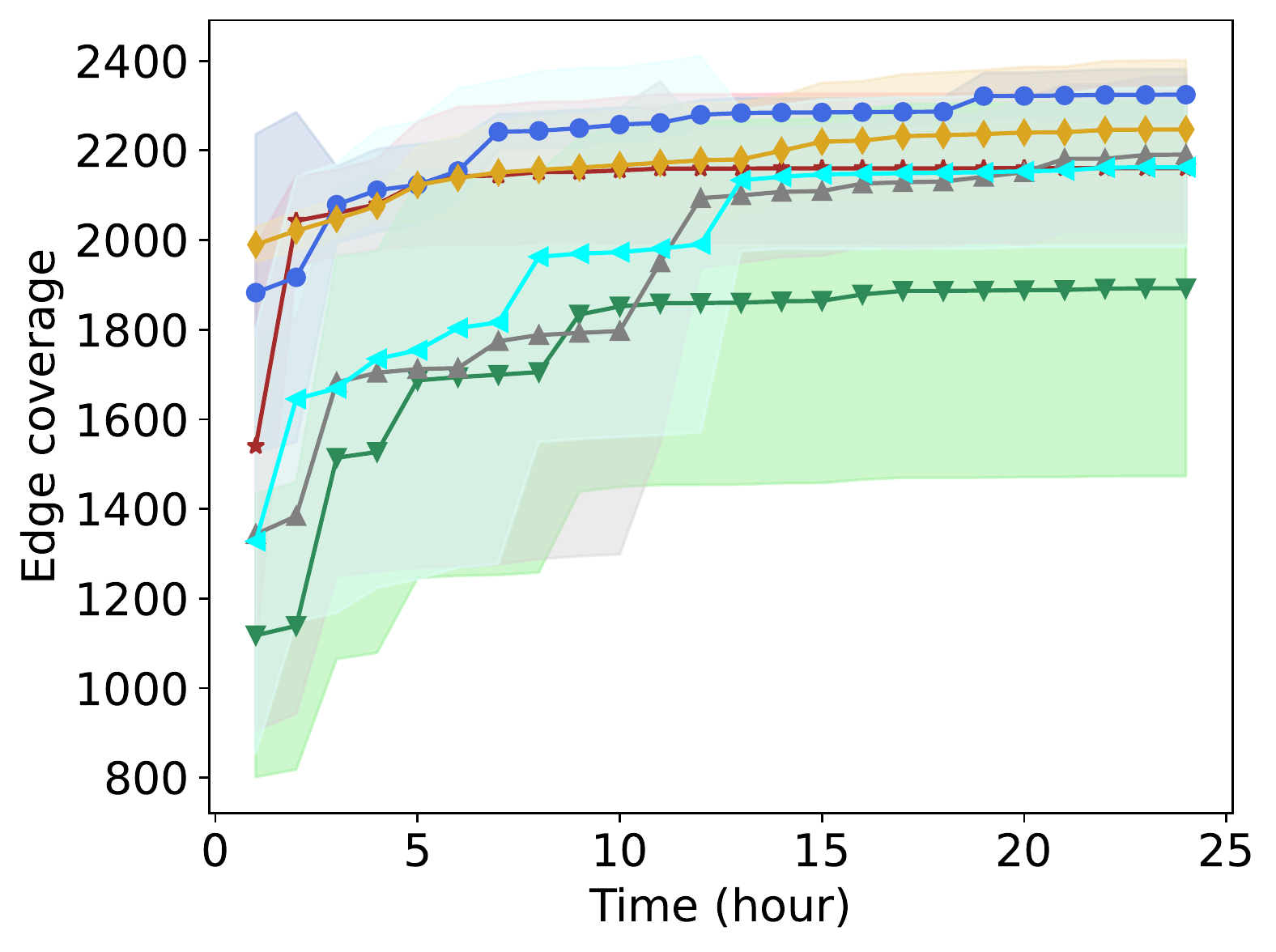}
\label{subfig:afl_lcms}}
\hspace{-.1cm}
\subfloat[libpng]{
\includegraphics[width=0.23\textwidth]{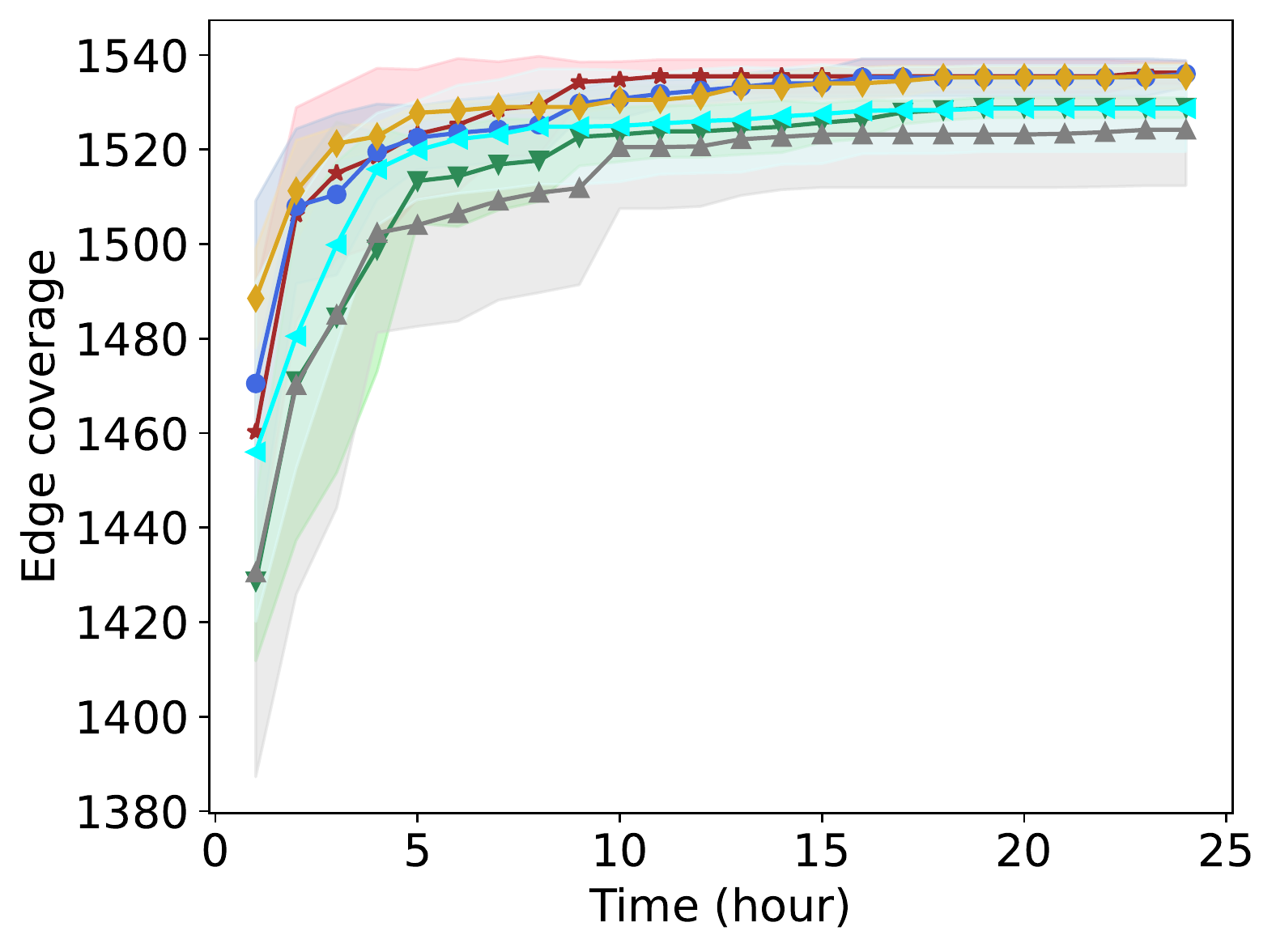}
\label{subfig:afl_png}}
\hspace{-.1cm}
\subfloat[re2]{
\includegraphics[width=0.23\textwidth]{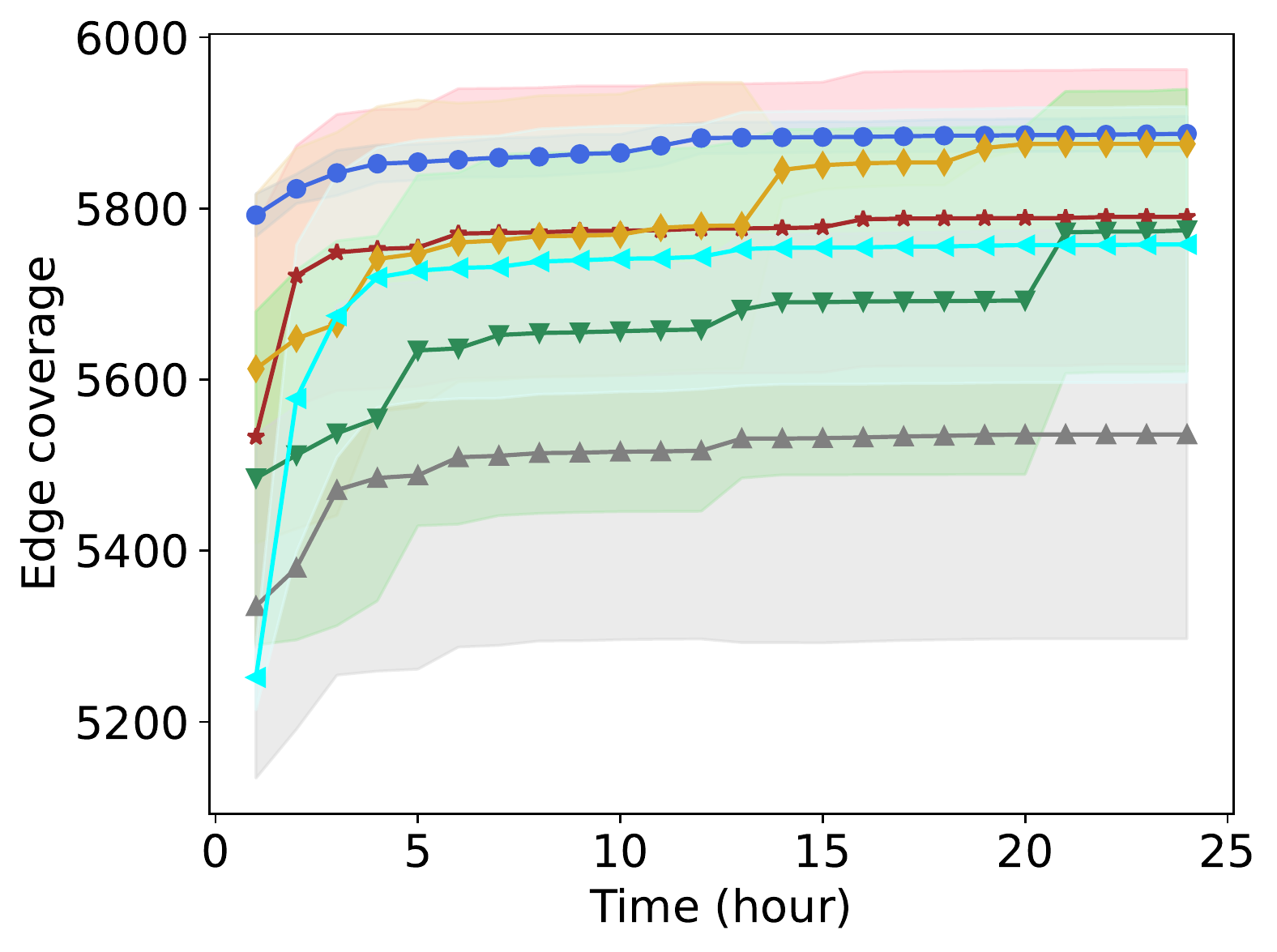}
\label{subfig:afl_re2}}
\hspace{-.1cm}
\subfloat[sqlite]{
\includegraphics[width=0.23\textwidth]{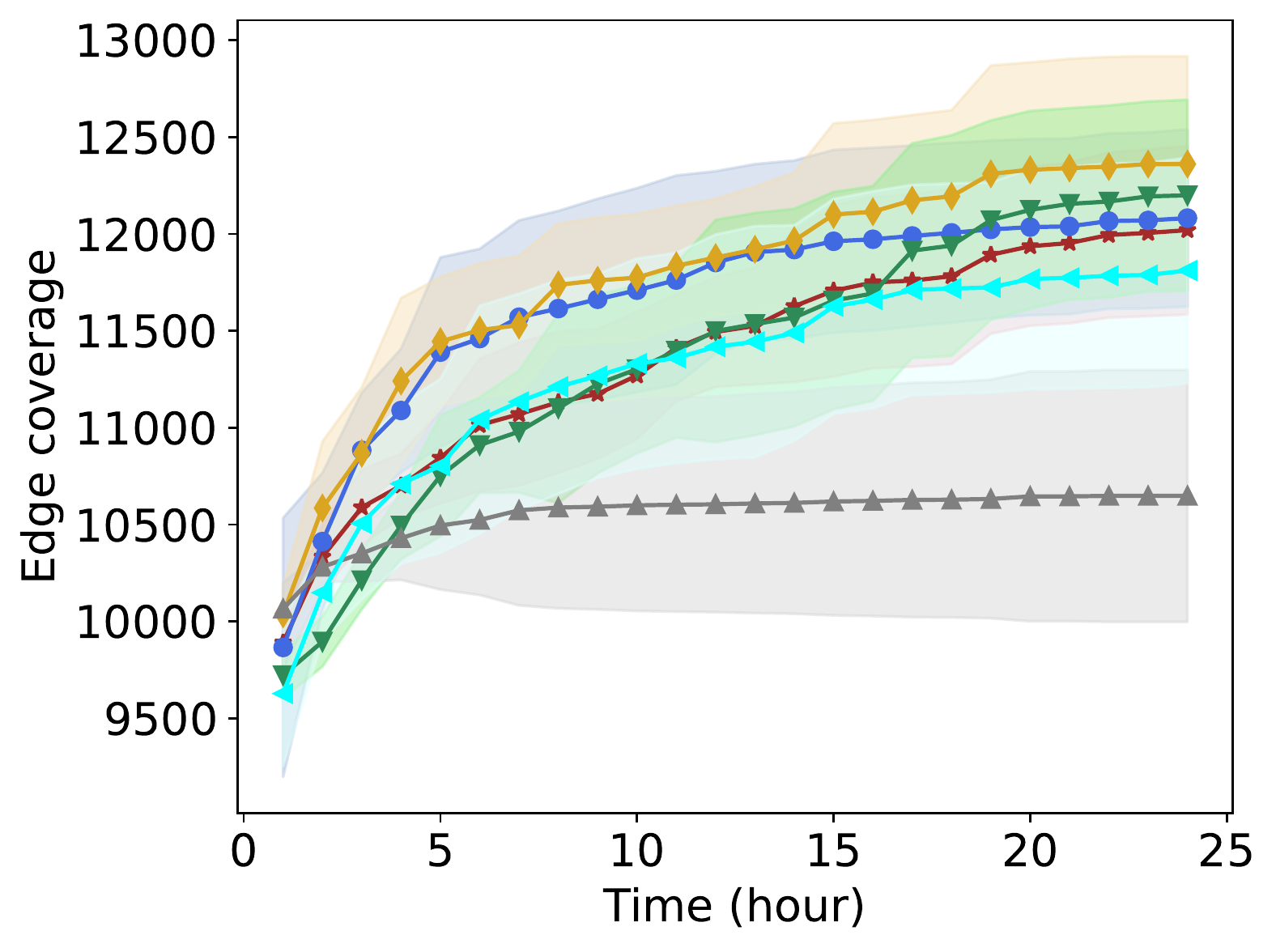}
\label{subfig:afl_sql}}

\subfloat[openssl]{
\includegraphics[width=0.23\textwidth]{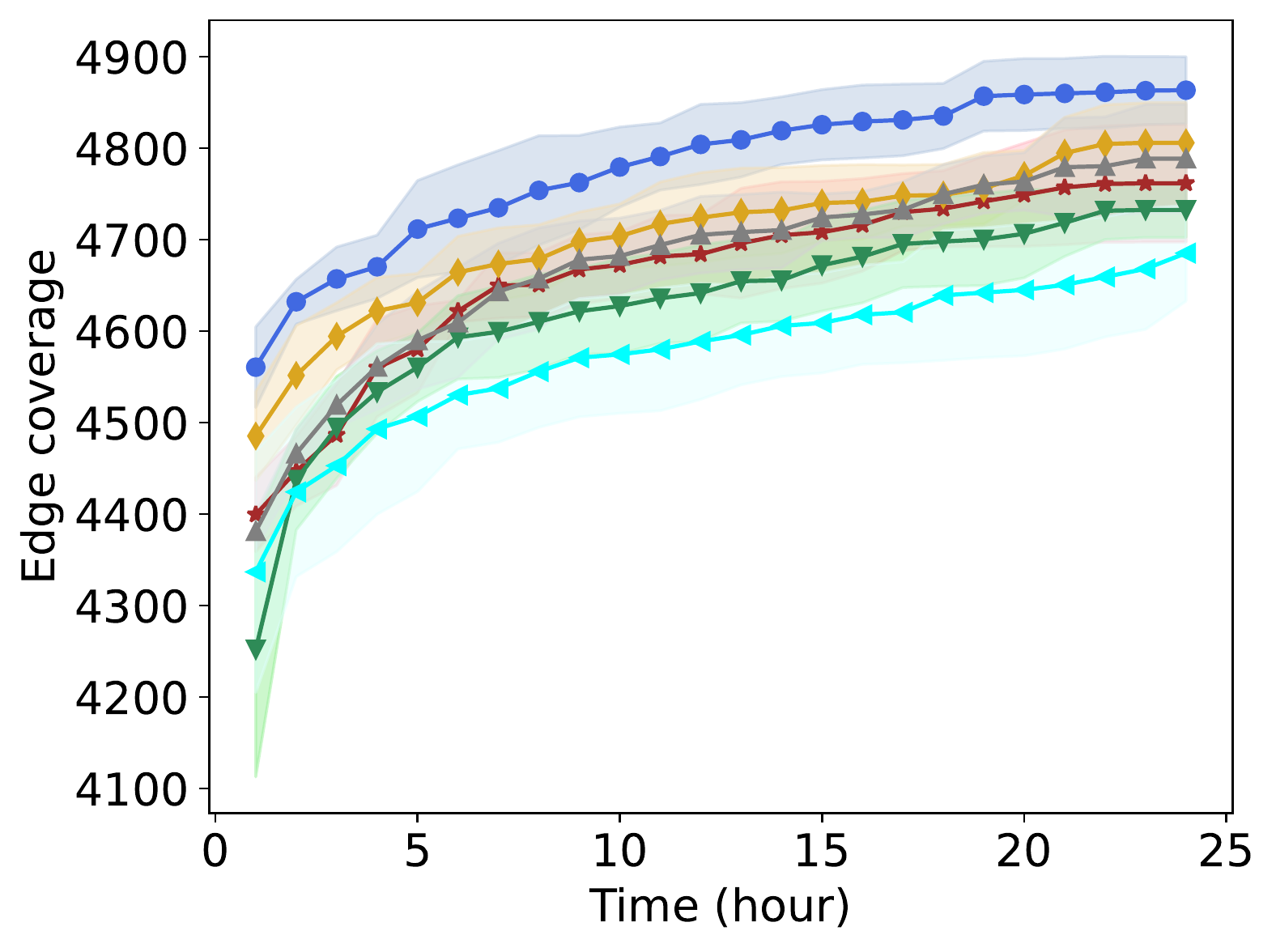}
\label{subfig:afl_ssl}}
\hspace{-.1cm}
\subfloat[vorbis]{
\includegraphics[width=0.23\textwidth]{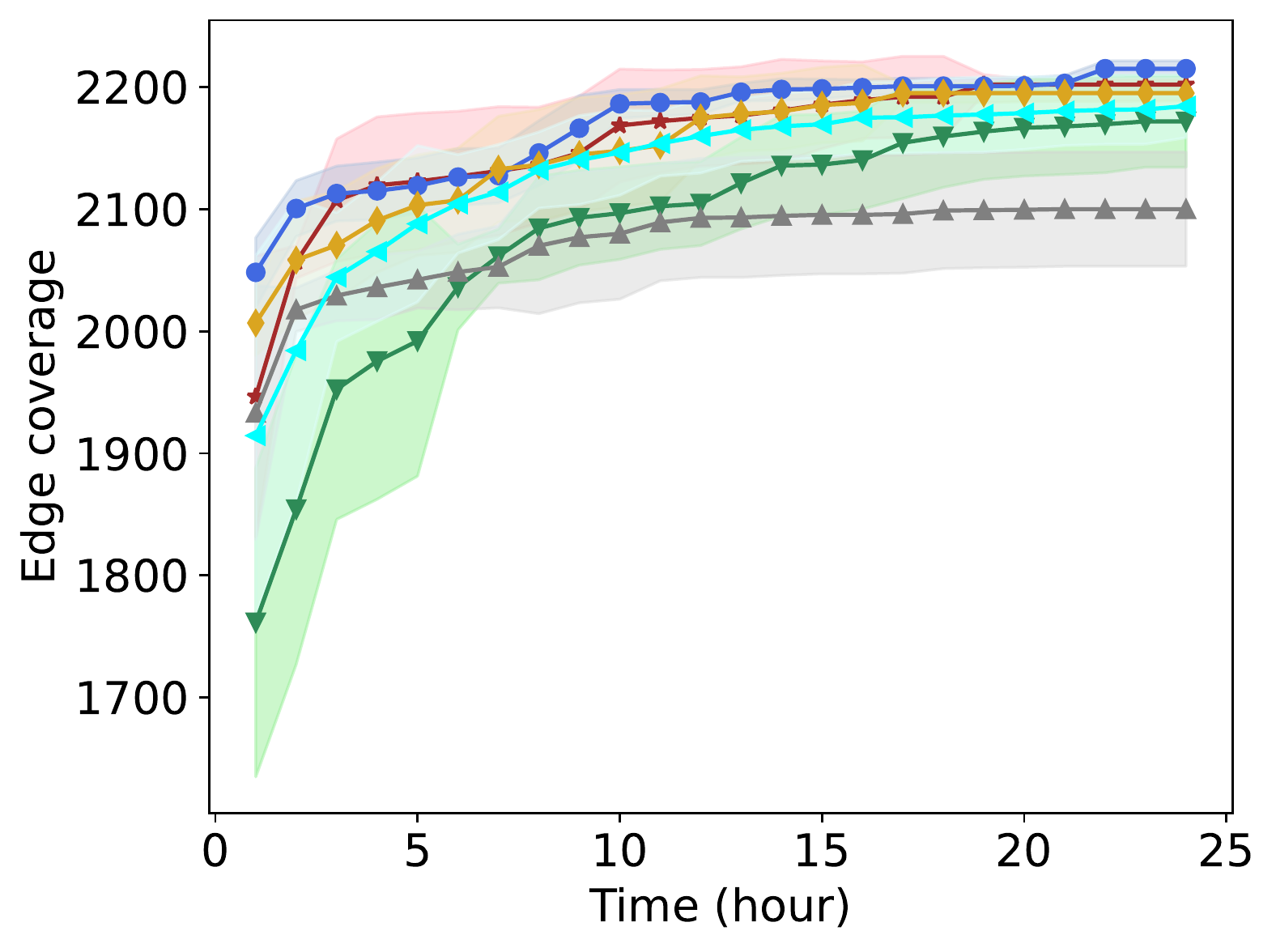}
\label{subfig:afl_vor}}
\hspace{-.1cm}
\subfloat[zlib]{
\includegraphics[width=0.23\textwidth]{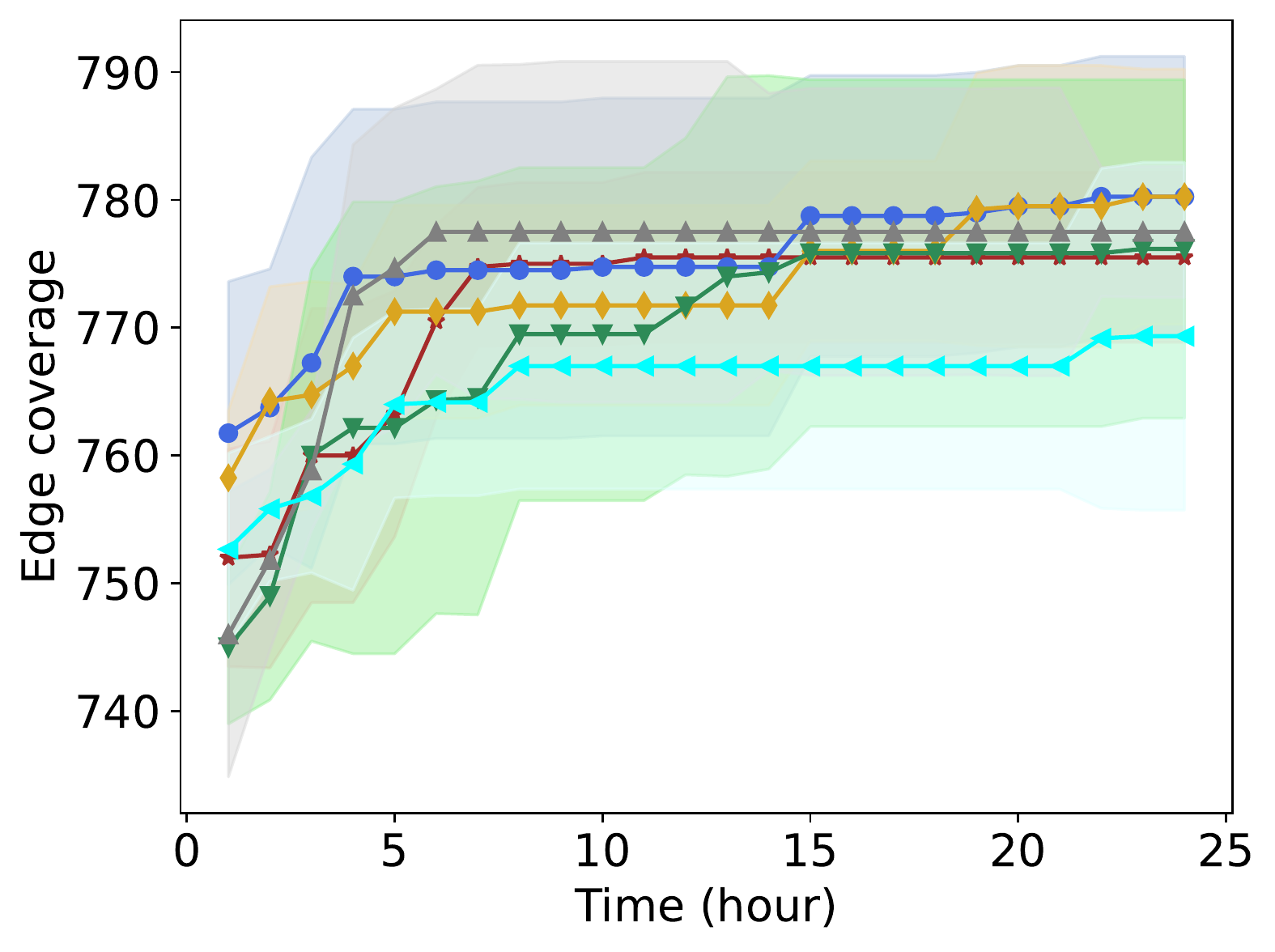}
\label{subfig:afl_zlib}}
\hspace{-.1cm}
\subfloat[libxml2]{
\includegraphics[width=0.23\textwidth]{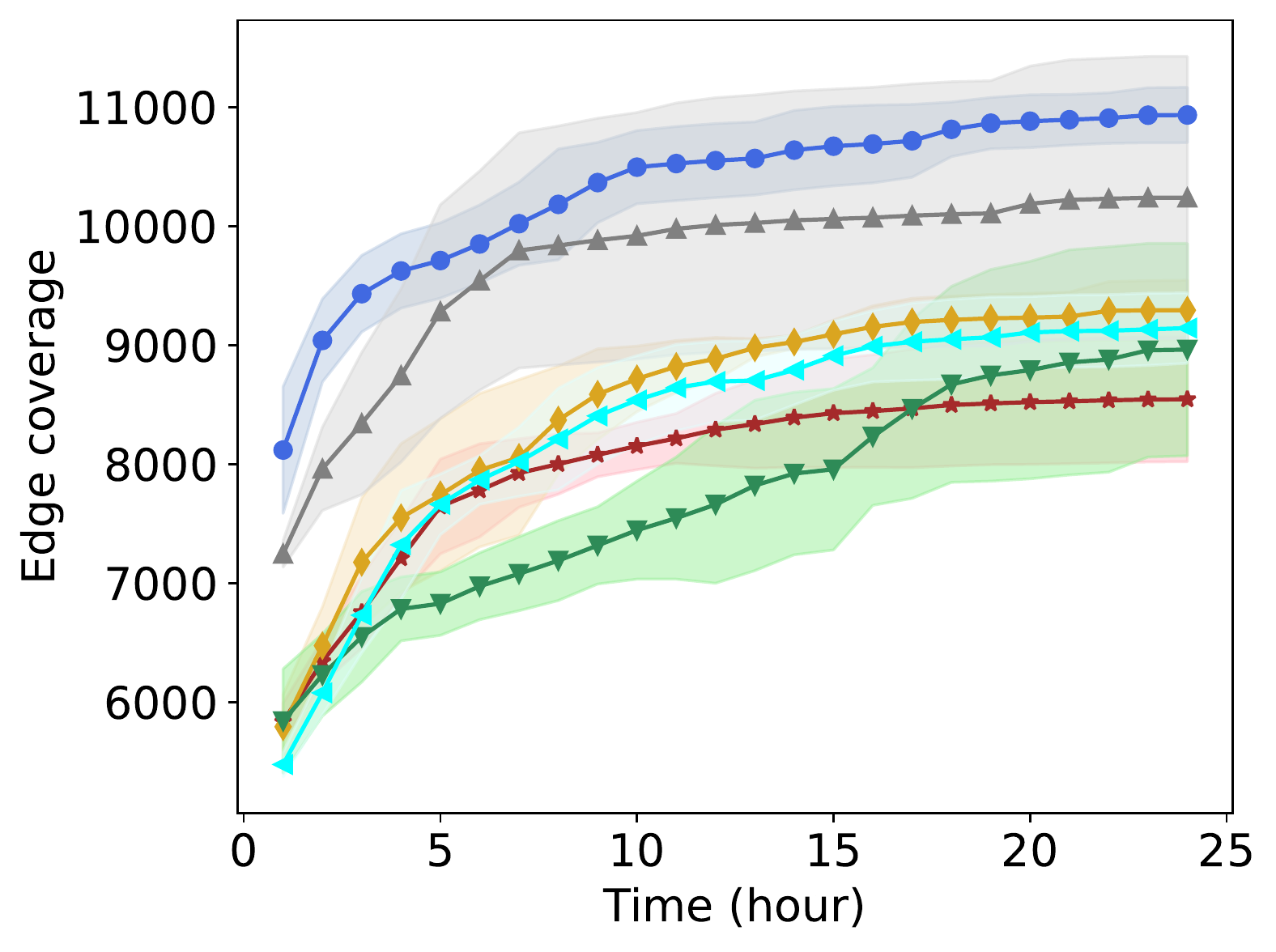}
\label{subfig:afl_libxml}}
\caption{\textbf{\small The \CamReady{arithmetic mean} edge coverage of of AFL-based seed schedulers running for 24 hours and one standard deviation error bars over 10 runs. Default refers to the default seed scheduler in AFL.}}
\label{fig:AFLtime}
\end{figure*}

The coverage plots over time also highlight the promise of \ToolName{}. Figure \ref{fig:libfuzzertime} and \ref{fig:AFLtime} show that \ToolName{} generally maintains its performance advantage during the lifetime of the fuzzing campaign.  
The consistency of \ToolName{}'s gain across many different seed schedulers show the promise of scheduling seeds based on CFG information. Moreover, it suggests \ToolName{} can be helpful independent of a fuzzer as we later explore.

\vspace{0.3cm}
\begin{longfbox}
\textbf{Result 1:} 
\ToolName{} increases feature coverage by 25.89\% compared to Entropic and edge coverage by 4.21\% compared to the next-best AFL-based seed scheduler (RarePath), in arithmetic mean on 12 Google FuzzBench programs. 
\end{longfbox}

\subsection{RQ2: Bug Finding}
In order to detect memory corruption bugs that do not necessarily lead to a crash, we compile program binaries with Address and Undefined Behavior Sanitizers. We then ran \ToolName{}, Default (i.e., AFL's default seed scheduler), RarePath, RareEdge, and NewPath on 12 real-world parsing programs for 24 hours, a total of 10 times. We could not run the Libfuzzer-based seed schedulers because the 12 parsing programs are not equipped with a Libfuzzer-compatible fuzzing harness (i.e., LLVMFuzzerTestOneInput is undefined). 

In our 24 hour runs, we found real-world bugs in binutils. Table \ref{tab:bug} shows the bug count for each seed scheduling strategy in terms of integer overflow, out of memory and memory leak bugs, in arithmetic mean over the 10 runs. 
We count bugs with the following procedure based on prior work~\cite{redqueen, angora, neuzz}. We first use AFL-CMin to reduce the number of crashing inputs. We then further deduplicate the crashing inputs by filtering them by unique stack traces. We lastly triage the remaining crashing inputs by manually reviewing their stack traces and corresponding source code. Our results show that \ToolName{} finds 3 more bugs than the next best seed scheduling strategy SecCov (i.e., TortoiseFuzz), which optimizes for bug-finding. 

\begin{table}[!]
    \centering
    \caption{\textbf{\small{{Tested Programs in Bug Finding Experiments.}}}}
     \begin{tabular}[t]{llcr}
        \toprule
        \textbf{Subjects} & \textbf{Version} & \textbf{Format} & \textbf{\# lines}\\
        \toprule
        xmllint & libxml2-2.9.7 & XML & 72,630 \\
        miniunz & zlib-1.2.11& ZIP &  1,895\\
        readpng & libpng-1.6.37& PNG & 3,205\\
        djpeg & libjpeg-9d & JPEG &  9,204\\
        size & binutils-2.36.1& ELF & 51,203 \\
        readelf -a & binutils-2.36.1& ELF & 29,954 \\
        nm -C & binutils-2.36.1& ELF & 52,763 \\
        objdump -D & binutils-2.36.1&  ELF & 78,610 \\
        strip & binutils-2.36.1 & ELF & 59,680 \\
        tiff2pdf & tiff-4.3.0 & TIFF &  20,387\\
        bsdtar -xf& libarchive-3.5.1 & TAR & 45,031 \\
        infotocap & ncurses-6.2 & TEXT&  23,145\\
        \bottomrule
    \end{tabular}
    \label{tab:studied_programs}
\end{table}

\begin{table}[t!]
    \caption{\small\textbf{Overview of bugs discovered in our AFL-based seed scheduling experiments categorized by type.  }}
    \centering
    \setlength{\tabcolsep}{1pt}
    \renewcommand{\arraystretch}{1.1}
    \begin{tabular}{lrrrrrr}
        \toprule
        \textbf{} & \textbf{\ToolNameShort{}}  & \textbf{Default} & \textbf{RarePath} & \textbf{RareEdge} & \textbf{NewPath} & \textbf{SecCov}  \\ 
        \midrule
        Fuzzer & AFL  & AFL &  
        AflFast & FairFuzz & EcoFuzz & Tortoise$\dagger$ \\ 
        \midrule
        out-of-memory & 21 & 14 &  19 & 17 & 18 & 21  \\
        memory leak & 24 & 20 & 21 & 19 & 20 & 22\\
        integer overflow & 3 & 2 & 3 & 3 & 2 & 2 \\
        \midrule
        Total & 48 & 36 & 43 & 39 & 40 & 45\\
        \bottomrule
    \end{tabular}
    \label{tab:bug}
    \hspace{-5.3cm}
    \footnotesize{$\dagger$ Tortoise denotes TortoiseFuzz.}\\
    \vspace{0.3cm}
\end{table}

\vspace{0.3cm}
\begin{longfbox}
\textbf{Result 2:} \ToolName{} discovers 3 more bugs than the next best seed-scheduling strategy (SecCov). 
\end{longfbox}

\subsection{RQ3: Runtime Overhead}

\CamReady{
In this experiment, we measure the overhead that \ToolName{} adds to a fuzzer. The runtime overhead can be classified into two components: a fuzzer maintenance (i.e., record hit count of edges and compute seeds' energy) and a fuzzer invoking \ToolName{} (i.e., construct edge horizon graph and perform Katz centrality analysis) for seed scheduling. To measure these overheads, we run our modified versions (see Section \ref{implementation}) of AFL and Libfuzzer against all 12 FuzzBench programs for 24 hours, recording the total time they spend in maintenance and separately the total time spent in computing Katz centrality over the edge horizon graph in the standalone process. We repeat this experiment 10 times to minimize variance. Table \ref{tab:libfuzzer_runtime} summarizes the runtime overhead added to AFL's and LibFuzzer fuzzing processes in terms of fuzzer maintenance and graph centrality analysis. }

\CamReady{
The overhead of fuzzer maintenance is $0.28\%$ for AFL and $1.74\%$ for Libfuzzer, in arithmetic mean over the 12 FuzzBench programs. 
The graph analysis overhead is minimal, adding $0.15\%$ in arithmetic mean over the 12 FuzzBench programs. We believe these small graph analysis overheads exist because Katz centrality can be efficiently computed with the power method (Section~\ref{background}) and the edge horizon graph is cached and updated instead of being constructed from scratch each time. 
For clarity, we did not report graph analysis overheads for AFL and Libfuzzer separately because they use the same standalone process, so the overheads were nearly indistinguishable. 
Moreover, the difference in overheads per-program is explained by the variance in the target program's CFG size (i.e., number of nodes).}
\vspace{0.3cm}
\begin{longfbox}
\textbf{Result 3:} \ToolName{} adds at most $1\%$ overhead from graph analysis and at most $2\%$ overhead for fuzzer maintenance.
\end{longfbox}

\subsection{RQ4: Impact of Design Choices}
\begin{table}[!]
    \caption{\small\textbf{Runtime overhead from \ToolName{} in Libfuzzer and AFL-based seed scheduling.}}
    \centering
    \setlength{\tabcolsep}{3pt}
    \renewcommand{\arraystretch}{1.1}
    \begin{tabular}{c|c|c|c|c}
    \toprule
\multirow{2}{*}{\textbf{Programs}} & \multirow{2}{*}{\textbf{Nodes  \#}} & \multirow{2}{*}{\textbf{Graph Analysis}} & \multicolumn{2}{c}{\textbf{Fuzzer Maintenance}} \\ \cline{4-5} 
   &   &   & \multicolumn{1}{c}{\textbf{LibFuzzer}} & \textbf{AFL} \\
        \midrule
        freetype & 38,352 & 0.20\% & 1.71\% & 0.23\% \\
        libxml2 &  96,732 & 0.22\%  & 2.53\%  & 0.39\% \\
        lcms & 13,081 &  0.06\%  &  0.92\% & 0.08\%\\
        harfbuzz & 21,066 & 0.11\% & 2.25\%  & 0.17\% \\
        libjpeg & 16,508 & 0.04\% & 0.79\%  & 0.06\%\\
        libpng & 7,215 &  0.02\% &  0.53\%  &  0.03\%\\
        openssl & 57,729 & 0.25\% &  2.43\% &  0.67\%\\
        openthread &27,263 & 0.09\% & 1.48\% &  0.24\% \\
        re2 & 12,020 &  0.03\% &  1.39\% &  0.26\%\\
        sqlite & 70,703 & 0.75\% &  3.12\% & 0.41\% \\
        vorbis & 9,494 & 0.04\% &  0.80\% & 0.55\%\\
        zlib & 1,882 & 0.02\% & 2.96\% & 0.29\%\\
        \midrule
        Arithmetic mean & 31,004 & 0.15\% &   1.74\% & 0.28\%  \\
        Median & 18,787  & 0.08\% & 1.60\%  & 0.25\% \\
        \bottomrule
    \end{tabular}
    \label{tab:libfuzzer_runtime}
        \vspace{0.3cm}
\end{table}
We conduct experiments to measure the performance effect of five design choices:  (i) centrality measure, (ii) $\mathbf{\beta}$ parameterization, (iii) visited node deletion, (iv) loop removal, and (v) $\alpha$ parameterization.
For each design choice experiment, we run \ToolName{} with Libfuzzer on the 12 Google FuzzBench programs for 1 hour, repeated 10 times, and compare their feature coverage. We run for 1 hour because the first hour of a fuzzing run often discovers more coverage than later hours and hence our results better measure the effect of the design choices. We also choose feature coverage because it provides more fine-grained information about a fuzzer's behavior than edge coverage. We describe each design choice experiment in more detail below. 


\subsubsection{Centrality measure}
We measure the effect of the centrality measure on seed scheduling in this experiment by varying the centrality measure used in \ToolName{}. We compare Eigenvector, Degree, Katz and PageRank centrality measures. Table~\ref{tab:aba_cent} shows the feature coverage results. Enabling Katz centrality improves the feature coverage by $16.54\%$, $23.69\%$, and $19.17\%$ in arithmetic mean over the 12 FuzzBench programs, relative to Pagerank, Eigenvector, and Degree centrality, respectively.  These results experimentally justify our claim from Section \ref{background} that Katz centrality is most desirable for seed scheduling. However, these results also show that for some programs, other forms of centrality are a better fit such as the superior performance of Pagerank on \texttt{re2} and Degree on \texttt{vorbis}.

\begin{table}[!]
    \caption{\small\textbf{\CamReady{Arithmetic mean} feature coverage of \ToolName{} with different centrality metrics.}}
    \centering
    \setlength{\tabcolsep}{1.4pt}
    \renewcommand{\arraystretch}{1.1}
    \begin{tabular}{lrrrr}
        \toprule
        \textbf{Programs} & \textbf{Katz}  & \textbf{Pagerank} & \textbf{Eigenvector} &\textbf{Degree}  \\ 
        \midrule
        freetype & \textbf{51,184} & 44,394 & 40,723 & 38,332 \\
        libxml2 &  \textbf{39,240} & 29,575  & 28,473  & 28,014 \\
        lcms & \textbf{2,886} & 2,071   & 1,557   & 2,054  \\
        harfbuzz & \textbf{35,017} & 28,563 & 26,253  & 27,485 \\
        libjpeg & \textbf{10,974} &  9,250  & 10,454  & 8,713 \\
        libpng & \textbf{5,001} &  4,804 &  4,505  &  4,923 \\
        openssl & \textbf{14,520} & 13,035 &  13,385 &  13,555 \\
        openthread & \textbf{6,525} & 5,201 & 5,380 &  5,298 \\
        re2 & 31,292 &  \textbf{32,309} &  29,648 &  29,595 \\
        sqlite & \textbf{73,532} & 68,328 &  65,538 & 63,997 \\
        vorbis & 9,106 & 8,129 &  7,470 & \textbf{9,363} \\
        zlib & \textbf{2,711} & 2,410 & 2,323 & 2,404 \\
        \midrule
        \multicolumn{2}{c|}{Arithmetic mean coverage gain} & 16.54\% &  23.69\% & 19.17\%  \\
        \multicolumn{2}{c|}{Median coverage gain} & 13.89\% & 18.99\% & 19.03\%  \\
        \bottomrule
    \end{tabular}
    \label{tab:aba_cent}
        \vspace{0.3cm}
\end{table}
\subsubsection{$\mathbf{\beta}$ parameterization}
In Section \ref{methodology}, we describe how we set $\mathbf{\beta}$ based on historical mutation data. In this comparison,
we see the effect of this technique by comparing \ToolName{} with uniform $\mathbf{\beta}$ against \ToolName{} with non-uniform $\mathbf{\beta}$. Table~\ref{tab:aba_beta} shows the feature coverage results. The non-uniform $\mathbf{\beta}$ technique increases feature coverage by $24.19\%$ in arithmetic mean over the 12 FuzzBench programs. These results show the utility of biasing $\mathbf{\beta}$.
\begin{table}[!]
    \caption{\small\textbf{\CamReady{Arithmetic mean} feature coverage from analyzing the effect of non-uniform $\mathbf{\beta}$.}}
    \centering
    \renewcommand{\arraystretch}{1.1}
    \begin{tabular}{lrr}
        \toprule
        \textbf{Programs} & \textbf{Non-uniform $\mathbf{\beta}$}  & \textbf{Uniform $\mathbf{\beta}$}  \\ 
        \midrule
        freetype & \textbf{51,184} & 40,396  \\
        libxml2 &  \textbf{39,240} & 31,733 \\
        lcms & \textbf{2,886} & 1,506     \\
        harfbuzz & \textbf{35,017} & 29,380 \\
        libjpeg & \textbf{10,974} &  8,834  \\
        libpng & \textbf{5,001} &  4,761  \\
        openssl & \textbf{14,520} & 12,542  \\
        openthread & \textbf{6,525} & 5,271  \\
        re2 & \textbf{31,292} &  28,263  \\
        sqlite & \textbf{73,532} & 64,893  \\
        vorbis & \textbf{9,106} & 7,679 \\
        zlib & \textbf{2,711} & 2,305 \\
        \midrule
        \multicolumn{2}{c|}{Arithmetic mean coverage gain} & 24.19\%\\
        \multicolumn{2}{c|}{Median coverage gain} & 18.88\%\\

        \bottomrule
    \end{tabular}
    \label{tab:aba_beta}
\end{table}
\begin{table}[t]
    \caption{\small\textbf{\CamReady{Arithmetic mean} feature coverage from analyzing the effect of $\alpha$.}}
    \centering
    \renewcommand{\arraystretch}{1.1}
    \begin{tabular}{lrrrr}
        \toprule
        \textbf{Programs} & \textbf{0.5} & \textbf{0.25} &\textbf{ 0.75} &\textbf{1}  \\ 
        \midrule
        freetype  & \textbf{51,184}& 38,369 & 41,777 & 40,723  \\
        libxml2 & \textbf{39,240}  & 28,644& 29,992  & 28,473   \\
        lcms   &\textbf{2,886}&  1,313 & 1,552   & 1,557    \\
        harfbuzz  & \textbf{35,017}&  27,250  & 28,276 & 26,253   \\
        libjpeg  & \textbf{10,974}& 9,542 &  10,336  & 10,454  \\
        libpng &\textbf{5,001}&  4,913  &  4,929 &  4,505  \\
        openssl& \textbf{14,520} &  13,420  & 13,302 &  13,385  \\
        openthread & \textbf{6,525} & 6,216 & 5,597 & 5,380 \\
        re2  & 31,292 &  29,590  &  \textbf{31,885} &  29,648 \\
        sqlite & \textbf{73,532} & 64,175 & 68,550 &  65,538 \\
        vorbis & \textbf{9,106}& 8,092   & 8,066 &  7,470  \\
        zlib  &\textbf{2,711}&  2,378 & 2,282 & 2,323\\
        \midrule
        \multicolumn{2}{c|}{Arithmetic mean coverage gain} & 24.53\% & 19.47\% &   23.69\%  \\
        \multicolumn{2}{c|}{Median coverage gain} & 14.29\%  & 14.74\% & 18.99\%   \\
        \bottomrule
    \end{tabular}
    \label{tab:aba_alpha}
\end{table}
\begin{table}[!]
    \caption{\small\textbf{\CamReady{Arithmetic mean} feature coverage from analyzing the effect of loop removal.}}
    \centering
    \renewcommand{\arraystretch}{1.1}
    \begin{tabular}{lrr}
        \toprule
        \textbf{Programs} & \textbf{loop removal}  & \textbf{no loop removal}  \\ 
        \midrule
        freetype & \textbf{51,184} & 38,646  \\
        libxml2 &  \textbf{39,240} & 28,737 \\
        lcms & \textbf{2,886} & 1,455     \\
        harfbuzz & \textbf{35,017} & 28,849 \\
        libjpeg & \textbf{10,974} &  10,142  \\
        libpng & \textbf{5,001} &  4,846  \\
        openssl & \textbf{14,520} & 13,300  \\
        openthread & \textbf{6,525} & 5,430  \\
        re2 & 31,292 &  \textbf{31,609}  \\
        sqlite & \textbf{73,532} & 64,560  \\
        vorbis &9,106 & \textbf{9,350}  \\
        zlib & \textbf{2,711} & 2,247 \\
        \midrule
        \multicolumn{2}{c|}{Arithmetic mean coverage gain} & 21.70\%\\
        \multicolumn{2}{c|}{Median coverage gain} & 17.03\%\\
        \bottomrule
    \end{tabular}
    \label{tab:aba_loop}
        \vspace{0.3cm}
\end{table}

\subsubsection{Visited node deletion}
In Section \ref{methodology}, we describe why we remove visited nodes from the edge horizon graph. In this comparison, we experimentally justify this choice. 
We compare \ToolName{} with visited node deletions from the edge horizon graph against \ToolName{} with no deletions from the edge horizon graph. 
Table \ref{tab:aba_dyn} shows the feature coverage results. The deleted edge horizon graph improves feature coverage by $24.13\%$ in arithmetic mean over the 12 FuzzBench programs. Therefore, this result justifies our deletion of visited nodes. 
\begin{table}[!]
    \caption{\small\textbf{\CamReady{Arithmetic mean} feature coverage from analyzing the effect of deleting visited nodes.}}
    \centering
    \renewcommand{\arraystretch}{1.1}
    \begin{tabular}{lrr}
        \toprule
        \textbf{Programs} & \textbf{Original}  & \textbf{Deleted}  \\ 
        \midrule
        freetype & \textbf{51,184} & 39,892  \\
        libxml2 &  \textbf{39,240} & 28,973 \\
        lcms & \textbf{2,886} & 1,493     \\
        harfbuzz & \textbf{35,017} & 24,667 \\
        libjpeg & \textbf{10,974} &  9,715  \\
        libpng & \textbf{5,001} &  4,827  \\
        openssl & \textbf{14,520} & 13,121  \\
        openthread & \textbf{6,525} & 5,712  \\
        re2 & \textbf{31,292} &  29,408  \\
        sqlite & \textbf{73,532} & 61,609  \\
        vorbis & \textbf{9,106} & 8,020  \\
        zlib & \textbf{2,711} & 2,470 \\
        \midrule
        \multicolumn{2}{c|}{Arithmetic mean coverage gain} & 24.13\%\\
        \multicolumn{2}{c|}{Median coverage gain} & 13.89\%\\
        \bottomrule
    \end{tabular}
    \label{tab:aba_dyn}
        \vspace{0.3cm}
\end{table}

\subsubsection{Loop Removal}
In Section \ref{methodology}, we introduce our loop removal transform as a technique to mitigate the effects of loops on computing centrality. In this experiment, we measure this effect by comparing \ToolName{} with and without the loop removal transform. Table~\ref{tab:aba_loop} shows that the loop removal transform improves edge coverage by $21.70\%$ in arithmetic mean over the 12 FuzzBench programs, justifying our loop removal transform. 


\subsubsection{$\alpha$ parameterization}
In this design choice experiment, we study how the choice of $\alpha$ affects the \ToolName{}'s performance.
Table \ref{tab:aba_alpha} summarizes our findings.  As described in Section \ref{methodology}, $\alpha=1$ treats far and close paths with equal contribution to centrality and its experimental results are worse compared to distinguishing them, showing the utility of the multipicative decay effect. We note that $\alpha=1$ is equivalent to Eigenvector centrality as seen by comparing the relevant column from Table~\ref{tab:aba_cent}. Given $\alpha=0.5$ performs best in arithmetic mean over the 12 FuzzBench programs, we pick it in our current implementation.

\vspace{0.3cm}
\begin{longfbox}
\textbf{Result 4:} 
Our results empirically support \ToolName{}'s design choices.
\end{longfbox}

\subsection{RQ5: Utility for non-evolutionary input generation}
In this experiment, we show the promise of \ToolName{} in non-fuzzing settings, we integrate \ToolName{} into concolic execution seed scheduling. 
Concolic execution is known to incur high overhead~\cite{qsym, poeplau2020symcc} during path constraint collection and solving. Hence, in concolic execution, scheduling promising seeds is crucial to its performance~\cite{meuzz, Zhao2019SendHP}. 
To perform this experiment, we use the concolic executor from QSYM's latest version~\cite{qsym}. QSYM, a hybrid fuzzer, consists of three components, a concolic executor, a fuzzer, and a coordinator that schedules seeds for the concolic executor. Since our goal is to show the utility of \ToolName{} for concolic execution seed scheduling, we disabled QSYM's fuzzer and only modified its coordinator’s seed scheduling algorithm to use \ToolName{}. \CamReady{We did not modify QSYM's concolic executor logic.}
We evaluate on the $3$ programs (\texttt{size}, \texttt{libarchive} and \texttt{tcpdump}). 
Note we did not run on SymCC because SymCC and QSYM have the same concolic execution scheduler~\cite{poeplau2020symcc}, so comparing against one is sufficient. We run \ToolName{} against the default seed scheduler in QSYM on the $3$ real world programs for $24$ hours and compare the total edge coverage. In arithmetic mean over the 10 runs, Table ~\ref{tab:qsym} shows that \ToolName{} improves edge coverage by $35.76\%$, in arithmetic mean over the 3 programs. Hence, this shows the potential promise \ToolName{} for seed scheduling in non-evolutionary fuzzing settings. 
\begin{table}[!]
    \caption{\small\textbf{Edge coverage of concolic-execution-based seed scheduling on 3 real-world programs for 24 hours over 5 runs. }}
    \centering
    \renewcommand{\arraystretch}{1.1}
    \begin{tabular}{lcc}
        \toprule
        \textbf{Scheduling} & \textbf{\ToolName{}}  & \textbf{Default}  \\ 
        \midrule
        libarchive &  \textbf{3,886} & 3,230    \\
        size &  \textbf{3,068} & 2,602   \\
        tcpdump & \textbf{3,552} & 2,101  \\        
        \midrule
        \multicolumn{2}{c|}{Arithmetic mean coverage gain} & 35.76\%\\
        \multicolumn{2}{c|}{Median coverage gain} & 20.31\%\\
        \bottomrule
    \end{tabular}
    \label{tab:qsym}
        \vspace{0.3cm}
\end{table}
However, we note that our results are preliminary and are inconclusive. We leave a detailed evaluation to future work.  

\vspace{0.3cm}
\begin{longfbox}
\textbf{Result 5:} \ToolName{} increases edge coverage by $35.76\%$, in arithmetic mean over 3 programs, compared to QSYM's default seed scheduling strategy. 
\end{longfbox}

%% file: related.tex
\section{Related Work}
\label{related_work}
\subsection{Graph Centrality}
Centrality is a commonly used measure in graph analysis. Researchers have proposed various centrality metrics including degree centrality~\cite{Shaw1954SomeEO}, semi-local centrality~\cite{CHEN20121777}, closeness centrality~\cite{Sabidussi1966TheCI}, betweenness centrality~\cite{between}, eigenvector centrality~\cite{STEPHENSON19891}, Katz centrality~\cite{Katz1953}, and PageRank~\cite{ilprints361}. These centrality measures has been applied to various fields such as social network analysis~\cite{gomez2013modeling, brown2008influencer}, biology~\cite{biocent}, finance~\cite{ruiz2012correlating} and geography~\cite{crucitti2006centrality}. To the best of our knowledge, we are the first to use centrality for seed selection in fuzzing. 

\subsection{Seed Scheduling}
While prior work has proposed a wide range of techniques to improve fuzzing such as symbolic execution~\cite{klee, dart, Godefroid2008AutomatedWF, angr, qsym, tfuzz, savior, stephens2016driller}, dynamic taint analysis~\cite{taintscope, angora, buzzfuzz, Gan2020GREYONEDF, vuzzer} and machine learning~\cite{Godefroid2017LearnFuzzML, neuzz, fuzzguard}, in this paper we focus on improving the seed scheduling component in a fuzzer. We describe prior work that has focused on improving fuzzing through seed scheduling. Seed scheduling consists of two main components: input prioritization~\cite{Wang2020NotAC, sensitive, Wang2021ReinforcementLH} and the input's corresponding mutation budget (i.e., power schedule)~\cite{entropic, aflfast}. Prior seed scheduling work has prioritized seeds based on edge or path coverage~\cite{lemieux2017fairfuzz,aflfast,entropic,ecofuzz} as well as more security-sensitive metrics such as execution time~\cite{slowfuzz, perffuzz}, exploitability~\cite{woo13}, memory accesses~\cite{memfuzz, memlock, Wang2020NotAC}, or a combination of them~\cite{sensitive, Wang2021ReinforcementLH}
Another line of work prioritizes seeds based on call graphs \cite{cerebro}. In contrast, we prioritize seeds based on the entire inter-procedural CFG. While AFLGo~\cite{aflgo} also uses the entire inter-procedural CFG, it computes the distance over the CFG for directed fuzzing and assigning a seed's mutation budget. In contrast, we approximate the count of reachable and feasible edges from a seed and use it for coverage-guided fuzzing. \texttt{SAVIOR}~\cite{savior} also approximates this count but uses it for bug-driven hybrid testing. Its approximation assumes all edges are equally likely to be reachable and feasible, independent of their distance from a seed's execution path, which does not hold true for many real-world programs as we showed in Section \ref{evaluation}. In contrast, we use the multiplicative decay property of Katz centrality to reflect this behavior in real-world programs and better approximate this count. Moreover, \texttt{SAVIOR}~\cite{savior}'s approximation is equivalent to setting $\alpha=1$ (i.e, no multiplicative decay) and our design choice experiments show this approximation performs worse than \ToolName{}'s default settings. Nonetheless, both \ToolName{} and \texttt{SAVIOR} utilize the mutation history information to improve their approximation.
Seed scheduling has also been a topic in other program testing techniques aside from fuzzing such as concolic execution~\cite{meuzz, Zhao2019SendHP}. Our preliminary experiments suggest that \ToolName{} can improve seed scheduling for concolic execution. 

%% file: conclusion.tex
\section{conclusion}
In this paper, we introduce a new approach to seed scheduling based on centrality analysis of seeds on the CFG. Centrality measures have several desirable properties that make them a natural fit for the seed scheduling problem. We implement our approach in \ToolName{} and show its effectiveness in seed scheduling: increasing feature coverage by 25.89\% compared to Entropic and edge coverage by 4.21\% compared to the next-best AFL-based seed scheduler, in arithmetic mean on 12 Google FuzzBench programs. 


%% file: appendix.tex
\newpage
\appendix
\subsection{Mann-Whitney U Test Results}
\begin{table}[h]
    \caption{\small\textbf{Mann-Whitney U test results over the feature and edge 
coverage of Libfuzzer-based seed schedulers on 12 FuzzBench programs for 1 hour over 10 runs \CamReady{(corresponding to Table \ref{tab:libfuzzer_cov_1h})}.}}
    \centering
    \setlength{\tabcolsep}{4pt}
    \renewcommand{\arraystretch}{1.1}
   \begin{tabular}{lcc|cc}
   \toprule
    \multicolumn{1}{c}{\multirow{2}{*}{\textbf{Programs}}}  & \multicolumn{2}{c}{\textbf{Entropic}} & \multicolumn{2}{c}{\textbf{Default}} \\ \cline{2-5} 
  & \multicolumn{1}{c}{\textbf {feature}} & \textbf{edge}  & \multicolumn{1}{c}{\textbf {feature}} & \textbf{edge}    \\
        \midrule
        freetype &  4.40E-4  &1.62E-2 &   7.69E-4 & 1.71E-3  \\
        libxml2 &  1.83E-4  & 1.82E-4 & 1.83E-4 & 1.83E-4   \\
        lcms & 3.61E-3 & 1.31E-3 &  1.83E-4  & 1.83E-4     \\
        harfbuzz & 1.82E-4 & 1.83E-4 & 1.83E-4 & 1.82E-4    \\
        libjpeg & 1.83E-4 & 1.82E-4 & 1.83E-4 &  1.82E-4 \\
        libpng & 1.82E-4 & 1.68E-4 & 1.81E-4 & 1.67E-4   \\
        openssl & 1.83E-4 & 1.82E-4 & 1.83E-4 & 1.82E-4   \\
        openthread & 1.83E-4 & 2.19E-3 & 1.83E-4 & 1.83E-4  \\
        re2 & 2.46E-4 & 3.28E-4 & 1.71E-3 & 2.47E-3   \\
        sqlite & 1.83E-4 & 1.83E-4 & 1.83E-4 & 1.73E-2  \\
        vorbis & 4.40E-4 & 7.69E-4 & 2.46E-4 & 2.46E-4  \\
        zlib & 8.90E-2 & 6.72E-2 & 1.31E-3 & 6.13E-2  \\        \bottomrule
    \end{tabular}
    \label{tab:libfuzzer_covp_1h}
\end{table}
\begin{table}[h]
    \caption{\small\textbf{Mann-Whitney U test results over the fuzzer and edge 
coverage of Libfuzzer-based seed schedulers on 12 FuzzBench programs for 24 hours over 10 runs \CamReady{(corresponding to Table \ref{tab:libfuzzer_cov})}.}}
    \centering
    \setlength{\tabcolsep}{4pt}
    \renewcommand{\arraystretch}{1.1}
   \begin{tabular}{lcc|cc}
   \toprule
    \multicolumn{1}{c}{\multirow{2}{*}{\textbf{Programs}}}  & \multicolumn{2}{c}{\textbf{Entropic}} & \multicolumn{2}{c}{\textbf{Default}} \\ \cline{2-5} 
  & \multicolumn{1}{c}{\textbf {feature}} & \textbf{edge}  & \multicolumn{1}{c}{\textbf {feature}} & \textbf{edge}    \\
        \midrule
        freetype &  1.70E-3  &7.56E-2 &   2.12E-1 & 3.12E-2  \\
        libxml2 &  1.83E-4  & 1.83E-4 & 1.83E-4 & 1.83E-4   \\
        lcms & 3.61E-3 & 9.11E-3 &  2.20E-3  & 3.61E-3     \\
        harfbuzz & 1.83E-4 & 1.82E-4 & 1.83E-4 & 1.82E-4    \\
        libjpeg & 1.83E-4 & 2.45E-4 & 1.83E-4 &  1.82E-4 \\
        libpng & 1.31E-3 & 2.89E-4 & 7.58E-4 & 2.74E-4   \\
        openssl & 1.82E-4 & 1.82E-4 & 1.83E-4 & 1.80E-4   \\
        openthread & 1.83E-4 & 1.83E-4 & 1.83E-4 & 1.83E-4  \\
        re2 & 3.30E-4 & 3.17E-3 & 7.65E-4 & 3.60E-3   \\
        sqlite & 1.83E-4 & 1.01E-3 & 1.31E-3 & 3.76E-2  \\
        vorbis & 1.83E-4 & 2.40E-4 & 1.83E-4 & 4.33E-4  \\
        zlib & 2.19E-3 & 5.65E-3 & 1.82E-4 & 3.84E-3  \\        \bottomrule
    \end{tabular}
    \label{tab:libfuzzer_covp}
\end{table}
\begin{table}[H]
    \caption{\small\textbf{Mann-Whitney U test results over the fuzzer and edge coverage of AFL-based seed schedulers on 12 FuzzBench programs for 1 hour over 10 runs \CamReady{(corresponding to Table \ref{tab:afl_cov_1h})}. }}
    \centering
    \setlength{\tabcolsep}{4pt}
    \renewcommand{\arraystretch}{1.1}
    \begin{tabular}{lrrrrr}
        \toprule
        \textbf{} &  \textbf{Default} & \textbf{RarePath} & \textbf{RareEdge} & \textbf{NewPath} &\textbf{SecCov}  \\ 
        \midrule
        Fuzzer   & AFL &  AflFast & FairFuzz & EcoFuzz & TortoiseFuzz \\ 
        \midrule
        freetype &   2.16E-3 & 2.16E-3 & 2.16E-3 & 2.16E-3 & 2.16E-3\\
        libxml2 &  2.16E-3 & 2.16E-3 & 2.16E-3 & 2.16E-3 & 2.16E-3  \\
        lcms &  8.18E-2  & 1.99E-2  & 1.52E-3  & 4.33E-4 & 9.31E-3 \\
        harfbuzz  &  2.16E-3  & 2.47E-2 & 2.60E-2 & 2.16E-3 & 8.13E-3 \\
        libjpeg  & 5.75E-2 & 6.87E-2 & 6.46E-3  & 4.99E-4 &  2.01E-3 \\
        libpng & 8.86E-2 &  8.85E-2  & 1.71E-2  & 1.71E-2 &  6.10E-2  \\
        openssl  & 1.14E-2 & 2.86E-3 & 9.52E-4 & 9.52E-4 & 9.52E-4  \\
        openthread  & 2.00E-2 & 1.14E-2 & 1.14E-2  & 3.81E-3 & 6.63E-3  \\
        re2  &  8.67E-3  & 9.31E-2 &  2.16E-3  & 2.16E-3 & 2.16E-3  \\
        sqlite & 5.89E-2 & 1.01E-1 & 3.10E-2 & 3.10E-2 & 3.94E-2 \\
        vorbis  & 2.45E-2 & 8.14E-3 & 6.63E-3  & 9.52E-4 & 1.14E-2 \\
        zlib & 8.82E-2 & 4.65E-2  & 1.99E-2 & 2.58E-3 & 3.34E-2 \\
        \bottomrule
    \end{tabular}
    \label{tab:afl_p_1h}
\end{table}
\begin{table}[h]
    \caption{\small\textbf{Mann-Whitney U test results over the fuzzer and edge coverage of AFL-based seed schedulers on 12 FuzzBench programs for 24 hours over 10 runs \CamReady{(corresponding to Table \ref{tab:afl_cov})}. }}
    \centering
    \setlength{\tabcolsep}{4pt}
    \renewcommand{\arraystretch}{1.1}
    \begin{tabular}{lrrrrr}
        \toprule
        \textbf{} &  \textbf{Default} & \textbf{RarePath} & \textbf{RareEdge} & \textbf{NewPath} &\textbf{SecCov}  \\ 
        \midrule
        Fuzzer   & AFL &  AflFast & FairFuzz & EcoFuzz & TortoiseFuzz \\ 
        \midrule
        freetype &  5.89E-2 & 8.18E-2 & 4.85E-2 & 2.16E-4 & 6.49E-3\\
        libxml2 &  2.16E-3 & 2.16E-3 & 1.80E-2 & 2.16E-3 & 2.16E-3  \\
        lcms &  3.10E-2  & 2.41E-2  & 6.49E-3  & 2.16E-4 & 2.41E-2 \\
        harfbuzz  & 1.32E-2  & 5.89E-2 & 2.16E-4 & 6.49E-3 & 1.52E-3 \\
        libjpeg  & 3.91E-2 & 8.20E-2 & 6.51E-3  & 2.16E-4 &  2.01E-3 \\
        libpng & 7.70E-2 &  1.12E-1  & 5.35E-3  & 2.39E-3 &  1.30E-2  \\
        openssl  & 3.43E-2 & 1.14E-2 & 1.14E-2 & 9.52E-4 & 9.52E-4  \\
        openthread  & 4.86E-2 & 2.87E-3 & 2.57E-2  & 9.52E-4 & 1.14E-2  \\
        re2  &  3.94E-2  & 1.09E-2 &  2.60E-3  & 1.29E-3 & 8.65E-4  \\
        sqlite & 6.99E-2 & 8.18E-2 & 1.51E-3 & 9.37E-2 & 3.94E-2 \\
        vorbis  & 2.01E-2 & 2.85E-3 & 1.87E-3  & 1.65E-2 & 3.92E-2 \\
        zlib & 9.35E-2 & 2.40E-2  & 1.34E-2 & 5.79E-2 & 1.93E-2 \\
        \bottomrule
    \end{tabular}
    \label{tab:afl_p_24h}
\end{table}

\subsection{Further-away Edges Are Harder to Reach by Mutations}
\label{appd:far}
\begin{figure}[h]
\centering
\includegraphics[width=0.75\columnwidth]{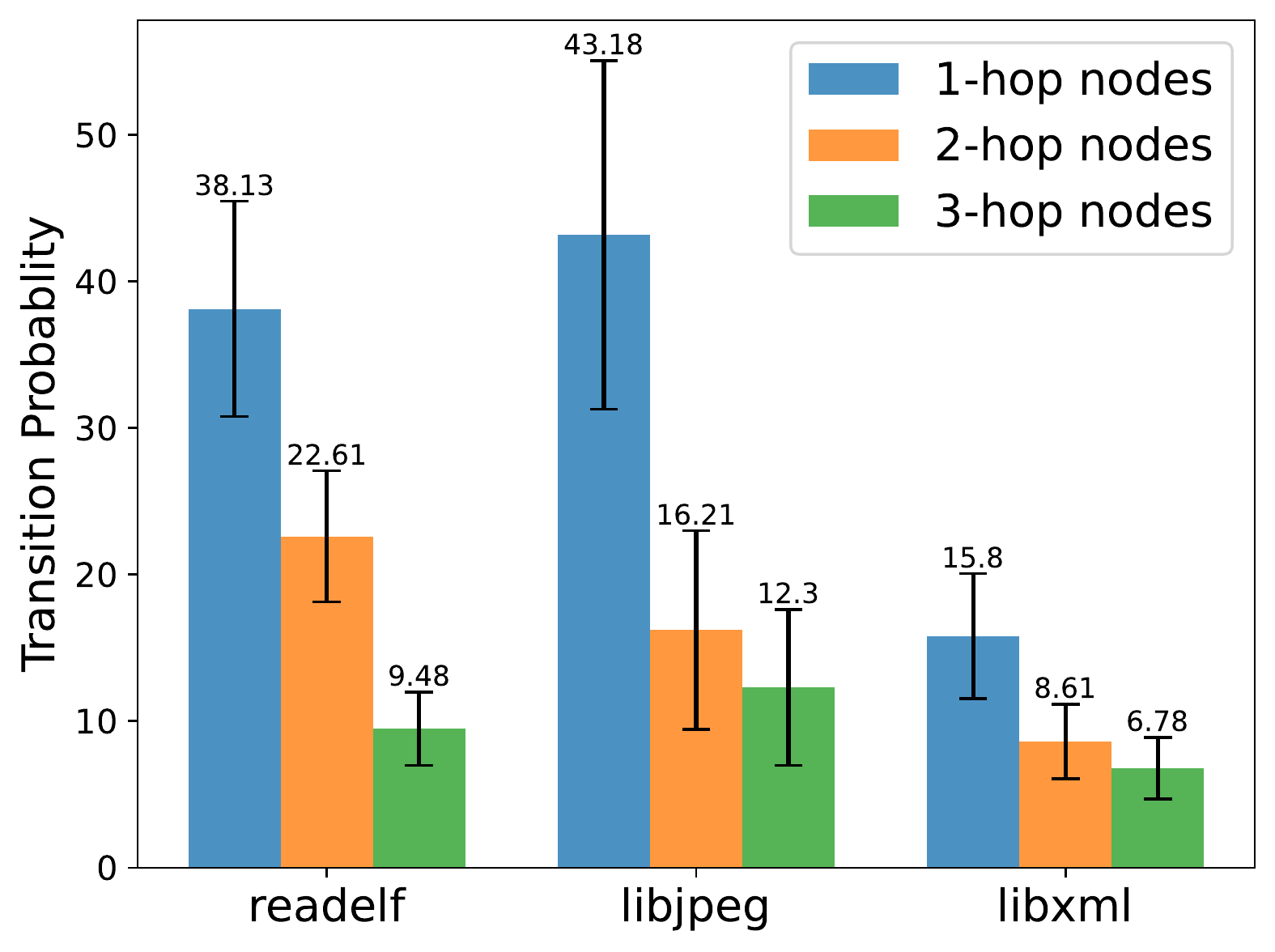}
\caption{\textbf{\small The transition probability on 3 real-world programs using 10 seeds with 10,000 mutations per seed. The 1-hop transition probability indicates the normalized amount of mutations that reached an edge of distance 1 from the current execution path, and similarly for 2-hop and 3-hop. }}
\label{fig:assumption}
\end{figure}

We run an experiment verifying our observation that further away edges in programs are harder to reach by mutations. In Section \ref{methodology}, we claimed that further away edges are harder to reach by mutations. This program property justified Katz centrality, which decays the contribution from further out edges.
To validate this claim, we measure the likelihood that a seed mutation will reach further-away edges on 3 real-world programs. For each program, we choose $10$ seeds and mutate each seed 10,000 times. We repeat this process 10 times to minimize variance. 
Figure~\ref{fig:assumption} shows the result, where n-hop indicates distance n from the original seed's execution path. This  experimentally shows that fewer mutations will reach farther away edges and hence further-away edges are harder to reach by mutations.

\subsection{\ToolName{}'s Approximation Accuracy}
In this section, we run an experiment to show the accuracy of Katz centrality in approximating the count of reachable and feasible edges. In Section \ref{methodology}, we claimed that an ideal seed scheduling strategy would prioritize seeds based on the count of all reachable and feasible edges from a seed by mutations. To better support this claim, we measure how much agreement exists between \ToolName{}'s centrality-based ranking with this ideal seed scheduler's ranking. We simulate the ideal seed scheduler's ranking by computing each CFG node's count of reachable and feasible edges based on graph traversal and covered edges (i.e., feasible) from 24 hour runs of Libfuzzer with \ToolName{} over all 12 FuzzBench programs, repeated 10 times.
We then use the Kendall tau independence test to measure the agreement between two rankings with a value between [-1, 1] and  report if the measured agreement is statistically significant. We note this Kendall tau independence test and its p-values are entirely separate from the Mann Whitney U test and its p-values from our edge coverage experiments. 

Table \ref{tab:aba_ranking} shows the results from the Kendall tau independence test. The absolute values of the correlation are expectedly small given the large size of the ranking lists (on the order of thousands). \ToolName{}'s centrality-based rankings and the ideal strategy's ranking strongly agree on 10 of the 12 programs (i.e., positive correlation values). On 8 of these 10 programs, this agreement is statistically significant with a significance level of $0.05$. This agreement suggests that \ToolName{}'s increased performance in our edge coverage experiments derives from approximating this ideal seed scheduling strategy and that improved approximations would lead to better seed scheduling strategies.

\begin{table}[!]
    \caption{\small\textbf{Using the Kendall tau independence test to measure the agreement between \ToolName{}'s per node rankings with the ideal seed scheduling ranking (i.e., the count of all reachable and feasible edges from a node). The correlation score ranges between $[-1, 1]$, with higher values indicating stronger agreement. Given the the absolute value of the correlation is small due to the large size of the ranking list (i.e., thousands of nodes), we also report the p-value and statistical significance under a 0.05 significance level. }}
    \centering
    \renewcommand{\arraystretch}{1.1}
    \begin{tabular}{cccc}
    \toprule
\multirow{2}{*}{\textbf{Programs}} & \multirow{2}{*}{\textbf{Correlation}} & \multirow{2}{*}{\textbf{p-value}} & \textbf{Statistical}                  \\  &   && \textbf{Significance} \\
        \midrule
        freetype & 0.01 & 8.9E-1  & $\times$ \\
        libxml2 &  \textbf{0.03} & 1.33E-58 &  \checkmark \\
        lcms & \textbf{0.06} & 4.10E-24 &  \checkmark \\
        harfbuzz & \textbf{0.09} &  4.58E-80 & \checkmark \\
        libjpeg & -0.03 &  4.72E-9 & \checkmark \\
        libpng & \textbf{0.05} &  2.24E-9 & \checkmark \\
        openssl & 0.01 & 3.4E-1 & $\times$\\
        openthread & -0.01 & 1.12E-5 &\checkmark \\
        re2 & \textbf{0.01} &  2.01E-2 & \checkmark \\
        sqlite & \textbf{0.06} & 4.24E-107 &\checkmark\\
        vorbis & \textbf{0.04} & 3.47E-6 & \checkmark \\
        zlib & \textbf{0.07} & 1.92E-5 & \checkmark \\
        \bottomrule
    \end{tabular}
    \label{tab:aba_ranking}
\end{table}
\subsection{Limitations}
\ToolName{} does not currently handle indirect function calls. We plan to handle them with static analysis techniques (e.g., Andersen’s points-to analysis) similar to prior work~\cite{savior}. Such a static analysis may produce imprecise CFGs which can affect the utility of a seed's centrality score for seed selection. However, \ToolName{} can mitigate the effects of imprecise CFGs on centrality by reducing the contributions from further away nodes (i.e. nodes in callee functions). Therefore, we believe \ToolName{} will still provide useful guidance despite the imprecision of the CFG. We also envision using $\beta$ for specific CFG nodes (i.e., nodes with indirect function calls) to further mitigate the effects of imprecise CFGs on centrality. We leave this to future work.